\documentclass[12pt,a4paper]{article}

\usepackage{ifthen} 
\newboolean{pdflatex}
\setboolean{pdflatex}{true} 

\newboolean{articletitles}
\setboolean{articletitles}{true} 

\newboolean{uprightparticles}
\setboolean{uprightparticles}{false} 

\newboolean{inbibliography}
\setboolean{inbibliography}{true} 

\usepackage[top=1in, bottom=1.25in, left=1in, right=1in]{geometry}

\usepackage{microtype}
\usepackage{lineno}  
\usepackage{xspace} 
\usepackage{caption} 

\usepackage{booktabs}

\usepackage{graphicx}  
\usepackage{grffile}
\usepackage{color}
\usepackage{colortbl}
\graphicspath{{./figs/}} 

\usepackage{amsmath} 
\usepackage{amssymb}
\usepackage{amsfonts}
\usepackage{upgreek} 

\newcommand*\patchAmsMathEnvironmentForLineno[1]{%
\expandafter\let\csname old#1\expandafter\endcsname\csname #1\endcsname
\expandafter\let\csname oldend#1\expandafter\endcsname\csname
end#1\endcsname
 \renewenvironment{#1}%
   {\linenomath\csname old#1\endcsname}%
   {\csname oldend#1\endcsname\endlinenomath}%
}
\newcommand*\patchBothAmsMathEnvironmentsForLineno[1]{%
  \patchAmsMathEnvironmentForLineno{#1}%
  \patchAmsMathEnvironmentForLineno{#1*}%
}
\AtBeginDocument{%
\patchBothAmsMathEnvironmentsForLineno{equation}%
\patchBothAmsMathEnvironmentsForLineno{align}%
\patchBothAmsMathEnvironmentsForLineno{flalign}%
\patchBothAmsMathEnvironmentsForLineno{alignat}%
\patchBothAmsMathEnvironmentsForLineno{gather}%
\patchBothAmsMathEnvironmentsForLineno{multline}%
\patchBothAmsMathEnvironmentsForLineno{eqnarray}%
}

\usepackage{hyperref}    
\usepackage[all]{hypcap} 

\usepackage{xspace} 
\usepackage{upgreek}

\def\lhcb {\mbox{LHCb}\xspace}

\def\MagUp {\mbox{\em Mag\kern -0.05em Up}\xspace}

\ifthenelse{\boolean{uprightparticles}}%
{

 \def\Ppsi        {\ensuremath{\uppsi}\xspace}

 \def\PDelta      {\ensuremath{\Delta}\xspace}                 
 \def\PXi      {\ensuremath{\Xi}\xspace}                 
 \def\PLambda      {\ensuremath{\Lambda}\xspace}                 
 \def\PSigma      {\ensuremath{\Sigma}\xspace}                 
 \def\POmega      {\ensuremath{\Omega}\xspace}                 
 \def\PUpsilon      {\ensuremath{\Upsilon}\xspace}                 
 
 \def\PB      {\ensuremath{\mathrm{B}}\xspace}                 
                  
 \def\PD      {\ensuremath{\mathrm{D}}\xspace}

 \def\PJ      {\ensuremath{\mathrm{J}}\xspace}                 
 \def\PK      {\ensuremath{\mathrm{K}}\xspace}

 \def\Pb      {\ensuremath{\mathrm{b}}\xspace}                 
 \def\Pc      {\ensuremath{\mathrm{c}}\xspace}

 \def\Pi      {\ensuremath{\mathrm{i}}\xspace}

}
{

 \def\Ppsi        {\ensuremath{\psi}\xspace}                 
                  
 \mathchardef\PDelta="7101
 \mathchardef\PXi="7104
 \mathchardef\PLambda="7103
 \mathchardef\PSigma="7106
 \mathchardef\POmega="710A
 \mathchardef\PUpsilon="7107
                  
 \def\PB      {\ensuremath{B}\xspace}                 
                  
 \def\PD      {\ensuremath{D}\xspace}

 \def\PJ      {\ensuremath{J}\xspace}                 
 \def\PK      {\ensuremath{K}\xspace}

 \def\Pb      {\ensuremath{b}\xspace}                 
 \def\Pc      {\ensuremath{c}\xspace}

 \def\Pi      {\ensuremath{i}\xspace}

}

\makeatletter
\ifcase \@ptsize \relax
  \newcommand{\miniscule}{\@setfontsize\miniscule{4}{5}}
\or
  \newcommand{\miniscule}{\@setfontsize\miniscule{5}{6}}
\or
  \newcommand{\miniscule}{\@setfontsize\miniscule{5}{6}}
\fi
\makeatother

\DeclareRobustCommand{\optbar}[1]{\shortstack{{\miniscule (\rule[.5ex]{1.25em}{.18mm})}
  \\ [-.7ex] $#1$}}

\def\cquark    {{\ensuremath{\Pc}}\xspace}

\def\bquark    {{\ensuremath{\Pb}}\xspace}

  \def\Kbar    {{\kern 0.2em\overline{\kern -0.2em \PK}{}}\xspace}

\def\KorKbar    {\kern 0.18em\optbar{\kern -0.18em K}{}\xspace}

  \def\Dbar    {{\kern 0.2em\overline{\kern -0.2em \PD}{}}\xspace}

\def\DorDbar    {\kern 0.18em\optbar{\kern -0.18em D}{}\xspace}

\def\Bbar    {{\ensuremath{\kern 0.18em\overline{\kern -0.18em \PB}{}}}\xspace}

\def\BorBbar    {\kern 0.18em\optbar{\kern -0.18em B}{}\xspace}

\def\jpsi     {{\ensuremath{{\PJ\mskip -3mu/\mskip -2mu\Ppsi\mskip 2mu}}}\xspace}
\def\psitwos  {{\ensuremath{\Ppsi{(2S)}}}\xspace}

  \def\Y#1S{\ensuremath{\PUpsilon{(#1S)}}\xspace}

\def\Lbar        {{\ensuremath{\kern 0.1em\overline{\kern -0.1em\PLambda}}}\xspace}
\def\LorLbar    {\kern 0.18em\optbar{\kern -0.18em \PLambda}{}\xspace}

\def\to                 {\ensuremath{\rightarrow}\xspace}

\def\AT#1     {\ensuremath{A_{\mathrm{T}}^{#1}}\xspace}           

\def\C#1      {\ensuremath{\mathcal{C}_{#1}}\xspace}                       
\def\Cp#1     {\ensuremath{\mathcal{C}_{#1}^{'}}\xspace}                    
\def\Ceff#1   {\ensuremath{\mathcal{C}_{#1}^{\mathrm{(eff)}}}\xspace}        
\def\Cpeff#1  {\ensuremath{\mathcal{C}_{#1}^{'\mathrm{(eff)}}}\xspace}       
\def\Ope#1    {\ensuremath{\mathcal{O}_{#1}}\xspace}                       
\def\Opep#1   {\ensuremath{\mathcal{O}_{#1}^{'}}\xspace}                    

\newcommand{\tev}{\ifthenelse{\boolean{inbibliography}}{\ensuremath{~T\kern -0.05em eV}}{\ensuremath{\mathrm{\,Te\kern -0.1em V}}}\xspace}
\newcommand{\gev}{\ensuremath{\mathrm{\,Ge\kern -0.1em V}}\xspace}
\newcommand{\mev}{\ensuremath{\mathrm{\,Me\kern -0.1em V}}\xspace}
\newcommand{\kev}{\ensuremath{\mathrm{\,ke\kern -0.1em V}}\xspace}
\newcommand{\ev}{\ensuremath{\mathrm{\,e\kern -0.1em V}}\xspace}
\newcommand{\gevc}{\ensuremath{{\mathrm{\,Ge\kern -0.1em V\!/}c}}\xspace}
\newcommand{\mevc}{\ensuremath{{\mathrm{\,Me\kern -0.1em V\!/}c}}\xspace}
\newcommand{\gevcc}{\ensuremath{{\mathrm{\,Ge\kern -0.1em V\!/}c^2}}\xspace}
\newcommand{\gevgevcccc}{\ensuremath{{\mathrm{\,Ge\kern -0.1em V^2\!/}c^4}}\xspace}
\newcommand{\mevcc}{\ensuremath{{\mathrm{\,Me\kern -0.1em V\!/}c^2}}\xspace}

\def\mum  {\ensuremath{{\,\upmu\mathrm{m}}}\xspace}

\newcommand{\chisq}{\ensuremath{\chi^2}\xspace}

\newcommand{\chisqip}{\ensuremath{\chi^2_{\text{IP}}}\xspace}

\def\gsim{{~\raise.15em\hbox{$>$}\kern-.85em
          \lower.35em\hbox{$\sim$}~}\xspace}
\def\lsim{{~\raise.15em\hbox{$<$}\kern-.85em
          \lower.35em\hbox{$\sim$}~}\xspace}

\def\pt         {\mbox{$p_{\mathrm{ T}}$}\xspace}

\def\evtgen     {\mbox{\textsc{EvtGen}}\xspace}

\def\geant      {\mbox{\textsc{Geant4}}\xspace}

\def\photos     {\mbox{\textsc{Photos}}\xspace}

\def\tell1  {TELL1\xspace}
\def\ukl1   {UKL1\xspace}

\newcommand{\ie}{\mbox{\itshape i.e.}\xspace}

\usepackage{cite} 
\usepackage{mciteplus}
\usepackage{bm}

\usepackage{longtable} 

\usepackage{multirow} 

\begin{document}

\renewcommand{\thefootnote}{\fnsymbol{footnote}}
\setcounter{footnote}{1}



\begin{titlepage}
\pagenumbering{roman}

\vspace*{-1.5cm}
\centerline{\large EUROPEAN ORGANIZATION FOR NUCLEAR RESEARCH (CERN)}
\vspace*{1.5cm}
\noindent
\begin{tabular*}{\linewidth}{lc@{\extracolsep{\fill}}r@{\extracolsep{0pt}}}
\ifthenelse{\boolean{pdflatex}}
{\vspace*{-2.7cm}\mbox{\!\!\!\includegraphics[width=.14\textwidth]{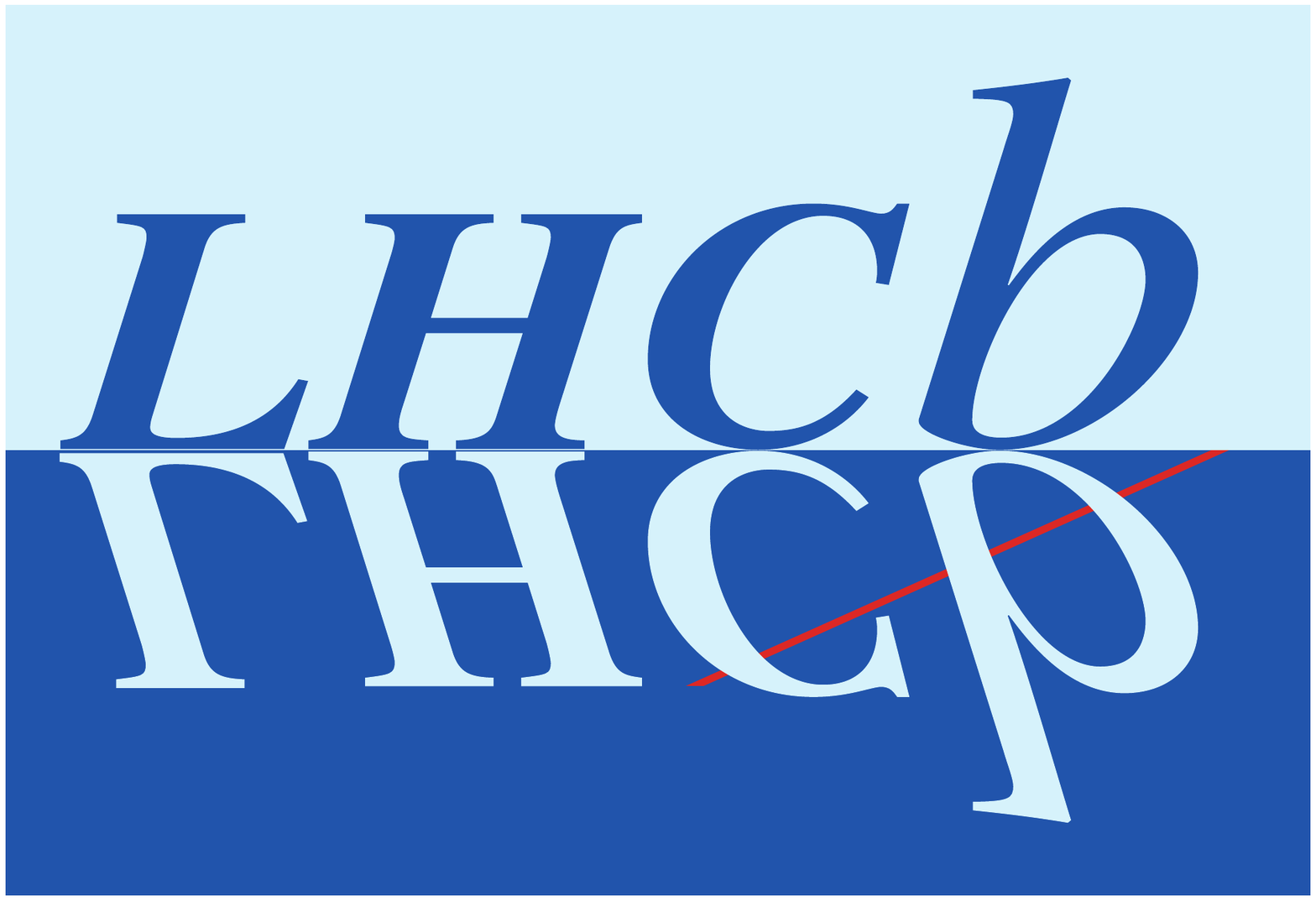}} & &}%
{\vspace*{-1.2cm}\mbox{\!\!\!\includegraphics[width=.12\textwidth]{lhcb-logo.eps}} & &}%
\\
 & & CERN-EP-2017-122 \\  
 & & LHCb-PAPER-2017-014 \\  
 & & \today \\ 
 & & \\
\end{tabular*}

\vspace*{4.0cm}

{\normalfont\bfseries\boldmath\huge
\begin{center}
      Prompt and nonprompt \jpsi production and nuclear modification in $p$Pb collisions at $\sqrt{s_{NN}}=8.16\,{\rm TeV}$ 
\end{center}
}

\vspace*{2.0cm}

\begin{center}
The LHCb collaboration\footnote{Authors are listed at the end of this paper.}
\end{center}

\vspace{\fill}

\begin{abstract}
  \noindent
  The production of \jpsi mesons is studied in proton-lead collisions at the centre-of-mass energy per nucleon pair $\sqrt{s_{NN}}=8.16\,{\rm TeV}$ 
  with the LHCb detector at the LHC. 
  The double differential cross-sections of prompt and nonprompt \jpsi production are measured as a function of the 
  \jpsi transverse momentum and rapidity in the nucleon-nucleon centre-of-mass frame.
  Forward-to-backward ratios and nuclear modification factors are determined. 
  The results are compared with theoretical calculations based on collinear factorisation using nuclear
  parton distribution functions, on the colour glass condensate or on coherent energy loss models.
\end{abstract}

\vspace*{2.0cm}

\begin{center}
  Published in  Phys. Lett. B774 (2017) 159
\end{center}

\vspace{\fill}

{\footnotesize 
\centerline{\copyright~CERN on behalf of the \lhcb collaboration, licence \href{http://creativecommons.org/licenses/by/4.0/}{CC-BY-4.0}.}}
\vspace*{2mm}

\end{titlepage}


\newpage
\setcounter{page}{2}
\mbox{~}
%
%
%
%

\cleardoublepage


\renewcommand{\thefootnote}{\arabic{footnote}}
\setcounter{footnote}{0}



\pagestyle{plain} 
\setcounter{page}{1}
\pagenumbering{arabic}


%

\section{Introduction}
\label{sec:intro}

The production of \jpsi mesons, and more generally of quarkonium states, has been considered as a sensitive probe of
colour screening in a hot and dense medium since the proposal by Matsui and Satz 
in 1986~\cite{Matsui:1986dk} of the suppression of the \jpsi meson production in heavy-ion collisions as a sign of deconfinement.
The theoretical understanding of the bound-state dynamics of quarkonium by means of lattice QCD and effective field theories 
has progressed substantially in the last 30 years. In heavy-ion collisions, the emerging picture indicates strong modifications of the 
quarkonium bound-state characteristics~\cite{Mocsy:2013syh}. 
Experimentally, measurements at the SPS, RHIC and LHC revealed interesting 
patterns~\cite{Andronic:2015wma}. In particular, an additional low transverse momentum (\pt) component of 
 \jpsi production was observed in PbPb collisions at the LHC~\cite{Abelev:2012rv,Abelev:2013ila,Adam:2015rba,Adam:2015isa,Adam:2016rdg}. 
This observation had been predicted as a sign of charmonium originating from unbound charm quarks, generated either during the lifetime of 
the deconfined medium~\cite{Thews:2000rj} or at the phase boundary~\cite{BraunMunzinger:2000px}. 

The limited understanding of nuclear phenomena unrelated to deconfinement, commonly called cold nuclear matter (CNM) effects, restricts 
the ability of phenomenological models to describe the experimental data on \jpsi production in PbPb collisions.
 The size of CNM effects can be quantified by measurements in proton-nucleus or deuteron-nucleus collisions, which have been pursued at fixed target experiments as well as at RHIC and LHC~\cite{Andronic:2015wma}.  
The feature of CNM drawing the highest attention for proton-lead collisions at the LHC is the modification of the gluon flux coupling to the 
charm quark pair. This modification is often treated within a collinear parton distribution framework employing nuclear parton distribution functions 
(nPDFs)~\cite{Hirai:2007sx,Eskola:2009uj,deFlorian:2011fp,Kovarik:2015cma,Eskola:2016oht}. At low longitudinal momentum fractions $x$ carried by the parton, calculations within the colour glass condensate (CGC) effective field theory, describing the saturation regime of QCD~\cite{Gelis:2010nm,Fujii:2006ab}, are  frequently employed. 
Several calculations have been pursued to quantify nuclear modifications of \jpsi 
production in the collinear framework~\cite{Vogt:2004dh,Ferreiro:2008wc,Vogt:2010aa,Lansberg:2016deg} or in the CGC 
framework~\cite{Fujii:2013gxa,Ma:2015sia,Ducloue:2015gfa}. It has to be noted that the low-$x$ gluon content of the nucleus is largely
unconstrained by experimental data at perturbative scales. In addition, small-angle gluon radiation taking into account interference between initial and final state radiation, called coherent energy loss,  was proposed as the dominant nuclear modification of quarkonium production 
in proton-lead collisions~\cite{Arleo:2012rs}. The discrimination between these phenomena is a strong motivation for the study 
of the production of quarkonium as a hard-scale probe of QCD at high density.

The experimental results on \jpsi production in proton-lead collisions based on the 2013 data samples at $\sqrt{s_{NN}}=5\,{\rm TeV}$ published by the
LHC experiments ALICE, ATLAS, CMS and LHCb~\cite{Abelev:2013yxa,LHCb-PAPER-2013-052,Adam:2015iga,Adam:2015jsa,Aad:2015ddl,Sirunyan:2017mzd}
can be qualitatively described by implementations of the approaches described above in the kinematic applicability range of the calculations~\cite{Vogt:2004dh,Ferreiro:2008wc,Vogt:2010aa,Lansberg:2016deg,Ma:2015sia,Ducloue:2015gfa,Arleo:2012rs}.
No conclusion on the dominant mechanism for nuclear modification of \jpsi production could be drawn.

The measurement of an additional suppression of the excited state \psitwos by ALICE~\cite{Abelev:2014zpa,Adam:2016ohd} and LHCb~\cite{LHCb-PAPER-2015-058} in proton-lead collisions at $\sqrt{s_{NN}}=5\,{\rm TeV}$ and by PHENIX at RHIC~\cite{Adare:2013ezl,Adare:2016psx} in various collision systems at $\sqrt{s_{NN}}=0.2\,{\rm TeV}$ cannot be explained by the modification of the gluon flux or by coherent energy loss because it would affect the \jpsi and the \psitwos states in a similar way.  These measurements motivated calculations involving hadronic and partonic interactions influencing the evolution of the $c\bar{c}$ pair after the first      interaction~\cite{Ferreiro:2014bia,Du:2015wha} for proton(deuteron)-nucleus collisions. Although the impact on \jpsi production is generally small in these models, it can be significant in  rapidity ranges with large particle densities.

The measurement of the nonprompt \jpsi production provides access to the production of beauty hadrons. The modification of their kinematic distributions in nucleus-nucleus collisions carries valuable information about the created matter~\cite{Andronic:2015wma}. Similarly to direct charmonium production, the production of beauty hadrons can be subject to CNM effects altering the interpretation of nucleus-nucleus collision data. Such effects can be precisely measured in proton-lead collisions. 

The measurements of the production of prompt \jpsi and nonprompt \jpsi mesons, called \jpsi-from-$b$-hadrons in the following,  presented in this letter are important ingredients 
for the understanding of the imprints of deconfinement in nucleus-nucleus collisions. They are based on larger integrated luminosities and on higher collision energies than the initial measurements with the 2013 proton-lead data sample by the LHCb experiment at $\sqrt{s_{NN}}=5\,{\rm TeV}$~\cite{LHCb-PAPER-2013-052}. 

\section{Detector, data sample and observables}
\label{sec:Detector}

The \lhcb detector~\cite{Alves:2008zz,LHCb-DP-2014-002} is a single-arm forward
spectrometer covering the \mbox{pseudorapidity} range $2<\eta <5$,
designed for the study of particles containing \bquark or \cquark
quarks. The detector includes a high-precision tracking system
consisting of a silicon-strip vertex detector surrounding the 
interaction region~\cite{LHCb-DP-2014-001}, a large-area silicon-strip detector located
upstream of a dipole magnet with a bending power of about
$4{\mathrm{\,Tm}}$, and three stations of silicon-strip detectors and straw
drift tubes~\cite{LHCb-DP-2013-003} placed downstream of the magnet.
The tracking system provides a measurement of momentum of charged particles with
a relative uncertainty that varies from 0.5\% at low momentum to 1.0\% at 200\gevc.
The minimum distance of a track to a primary vertex (PV), the impact parameter, 
is measured with a resolution of $(15+29/\pt)\mum$,
where \pt is the transverse momentum in the LHCb frame, in\,\gevc.
Different types of charged hadrons are distinguished using information
from two ring-imaging Cherenkov detectors~\cite{LHCb-DP-2012-003}. 
Photons, electrons and hadrons are identified by a calorimeter system consisting of
scintillating-pad and preshower detectors, an electromagnetic
calorimeter and a hadronic calorimeter. Muons are identified by a
system composed of alternating layers of iron and multiwire
proportional chambers~\cite{LHCb-DP-2012-002}.

This analysis is based on data acquired during the 2016 LHC heavy-ion run, where protons and 
$^{208}$Pb ions were colliding at a centre-of-mass energy per nucleon pair of $\sqrt{s_{NN}}=8.16\,{\rm TeV}$. Since the energy per nucleon in the proton beam is larger than in the lead beam, the nucleon-nucleon centre-of-mass system has a rapidity in the laboratory frame of 0.465 ($-0.465$), when the proton (lead) beam travels from the vertex detector towards the muon chambers.
Consequently, the LHCb detector covers two different acceptance regions:
\begin{enumerate}
\item $1.5<y^*<4.0$ when the proton beam travels from the vertex detector towards the muon chambers,
\item $-5.0<y^*<-2.5$ when the proton beam travels from the muon chambers towards the vertex detector,
\end{enumerate}
where $y^*$ is the rapidity in the centre-of-mass frame of the colliding nucleons, with respect to the proton beam 
direction. In this letter, the first configuration is denoted  $p$Pb  and the second one Pb$p$.
The data samples correspond to an integrated luminosity of $13.6\pm 0.3\,{\rm nb}^{-1}$ of $p$Pb collisions and
$20.8\pm0.5\,{\rm nb}^{-1}$ of Pb$p$ collisions.
The instantaneous luminosity for the majority of the recorded events ranges between 0.5 and $1.0\times 10^{29}\,{\rm cm}^{-2}{\rm s}^{-1}$. 
This luminosity corresponds on average to about 0.1 or fewer collisions  per bunch crossing.

In this letter, we describe the measurement of the double-differential production cross-sections of \jpsi mesons as a function of \pt and $y^*$   in the ranges $0<\pt<14\gevc$ and 
$1.5<y^*<4.0$ for $p$Pb and $-5.0<y^*<-2.5$ for Pb$p$.
The measurement is performed separately for 
prompt \jpsi mesons, \ie produced directly in the initial hard scattering or from the decay of an excited charmonium state
produced directly, and for \jpsi mesons coming from the decay of a long-lived $b$-hadron, either directly or via an excited charmonium state. 

Nuclear effects are quantified by the nuclear modification factor, $R_{p{\rm Pb}}$,
\begin{equation}\label{eq:rpa}
R_{p{\rm Pb}} (\pt,y^*) \equiv \frac{1}{A} \frac{{\rm d}^2 \sigma_{p{\rm Pb}}(\pt,y^*)/{\rm d}\pt{\rm d}y^*}{{\rm d}^2\sigma_{pp}(\pt,y^*)/{\rm d}\pt{\rm d}y^*},
\end{equation}
where 
$A=208$ is the mass number of the Pb ion, 
${\rm d}^2 \sigma_{p{\rm Pb}}(\pt,y^*)/{\rm d}\pt{\rm d}y^*$ the \jpsi production cross-section in $p$Pb or Pb$p$ collisions and
${\rm d}^2 \sigma_{pp}(\pt,y^*)/{\rm d}\pt{\rm d}y^*$ the \jpsi reference production cross-section in $pp$ collisions at the same nucleon-nucleon centre-of-mass
energy. The determination of the reference cross-section is described in Sec.~\ref{sec:refxsection}.
In the absence of nuclear effects, the nuclear modification factor is equal to unity. 

In addition to the nuclear modification factor, the observable $R_{\rm FB}$ quantifies the relative forward-to-backward production rates. The forward-to-backward ratio is measured as the ratio of cross-sections in the positive and negative $y^*$ acceptances evaluated in the same absolute $y^*$  value ranges,
\begin{equation}
R_{\rm FB} (\pt,y^*) \equiv \frac{{\rm d}^2 \sigma_{p{\rm Pb}}(\pt,+y^*)/{\rm d}\pt{\rm d}y^*}{{\rm d}^2 \sigma_{p{\rm Pb}}(\pt,-y^*)/{\rm d}\pt{\rm d}y^*}.
\end{equation}


\section{Event selection and cross-section determination}
\label{sec:selection}

The \jpsi production cross-section measurement follows the approach described in Ref.~\cite{LHCb-PAPER-2015-037}. 
The double differential \jpsi production cross-section in each kinematic bin of \pt and $y^*$ is computed as 
\begin{align}
\frac{{\rm d}^2 \sigma}{{\rm d}\pt{\rm d}y^*} &= \frac{N(\jpsi \to \mu^+ \mu^-)}{{\cal L}\times \epsilon_{\text{tot}}\times {\cal B(}\jpsi \to \mu^+\mu^-)  \times \Delta \pt \times \Delta y^*},
\end{align}
where $N(\jpsi \to \mu^+ \mu^-)$ is the number of reconstructed prompt $\jpsi$ or $\jpsi$-from-$b$-hadrons signal mesons, 
$\epsilon_{\text{tot}}$ is the total detection efficiency in the given kinematic bin,  
\mbox{${\cal B}(\jpsi \to \mu^+ \mu^-)=(5.961 \pm 0.033)\%$~\cite{PDG2016}} is the branching fraction of the decay \mbox{$\jpsi \to \mu^+ \mu^-$,} 
$\Delta \pt= 1 \gevc$ and $\Delta y^* = 0.5$ are the bin widths and ${\cal L}$ is the integrated luminosity.
The luminosity is determined with a van der Meer scan, which was performed for both beam configurations. 
The luminosity determination follows closely the approach described in Ref.~\cite{LHCb-PAPER-2014-047}.

\subsection{Selection}

An online event selection is performed by a trigger system consisting
of a hardware stage, which, for this analysis, selects events containing at least one muon with \pt larger than 500\mevc,
followed by a software stage. In the first stage of the software trigger, two muon
tracks with $\pt>500\mevc$ are required to form a \jpsi candidate with invariant mass 
$M_{\mu^+\mu^-}>2.5 \gevcc$.  In the second stage, \jpsi candidates with an
invariant mass within 120\mevcc of the known value of the \jpsi mass~\cite{PDG2016} are selected. 

In between the two software stages, the alignment and calibration of the detector is performed in
near real-time~\cite{LHCb-PROC-2015-011}. The same alignment and calibration is propagated
to the offline reconstruction, ensuring consistent and high-quality particle identification (PID) information 
between the online and offline processings. The identical performance of the online and offline
reconstructions offers the opportunity to perform physics analyses
directly using candidates reconstructed in the trigger~\cite{LHCb-DP-2012-004,LHCb-DP-2016-001} as well as storing 
all reconstructed particles in the event~\cite{LHCb-PAPER-2016-064}. The present analysis exploits this feature for the first 
time in proton-lead collisions and is using the online reconstruction. 

At the analysis stage,  each event is required to have at least one PV 
reconstructed from at least four tracks measured in the vertex detector. For events with multiple PVs, the PV that has the smallest \chisqip 
with respect to the \jpsi candidate is chosen. Here, \chisqip is defined as the difference between the vertex-fit \chisq calculated with the \jpsi meson candidate included in or excluded from the PV fit. Each identified muon track is required to have $\pt>750 \mevc$, $2<\eta<5$ and to have a good-quality track fit. The two
muon tracks of the \jpsi candidate must form a good-quality vertex, representing a tighter selection compared to the software trigger requirement.  

\subsection{Determination of signal yields}

The reconstructed vertex of the \jpsi mesons originating from $b$-hadron decays tends to be separated from the PVs. 
These \jpsi mesons can thus be distinguished from prompt \jpsi mesons by exploiting the pseudo proper time defined as 
\begin{equation}
t_{z} \equiv \frac{(z_{\jpsi}-z_{\rm PV})\times M_{\jpsi} }{ p_{z}},
\end{equation}
where $z_{\jpsi}$ and $z_{\rm PV}$ are the coordinates along the beam axis of the \jpsi decay vertex position and of the PV
position,  $p_z$ is the $z$ component of the \jpsi momentum and $M_{\jpsi}$ the known \jpsi mass. 
The yields of \jpsi signal candidates, for the prompt and \jpsi-from-$b$-hadrons categories,
are determined from a simultaneous two-dimensional unbinned
maximum likelihood fit to their invariant mass and pseudo proper time distributions, performed independently for each $(\pt, y^*)$ bin.

\begin{figure}[t]
\includegraphics[width=0.5\textwidth]{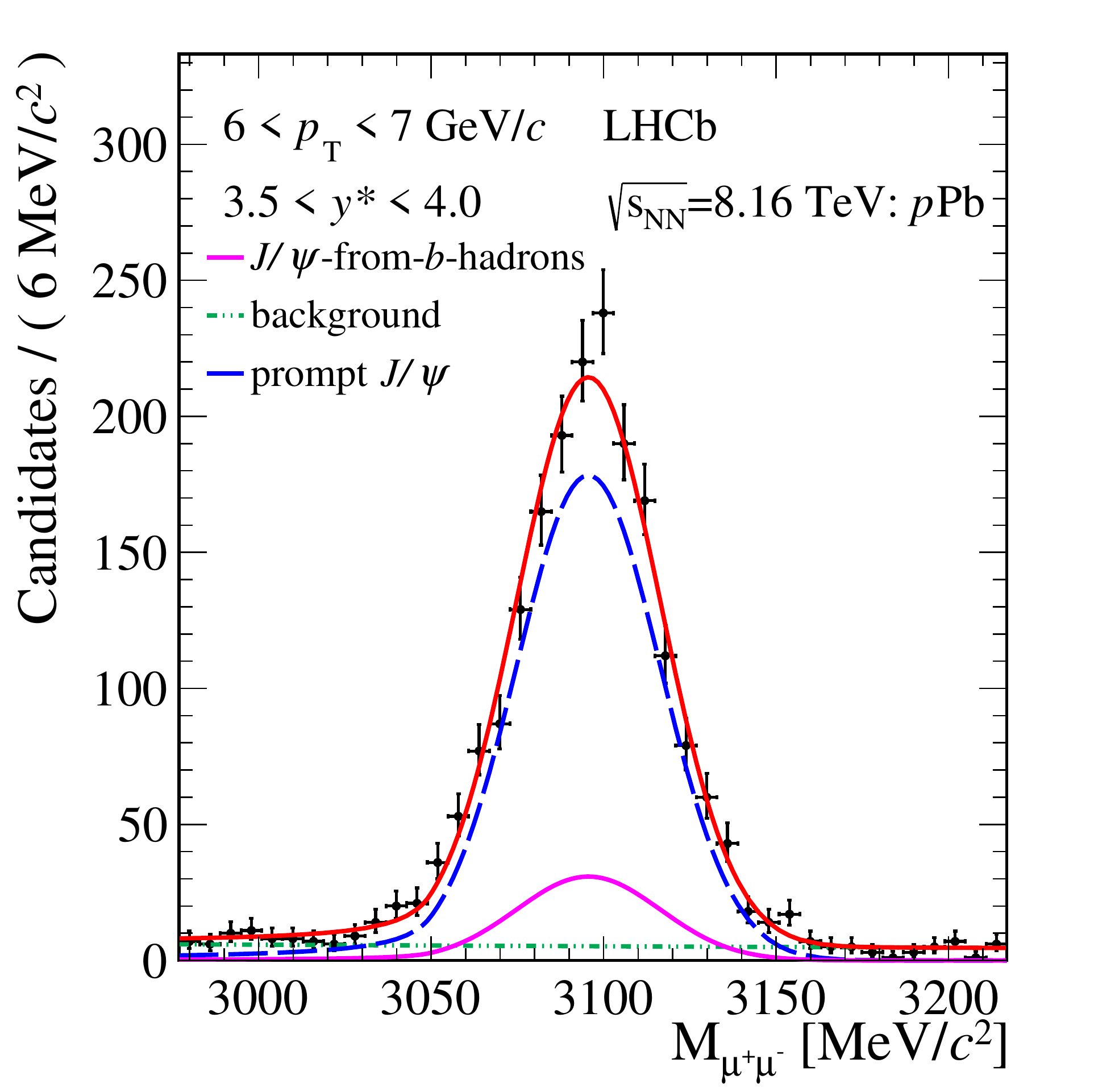}
\includegraphics[width=0.5\textwidth]{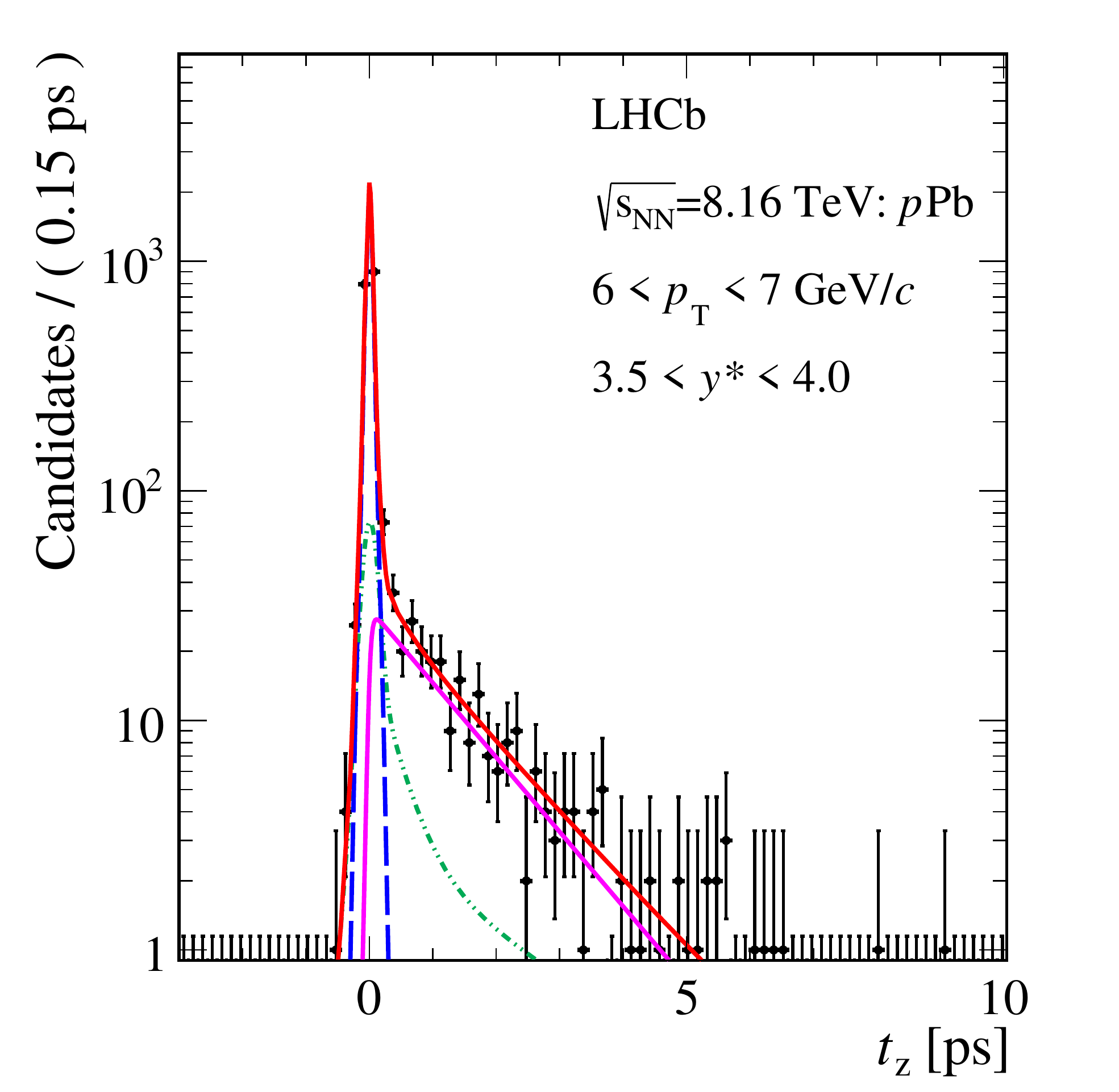}
\includegraphics[width=0.5\textwidth]{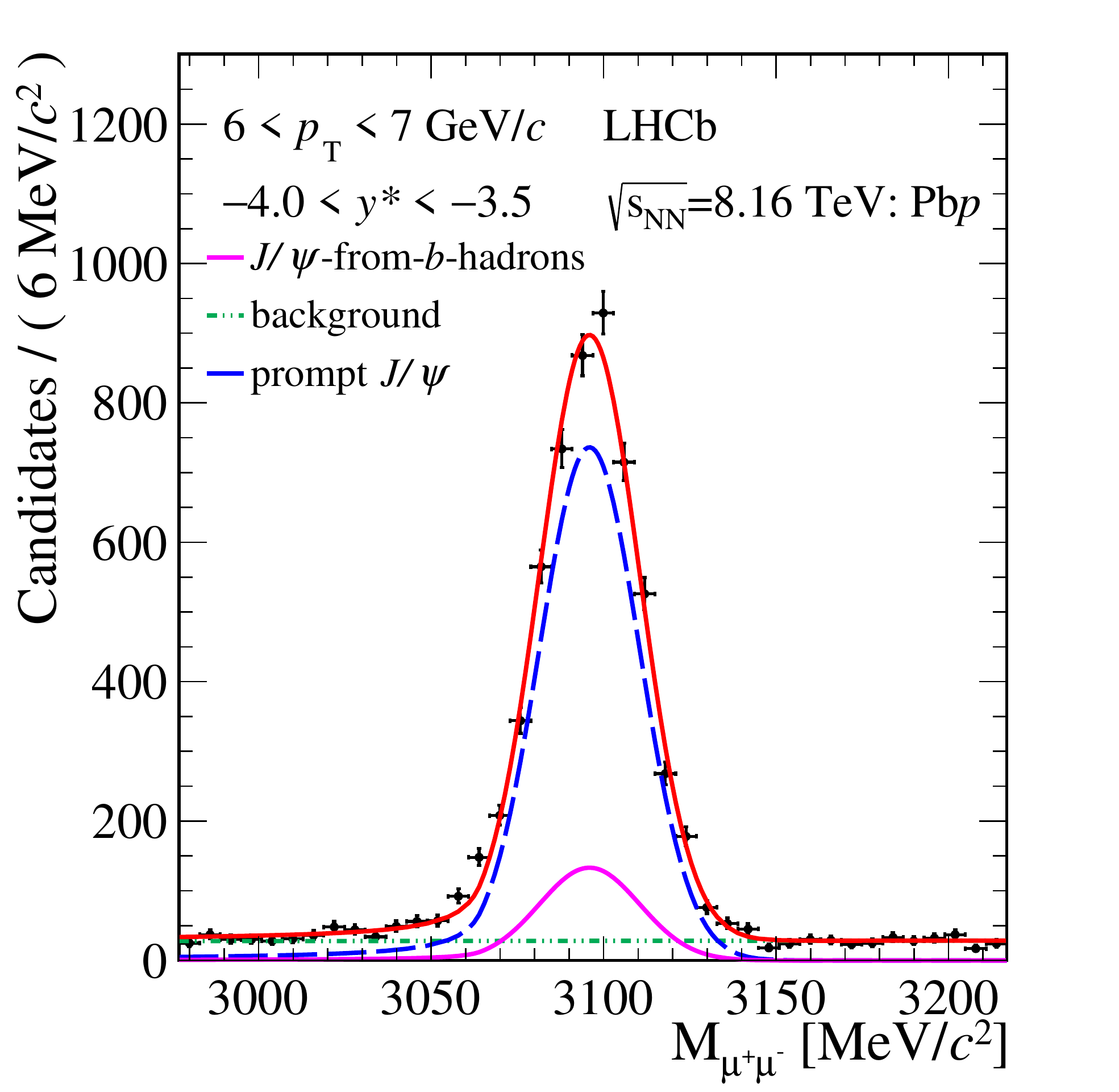}
\includegraphics[width=0.5\textwidth]{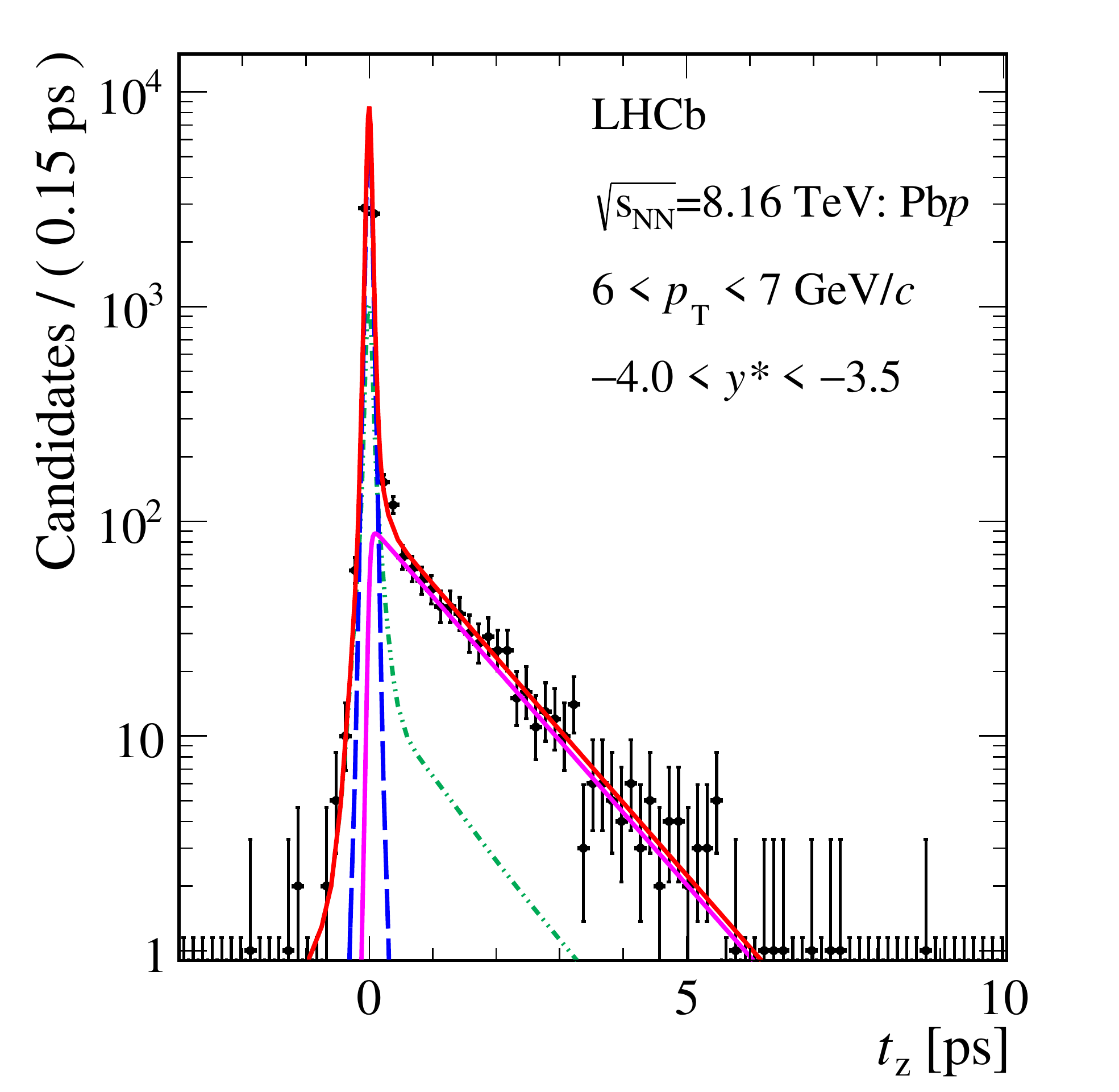}
\caption{(left) Invariant mass and (right) pseudo proper time distributions for \jpsi candidates in the bin $6<\pt<7\gevc$ and $3.5<|y^*|<4.0$ for the (top) $p$Pb and (bottom) Pb$p$ samples respectively. The black 
circles with error bars represent the LHCb data. The projection of the result
of the fit described in the text is drawn on each distribution: the red solid line is the total fit function, the blue dashed line is the prompt \jpsi signal
component, the purple solid line is the \jpsi-from-$b$-hadrons signal component and the green dashed line is the combinatorial
background component.}
\label{fig:masstzfit}
\end{figure}

In the fit function, the invariant-mass distribution of the signal is described by a Crystal Ball function~\cite{Skwarnicki:1986xj}, 
and the combinatorial background by an exponential function. The $t_z$ distribution of prompt \jpsi is described by a Dirac  $\delta$-function $\delta(t_z)$, and that of \jpsi-from-$b$-hadrons by an exponential function for $t_z>0$. Both of them are convolved with a triple-Gaussian resolution function, modelled from  simulation samples to take into account the vertex resolution. The background $t_z$ distribution is described by an empirical function derived from the shape 
observed in the \jpsi upper mass sideband, $3200<M_{\mu^+\mu^-}<3250\mevcc$. This background comes from muons of semileptonic $b$- and $c$-hadron decays and
from pions and kaons decaying in the detector. The distribution is parameterised as a sum of a Dirac $\delta$-function and of five exponential functions, 
three for positive $t_z$ values and two for negative $t_z$ values, convolved with the sum of two 
Gaussian functions.
An example of the invariant mass and the pseudo proper time distributions for one $(\pt,y^*)$ bin is shown in Fig.~\ref{fig:masstzfit} for the $p$Pb and Pb$p$ samples, where the one-dimensional projections of the fit result are drawn on the distributions.
The width of the Gaussian part of the Crystal Ball function varies as a function of \pt between $10\mevcc$ ($15\mevcc$) and $15\mevcc$ ($33\mevcc$) in the lowest (highest) rapidity bins in the laboratory frame in both beam configurations.
Due to the rapidity shifts between the laboratory frame and the nucleon-nucleon centre-of-mass frames, the two examples do not correspond to the same rapidity range 
in the laboratory while they are in the same $|y^*|$ range and, in this example, the mass resolution in the Pb$p$ configuration is different from the one in the $p$Pb configuration. 

\subsection{Efficiencies}

The total detection efficiency, $\epsilon_{\rm tot}$, is the product of the geometrical acceptance, and the efficiencies for 
charged track reconstruction,
 particle identification, candidate and trigger selections. Samples of simulated events are used to evaluate 
these efficiencies except for the particle identification, which is determined in a data-driven approach. 
In the simulation, $p$Pb and Pb$p$ minimum-bias collisions are generated using the \textsc{Epos} event generator tuned with 
the LHC model~\cite{Pierog:2013ria}. The $\jpsi\to\mu^+\mu^-$ signal candidates are generated separately, with the {\sc Pythia}8  
generator~\cite{Sjostrand:2007gs} in $pp$ collisions with beams having momenta equal to the momenta per nucleon 
of the $p$ and Pb beams. They are then merged with the \textsc{Epos} minimum bias collisions to build the samples out of which the
efficiencies are computed.  The decays of hadrons are generated by \evtgen~\cite{Lange:2001uf}, in which final-state electromagnetic
radiation is generated with \photos~\cite{Golonka:2005pn}. The interaction of the particles with the detector, and the detector response, are 
implemented using the \geant toolkit~\cite{Allison:2006ve, *Agostinelli:2002hh} as described in Ref.~\cite{LHCb-PROC-2011-006}. 

The charged-track reconstruction efficiency is first evaluated in simulation and is corrected using a data-driven tag-and-probe approach. 
For this purpose, \jpsi candidates are formed with one fully-reconstructed ``tag'' track and one ``probe'' track reconstructed partially 
with a subset of the tracking sub-detectors and both identified as muons~\cite{LHCB-DP-2013-002} in data and in simulation. The ratio of the single track efficiencies from this tag-and-probe approach is used as a correction factor. These correction factors for each track
are then applied to the signal candidates in the simulation to obtain the integrated efficiency in every kinematic bin.  The tag-and-probe correction evaluation is relying on the $p$Pb and the Pb$p$ data samples, since the larger tracking calibration samples in $pp$ collisions are limited in detector occupancy by an additional selection criterion on trigger level.

The muon identification efficiency is determined for each track in data with a tag-and-probe method~\cite{Anderlini:2202412}  taking into account the efficiency variations as function of track momentum, pseudorapidity and detector occupancy. 
Calibration samples of \jpsi mesons are selected applying a tight identification criterion on one of the muons  and no identification requirements to the 
second muon.
However, the sizes of the calibration samples collected in $p$Pb and Pb$p$ collisions are limited. The efficiency is thus evaluated 
using the calibration samples collected in $pp$ collisions, taking into account the different detector occupancies between $pp$, $p$Pb and 
Pb$p$ collisions, since this parameter affects the muon identification performance. The \jpsi simulation is weighted with the 
efficiencies determined per track in data in order to compute the muon identification efficiency in bins of \jpsi \pt and $y^*$.

The hardware and software trigger efficiencies obtained from the simulation are validated by 
comparing them with the efficiencies measured in  control data samples recorded with minimum and unbiased trigger
requirements, and containing \jpsi candidates.

\begin{figure}[t]
\includegraphics[width=0.46\textwidth]{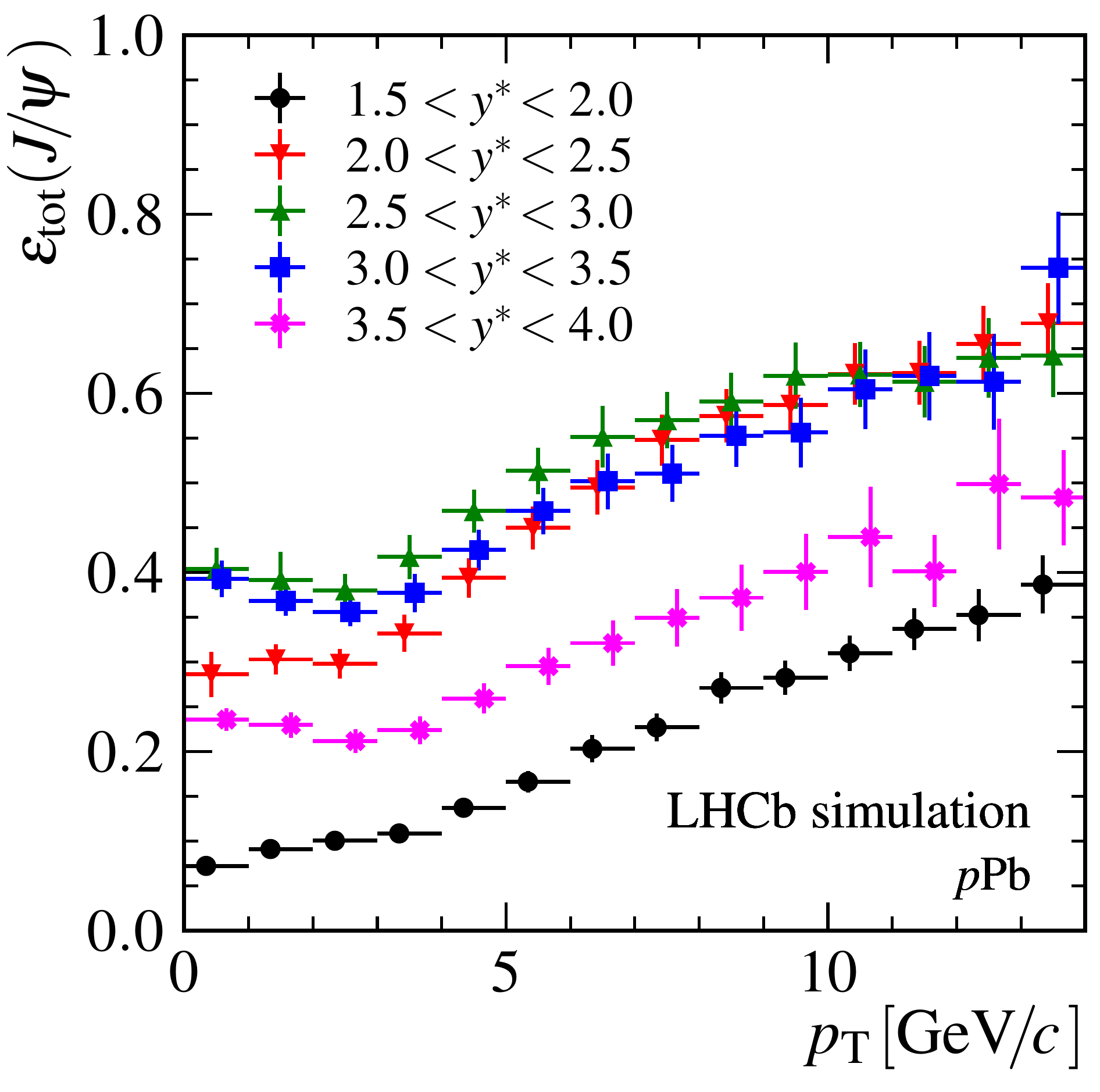}
\includegraphics[width=0.46\textwidth]{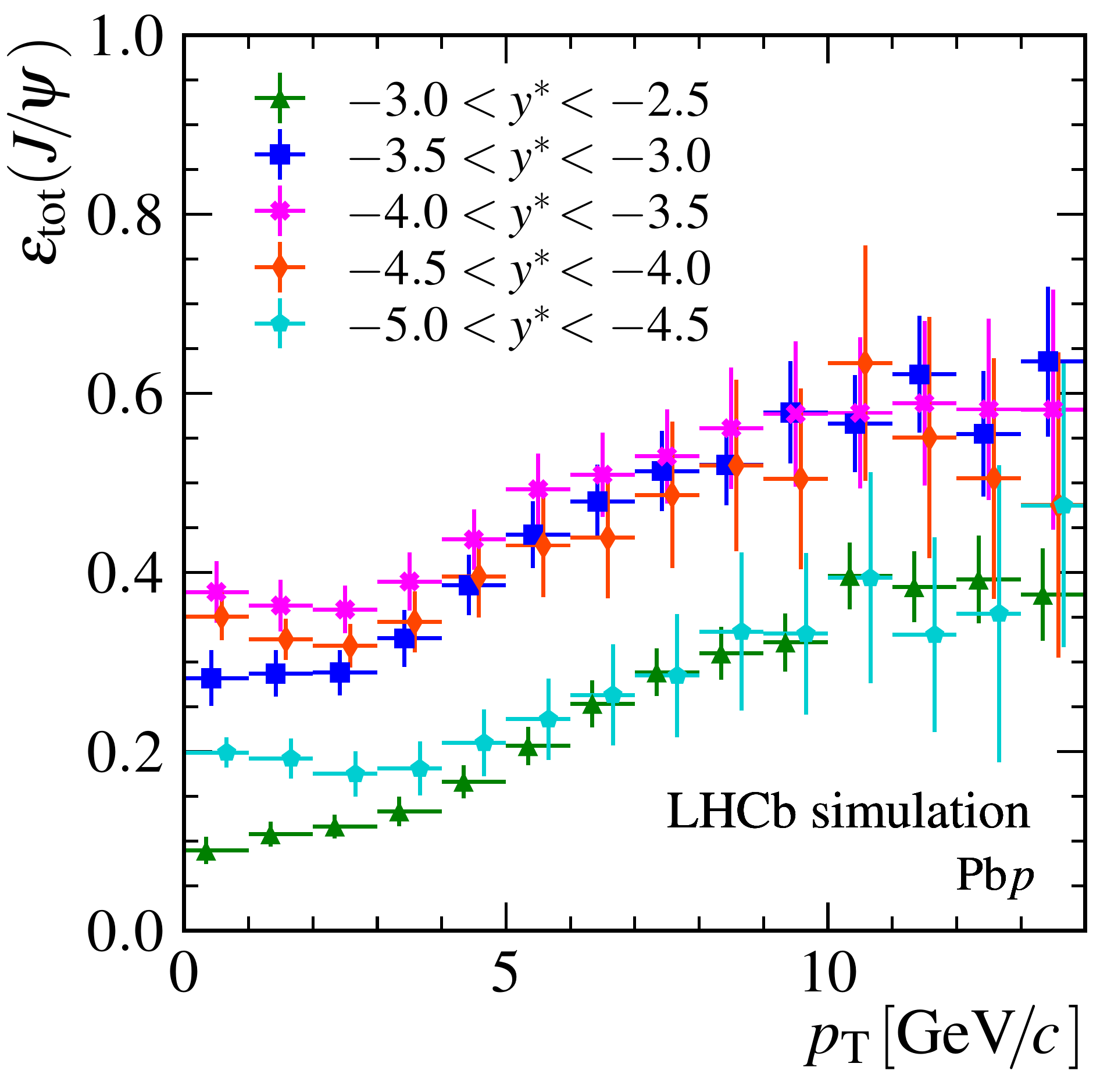}
\caption{Total \jpsi detection efficiency, $\epsilon_{\text{tot}}$, as a function of the \jpsi \pt in different $y^*$ bins for (left) $p$Pb and
(right) Pb$p$.
\label{fig:efftot}
}

\end{figure}

The total efficiency in each $(\pt,y^*)$ bin, $\epsilon_{\text{tot}}$, is found to be the same for prompt \jpsi and \jpsi-from-$b$-hadrons within 
uncertainties and is taken to be identical for the two components. It is shown in Fig.~\ref{fig:efftot} for $p$Pb and Pb$p$ collision data, as a function of the \jpsi \pt in the different rapidity
bins. The uncertainties are the quadratic sums of the statistical uncertainties and the
uncertainties associated to the data-driven corrections and validations, described in the following section.

\section{Systematic uncertainties}

The systematic uncertainties on the cross-section of prompt \jpsi and \jpsi-from-$b$-hadrons are summarised in Table~\ref{tab:syst} and 
described in the following. The total detection efficiency $\epsilon_{\text{tot}}$ for prompt \jpsi and \jpsi-from-$b$ is found to be equal within the statistical precision of the simulation and all
systematic uncertainties apply both for prompt \jpsi and \jpsi-from-$b$.
Acceptance and reconstruction efficiencies of the \jpsi  vector meson depend on its polarisation at production. The ALICE and the LHCb
 measurements in $pp$ collisions~\cite{Abelev:2011md,LHCb-PAPER-2013-008} indicate a polarisation consistent with zero in most of the kinematic 
 region of the analysis presented in this letter. 
 In this analysis, it is assumed that the \jpsi mesons 
 are produced with no polarisation in $p$Pb and Pb$p$ collisions at $\sqrt{s_{NN}}=8.16\,{\rm TeV}$. 
 No systematic uncertainty is assigned for the effects of polarisation. 

\begin{table}[!t]
\caption{\small Summary of relative systematic uncertainties in $p$Pb and Pb$p$ on the cross-section of prompt 
  \jpsi and \jpsi-from-$b$-hadrons. Uncertainties that are computed bin-by-bin are expressed as ranges giving the minimum
  to maximum values. The last column indicates the correlation 
between bins within the same beam configuration. }
\label{tab:syst}
\centering
\begin{tabular}{@{}llll@{}}\toprule
Source & $p$Pb & Pb$p$ & Comment \\ \midrule
Signal model & 1.3\% & 1.3\%& correlated \\
Muon identification & 2.0\% $-$ 11.0\% & 2.1\% $-$ 15.3\%& correlated \\
Tracking  & 3.0\% $-$ 8.0\% & 5.9\% $-$ 26.5\% & correlated \\
Hardware trigger & 1.0\% $-$ 10.9\% & 1.0\% $-$ 7.4\% & correlated \\
Software trigger & 2.0\% & 2.0\% & correlated \\
Simulation statistics & 0.4\% $-$ 7.0\% & 0.4\% $-$ 26.2\% & uncorrelated \\
${\cal B}(\jpsi\to\mu^+\mu^-)$ & 0.05\% & 0.05\% & correlated \\
Luminosity & 2.6\% & 2.5\% & correlated \\
Polarisation & --  & --    & not considered \\
 \bottomrule
\end{tabular}
\end{table}

The uncertainty on the \jpsi-meson yields, related to the modelling of the signal mass shape in the simultaneous mass
and $t_z$ fit, is studied using an alternative fit model. In this model, the signal mass shape is described by the sum of a Crystal Ball function 
and of a Gaussian function. The relative difference of the signal yields between the nominal and alternative fits amounts to 1.3\%, 
which is taken as a fully correlated systematic uncertainty between bins. The uncertainty associated to the shape of the $t_z$
distribution is negligible.
 
The uncertainty on the muon identification has multiple contributions. The statistical uncertainty of the efficiencies is derived from the
calibration sample. The impact of the finite binning in muon momentum, pseudorapidity and detector occupancy
on the efficiencies is estimated by varying the binning scheme. Finally, an uncertainty due to the method to determine the number of signal
candidates in the calibration samples is also considered. The total systematic uncertainty due to these three sources varies between 
2\% and 15\%. It is assumed to be fully correlated between bins. This assumption is valid for neighbouring bins in acceptance. The bias introduced by this assumption in the evaluation of the total systematic uncertainty on integrated quantities is negligible.

The data-driven corrections to the track reconstruction efficiency carry uncertainties related to the statistical uncertainties of the data, dominating in most bins. In addition, a systematic uncertainty is related to a potential bias of the selection criteria  which 
are necessary to obtain a good signal over background ratio for the determination of the efficiency corrections. 
A systematic uncertainty related to the method is applied similarly to $pp$ collisions and amounts to 0.8\% per 
track~\cite{LHCB-DP-2013-002}. The total uncertainty related to charged track reconstruction varies from 3.0\% to 8.0\% 
for $p$Pb and 5.9\% to 26.5\% for Pb$p$, correlated between bins.
The uncertainty in the Pb$p$ case is larger due to the smaller signal over background ratio for the partially reconstructed candidates used in the data-driven tag-and-probe method compared to the $p$Pb case.
The assumption on the correlation is valid for neighbouring bins. The introduced bias in the evaluation of the total systematic uncertainty on integrated quantities is negligible. The largest uncertainties appear 
at low track momenta and hence low \jpsi \pt. 

The trigger efficiency is determined in data and in simulation by the data-driven method described in the previous section and in Ref.~\cite{LHCb-DP-2012-004}.  The uncertainties related to the trigger are estimated by comparing the results in simulation and in data. The uncertainty on the hardware trigger efficiency is found to vary between 1\% and 11\%, and the uncertainty on the software trigger efficiency is estimated to amount to 2\%. The trigger uncertainties are assumed to be fully correlated between bins. 

The finite size of the simulation event sample used for the efficiency determination introduces a systematic uncertainty, which varies between 0.4\% and 26.2\% between the kinematic bins of the $p$Pb and the Pb$p$ simulation.
The largest relative values appear at high \pt and large rapidities and do not dominate the overall uncertainties. They differ between the $p$Pb and Pb$p$ case due to the different rapidity coverage in the centre-of-mass system. 
The branching fraction contributes to the cross-section uncertainty with 0.05\%.   The luminosity measurement uncertainty amounts to 2.6\% in $p$Pb and to 2.5\% in Pb$p$ collisions.
The uncertainty on all other applied selections is found to be negligible based on comparisons between data and simulation signal distributions of selection and kinematics variables.

\section{Results}

\subsection{Cross-sections}
\label{sec:refxsection}

\begin{figure}[!t]
\begin{center}
\includegraphics[width=0.49\textwidth]{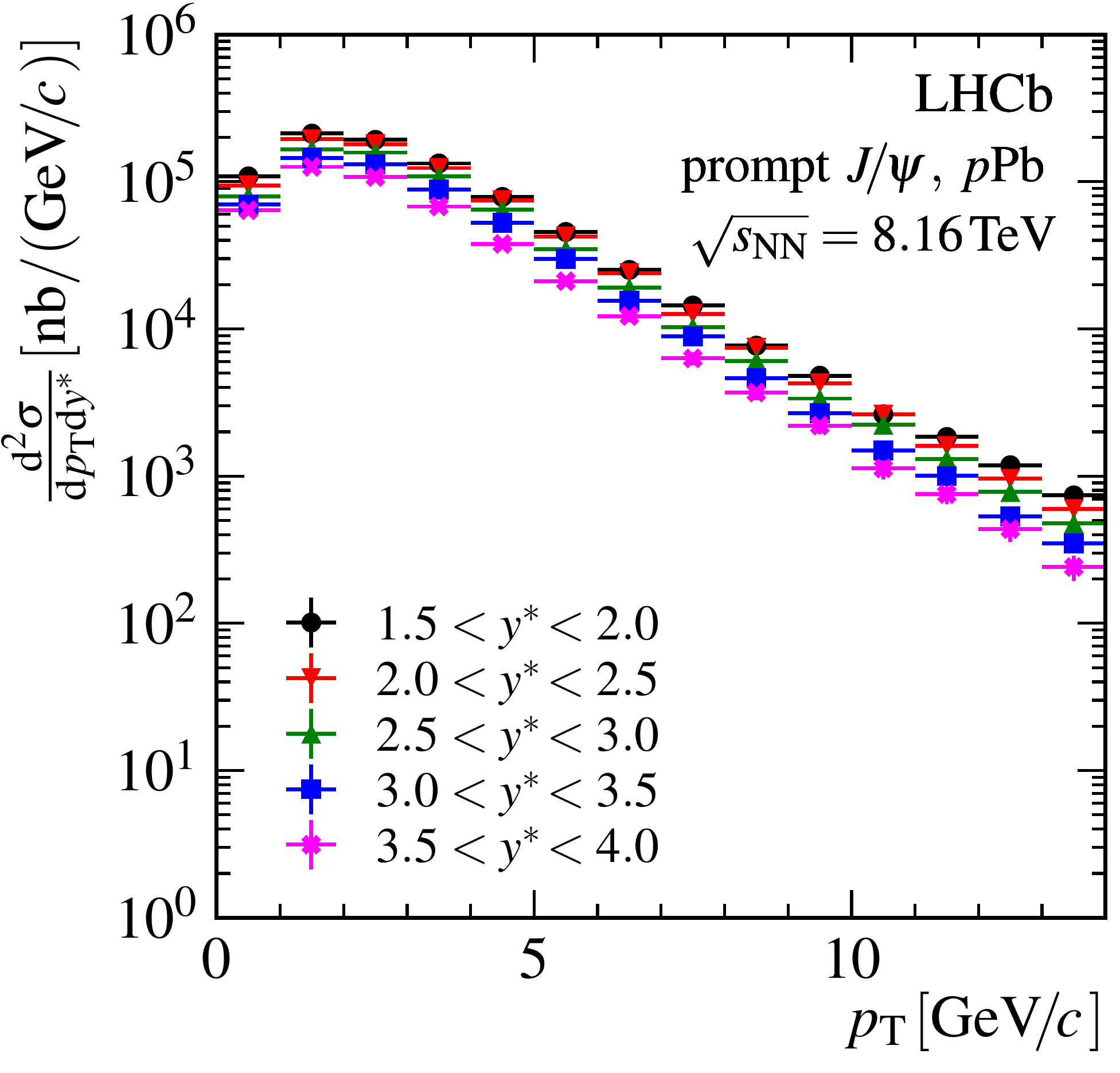}
\includegraphics[width=0.49\textwidth]{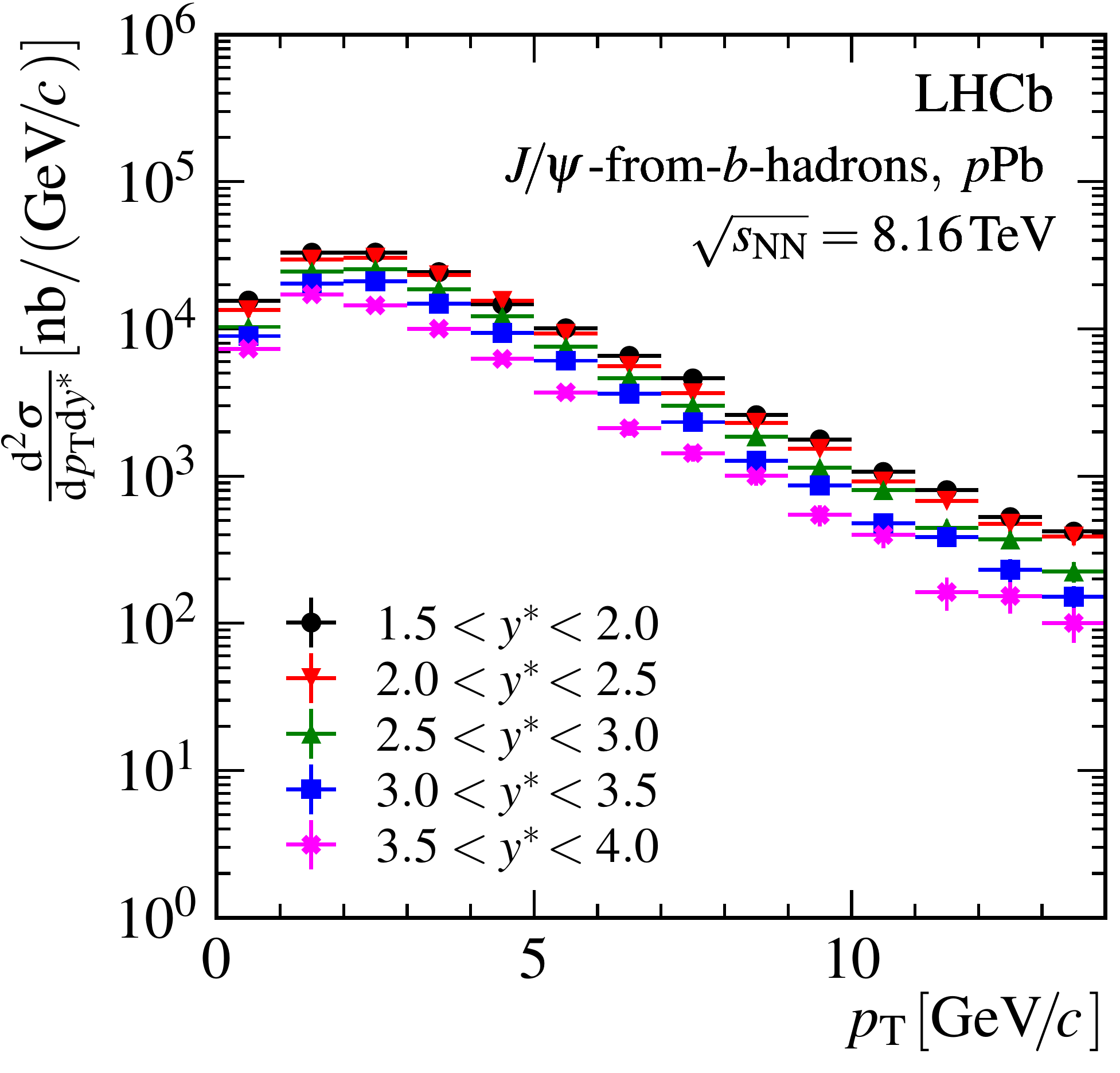}
\includegraphics[width=0.49\textwidth]{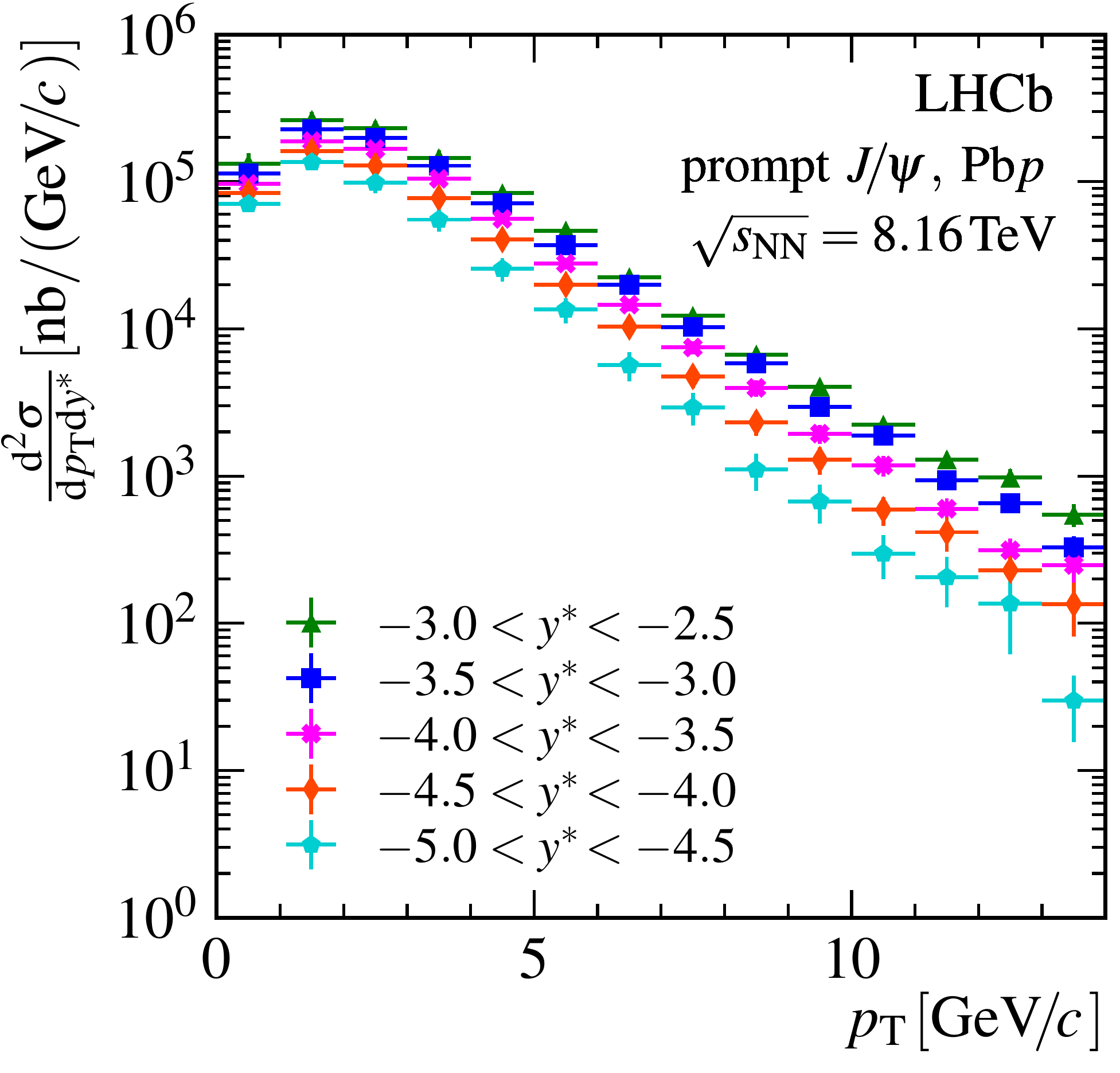}
\includegraphics[width=0.49\textwidth]{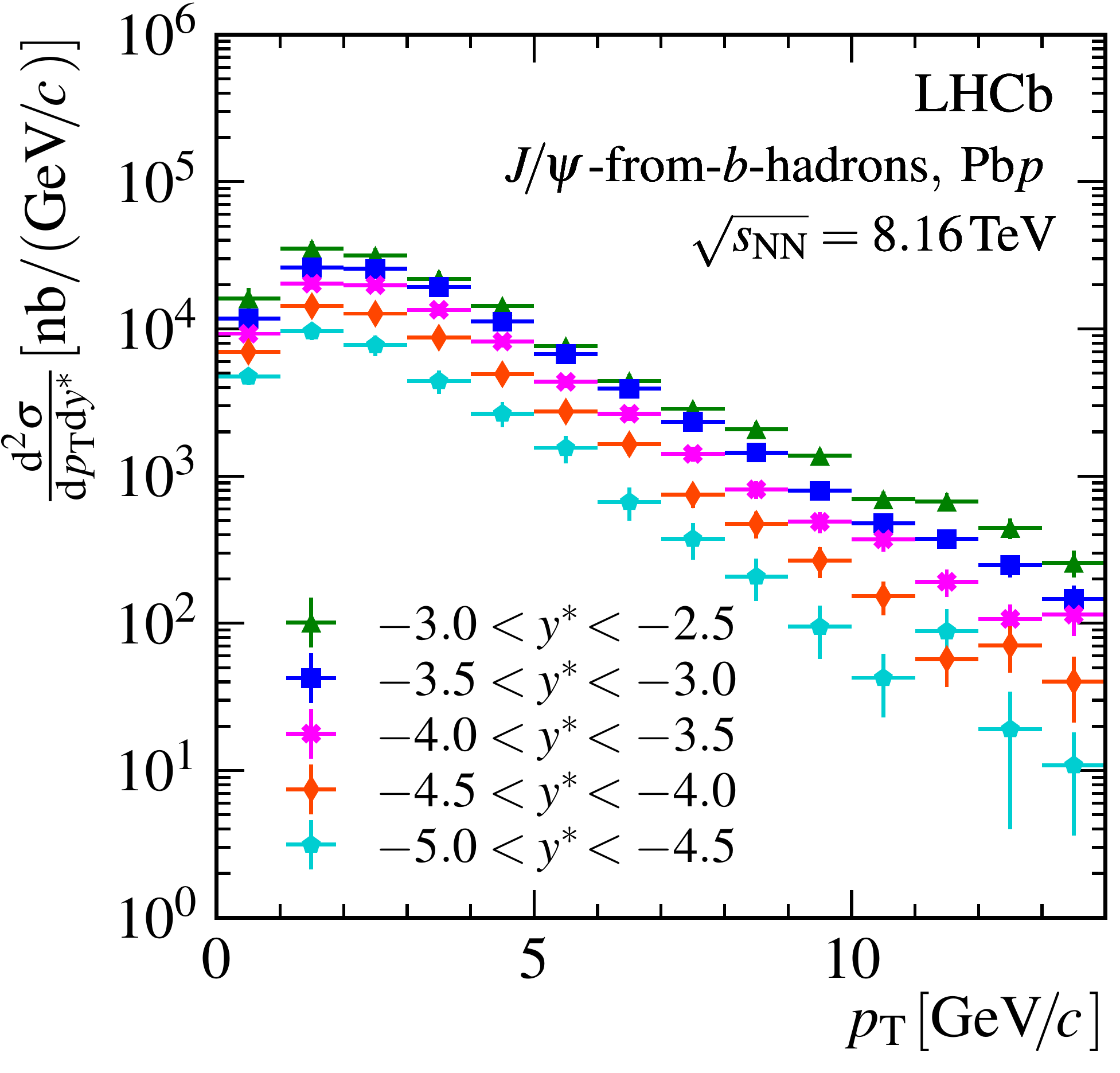}
\caption{\small Production cross-section for (top left) prompt \jpsi in $p$Pb, (top right) \jpsi-from-$b$-hadrons in $p$Pb, 
(bottom left) prompt \jpsi in Pb$p$ and (bottom right) \jpsi-from-$b$-hadrons in Pb$p$. The data points are placed at the centre of the
\pt bins, the horizontal error bars indicate the bin widths and the vertical error bars the total uncertainties, calculated as quadratic sums of the
statistical and systematic uncertainties.}
\label{fig:crosssection}
\end{center}
\end{figure}

The measured double-differential cross-sections of prompt \jpsi and \jpsi-from-$b$-hadrons in the $p$Pb and Pb$p$ data samples 
are shown in Fig.~\ref{fig:crosssection}, as a function of \pt for the considered $y^*$ bins. The numerical values
are presented in Appendices~\ref{app:crosssectionpPbprompt}--\ref{app:crosssectionPbpb}. The total cross-sections, integrated over the measurement ranges, amount to
\begin{align*}
\sigma_{\text{prompt}\ \jpsi} (1.5<y^*<4.0 ,\ \pt < 14 \gevc )& = 1625 \pm 4 \pm 117\,\upmu{\rm b}, \\ 
\sigma_{\jpsi\text{-from-}b\text{-hadrons}} (1.5<y^*<4.0 ,\ \pt<14 \gevc )& =  \phantom{1}276 \pm 2 \pm  \phantom{1}20\,\upmu{\rm b},\\
\sigma_{\text{prompt}\ \jpsi} (-5.0<y^*<-2.5,\ \pt < 14 \gevc )& = 1692 \pm 4 \pm 182\,\upmu{\rm b},\\
\sigma_{\jpsi\text{-from-}b\text{-hadrons}}  (-5.0<y^*<-2.5,\ \pt < 14 \gevc )& = \phantom{1}209 \pm 1 \pm  \phantom{1}22\,\upmu{\rm b}, 
\end{align*}
where the first uncertainties are statistical and the second systematic.

The fraction of \jpsi-from-$b$-hadrons, $f_b$, is derived from the cross-section measurements. The fraction $f_b$ is
defined as
\begin{equation}
f_b (\pt,y^*) \equiv \frac{\mathrm{d}^2\sigma_{\jpsi\text{-from-}b\text{-hadrons}}/\mathrm{d}\pt\mathrm{d}y^*}{\mathrm{d}^2\sigma_{\text{prompt}\ \jpsi}/\mathrm{d}
\pt\mathrm{d}y^*
+\mathrm{d}^2\sigma_{\jpsi\text{-from-}b\text{-hadrons}}/\mathrm{d}\pt\mathrm{d}y^*}.
\end{equation} 
Most of the systematic uncertainties cancel in the determination of $f_b$, which can thus be measured precisely.
The values of $f_b$ as a function of \pt in the different $y^*$ bins are shown in Fig.~\ref{fig:fb} for $p$Pb and Pb$p$ and listed
in Appendices~\ref{app:fb} and \ref{app:fb1}.
The values of $f_b$ measured in $pp$ collisions at a centre-of-mass energy of $8\,\rm{TeV}$~\cite{LHCb-PAPER-2013-016}, 
are shown on the same figure for comparison. The differences that appear between the measurements performed in the two 
collision systems indicate, particularly at low \pt, different nuclear modifications for prompt \jpsi and $b$-quark production.

\begin{figure}[!t]
\begin{center}
\includegraphics[width=0.76\textwidth]{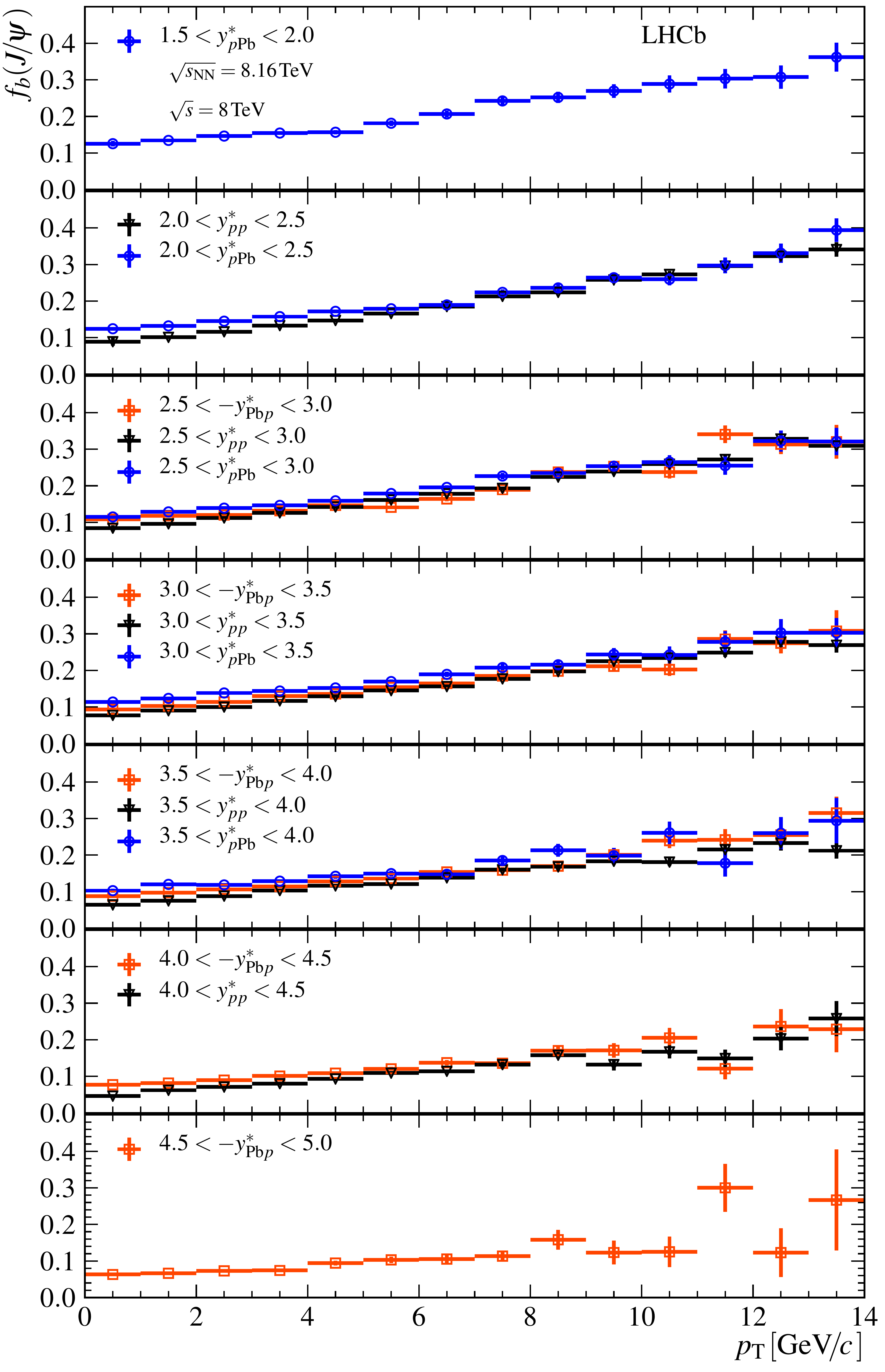}
\caption{\small Fraction of \jpsi-from-$b$-hadrons, $f_b$, as a function of \pt for (from top to bottom) $1.5<|y^*|<2.0$, 
$2.0<|y^*|<2.5$, $2.5<|y^*|<3.0$, $3.0<|y^*|<3.5$, $3.5<|y^*|<4.0$, $4.0<|y^*|<4.5$ and 
$4.5<|y^*|<5.0$. 
The data points are placed at the centre of the \pt bins, the horizontal error bars indicate the bin widths and the 
vertical error bars the total uncertainties, calculated as quadratic sums of the statistical and systematic uncertainties.
Blue circles are for $p$Pb collisions, red squares for Pb$p$ collisions and black triangles for $pp$ collisions at 
$8\,{\rm TeV}$ taken from Ref.~\cite{LHCb-PAPER-2013-016}.
}
\label{fig:fb}
\end{center}
\end{figure}

The focus of this publication is the quantification of the nuclear effects, comparing in particular the \jpsi production
in proton-lead collisions with that in $pp$ collisions at the same energy. 
Following the same approach as in the previous LHCb publication on \jpsi production in $p$Pb collisions at $\sqrt{s_{NN}}=5\,{\rm TeV}$~\cite{LHCb-PAPER-2013-052}, a $pp$ reference cross-section 
at $\sqrt{s}=8.16\,{\rm TeV}$ is determined from an interpolation of the LHCb cross-section measurements 
at $7\,{\rm TeV}$~\cite{LHCb-PAPER-2011-003}, $8\,{\rm TeV}$~\cite{LHCb-PAPER-2013-016} and 
$13\,{\rm TeV}$~\cite{LHCb-PAPER-2015-037}.  The extracted reference cross-section is in agreement with the measured cross-section at $\sqrt{s}=8\,{\rm TeV}$. For the edges of the rapidity range in $p$Pb collisions ($1.5<y^*<2.0$) and in 
Pb$p$ collisions ($4.5<y^*<5.0$), which are not covered by the measurements in $pp$ collisions, an extrapolation is used based
on the experimental measurements. The interpolation and the extrapolation methods were validated with ALICE and LHCb data and are described in Ref.~\cite{LHCb-CONF-2013-013}.

\begin{figure}[!t]
\begin{center}
\includegraphics[width=7.5cm]{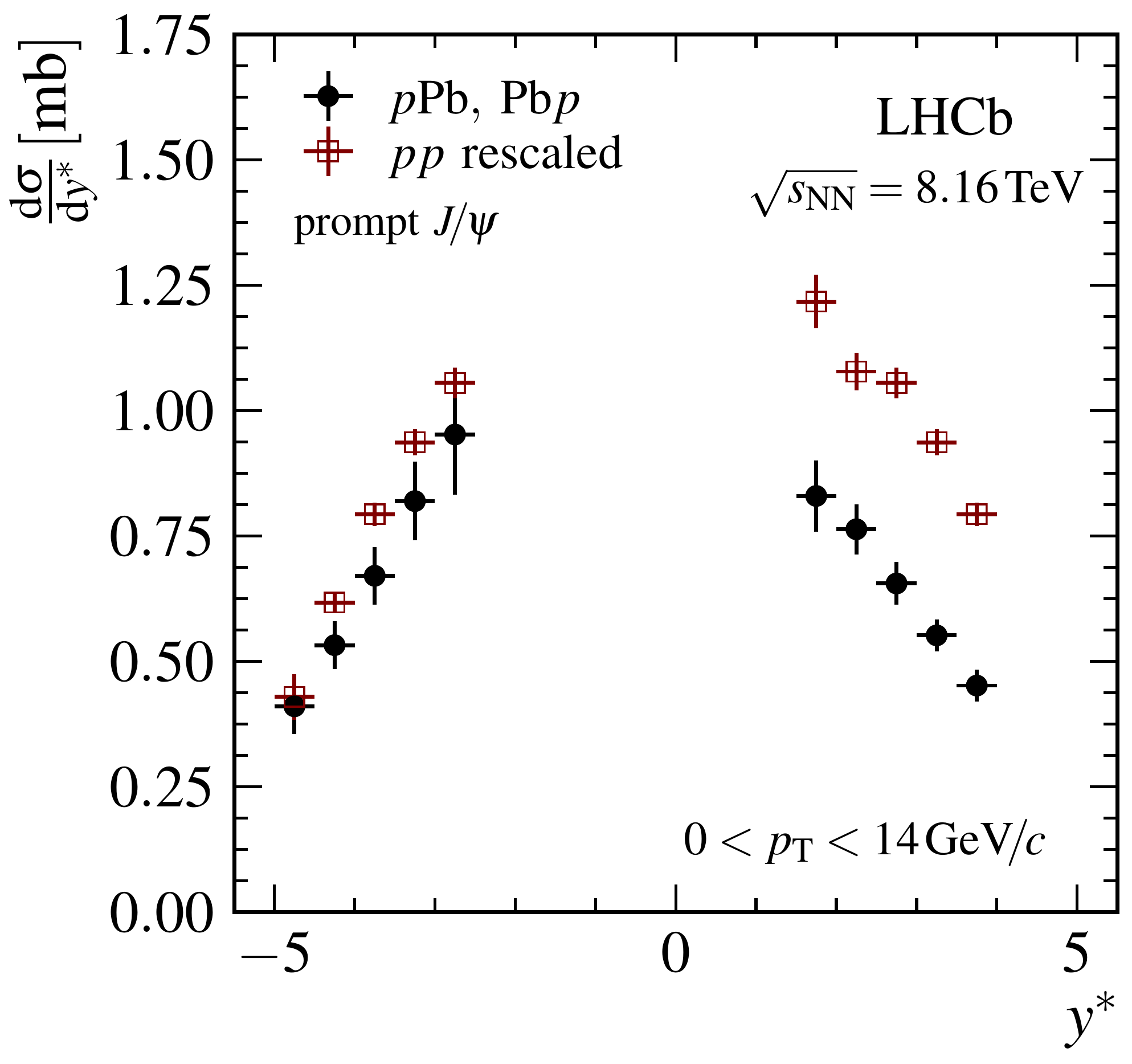}
\includegraphics[width=7.5cm]{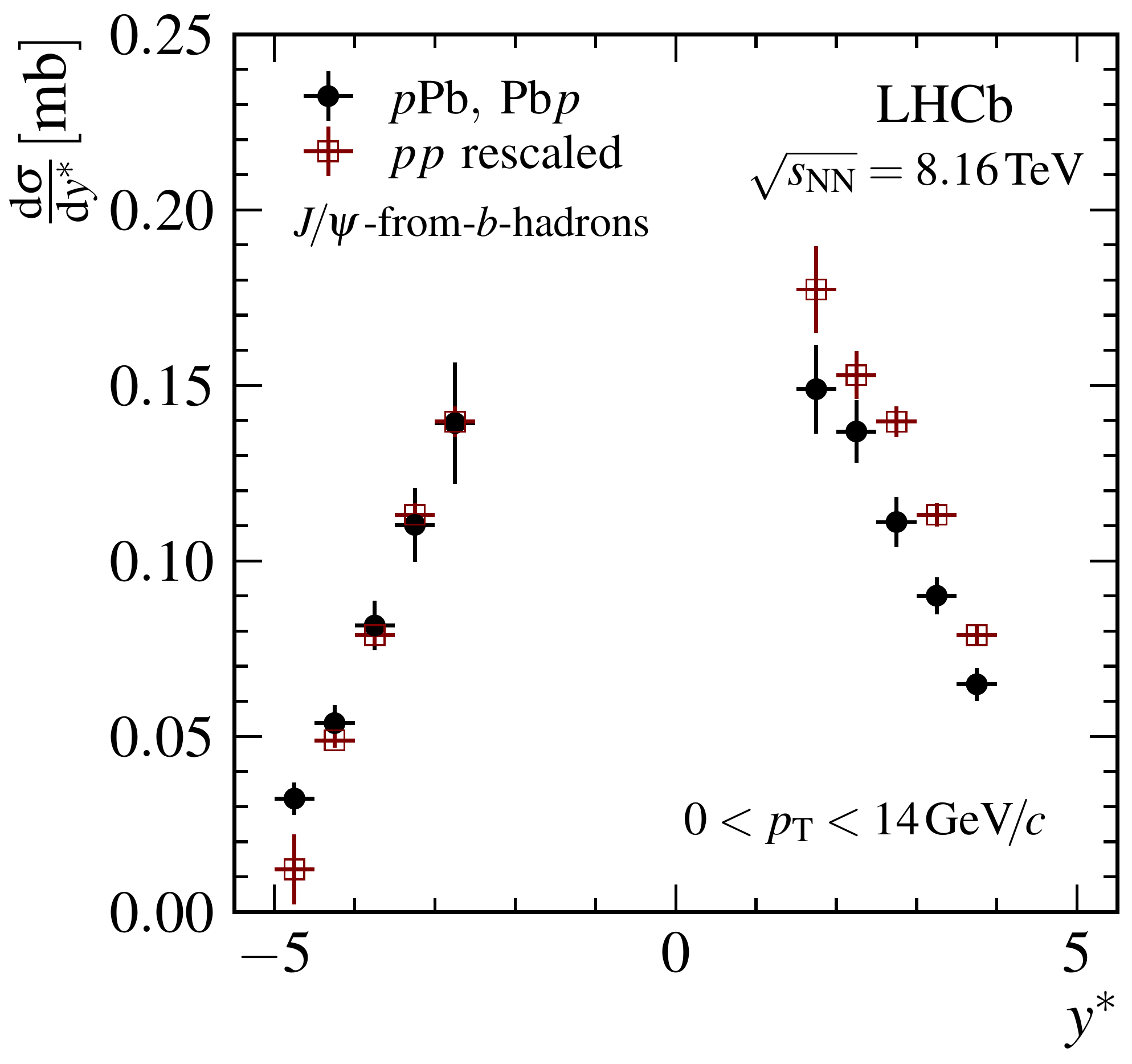}
\caption{\small Absolute production cross-sections of (left) prompt \jpsi and (right) \jpsi-from-$b$-hadrons, as a function of $y^*$, integrated
over the range $0<\pt<14\gevc$. The black circles are the $p$Pb and Pb$p$ values and the red open squares the values for $pp$ collisions
at the same energy, multiplied by the Pb  mass number $A=208$. The horizontal error bars are the bin widths and vertical error bars the total uncertainties.}
\label{fig:integrated_y_withpp}
\end{center}
\end{figure}

\begin{figure}[!t]
\begin{center}
\includegraphics[width=7.8cm]{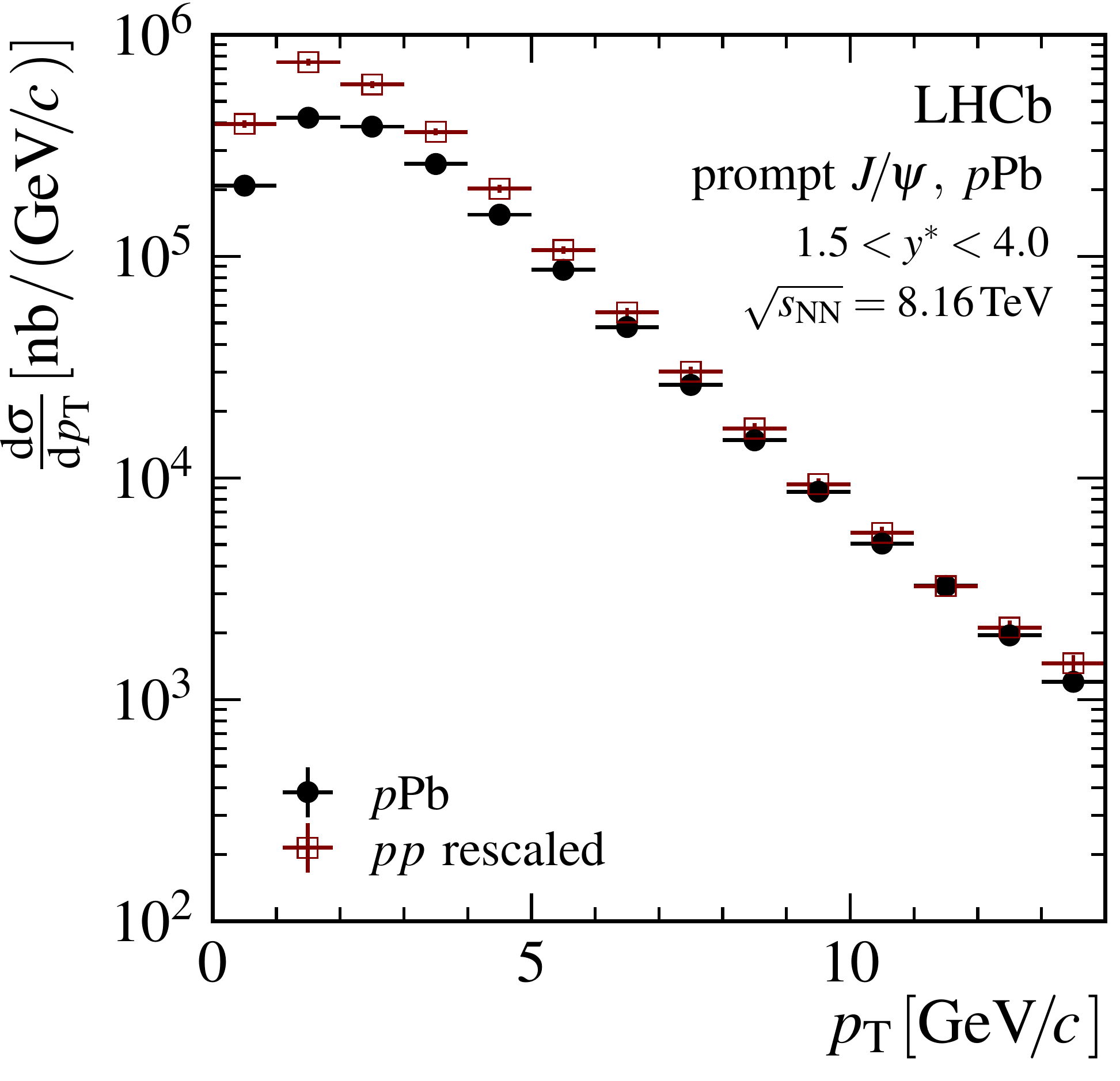}
\includegraphics[width=7.8cm]{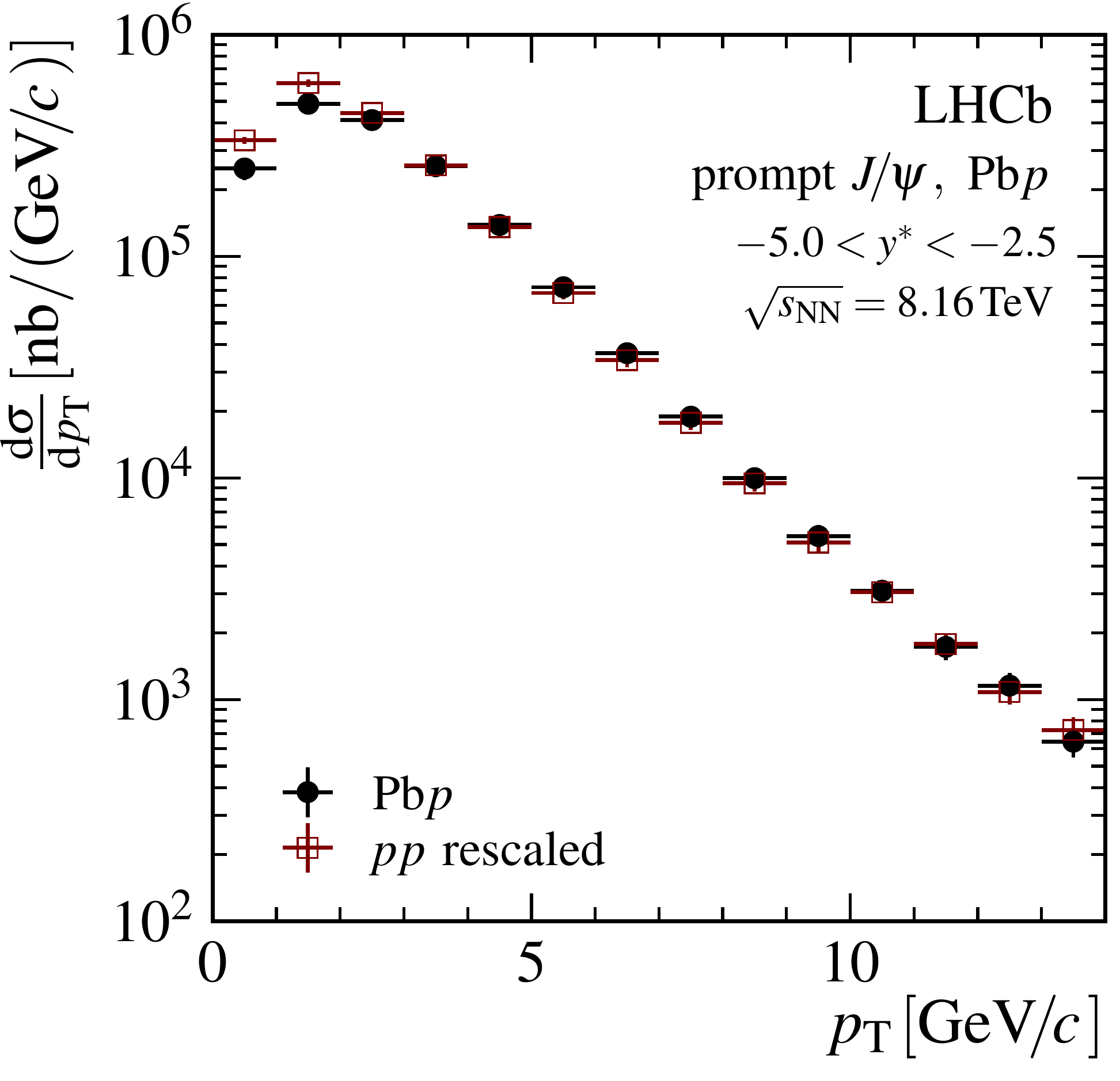}
\includegraphics[width=7.8cm]{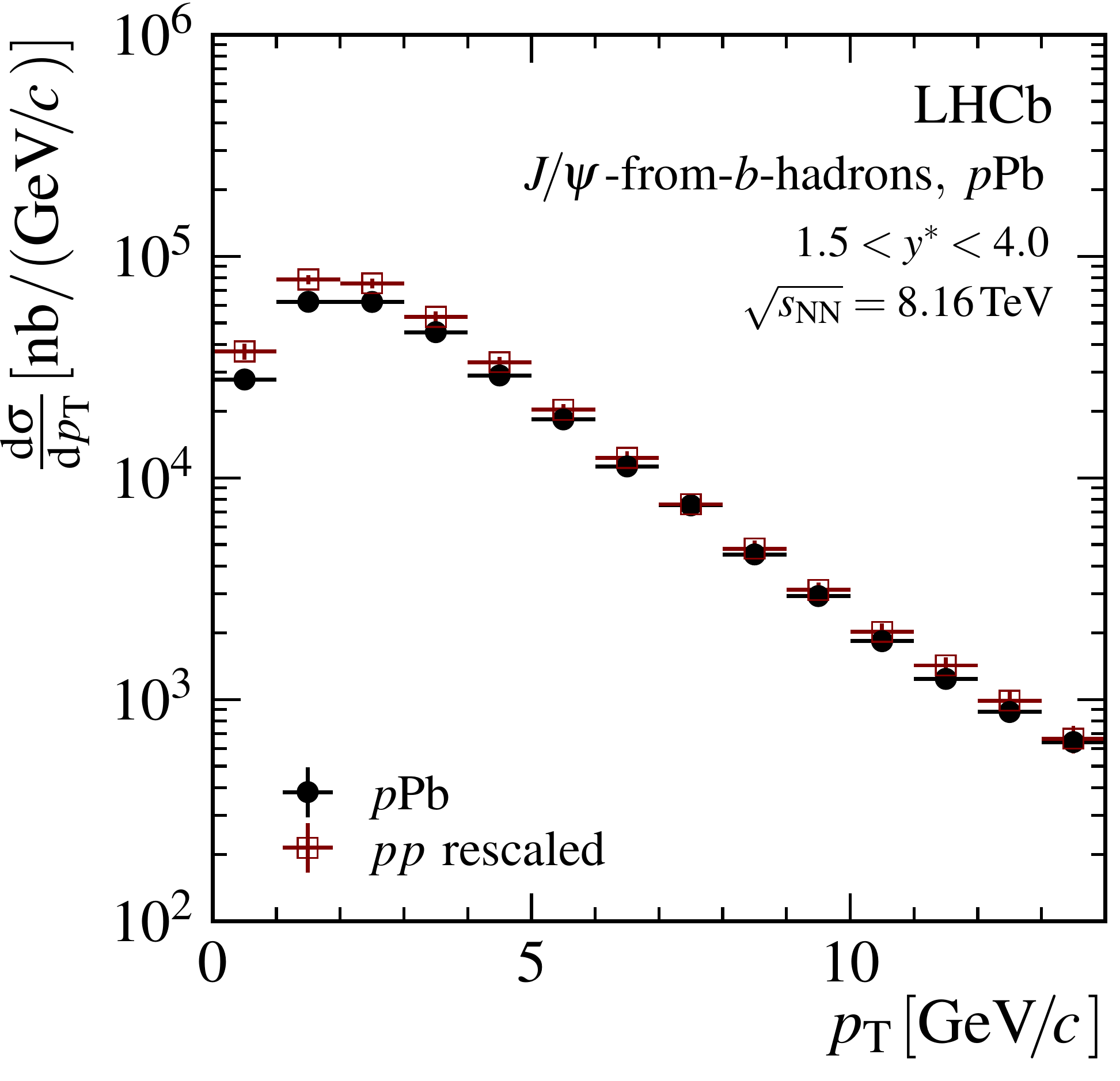}
\includegraphics[width=7.8cm]{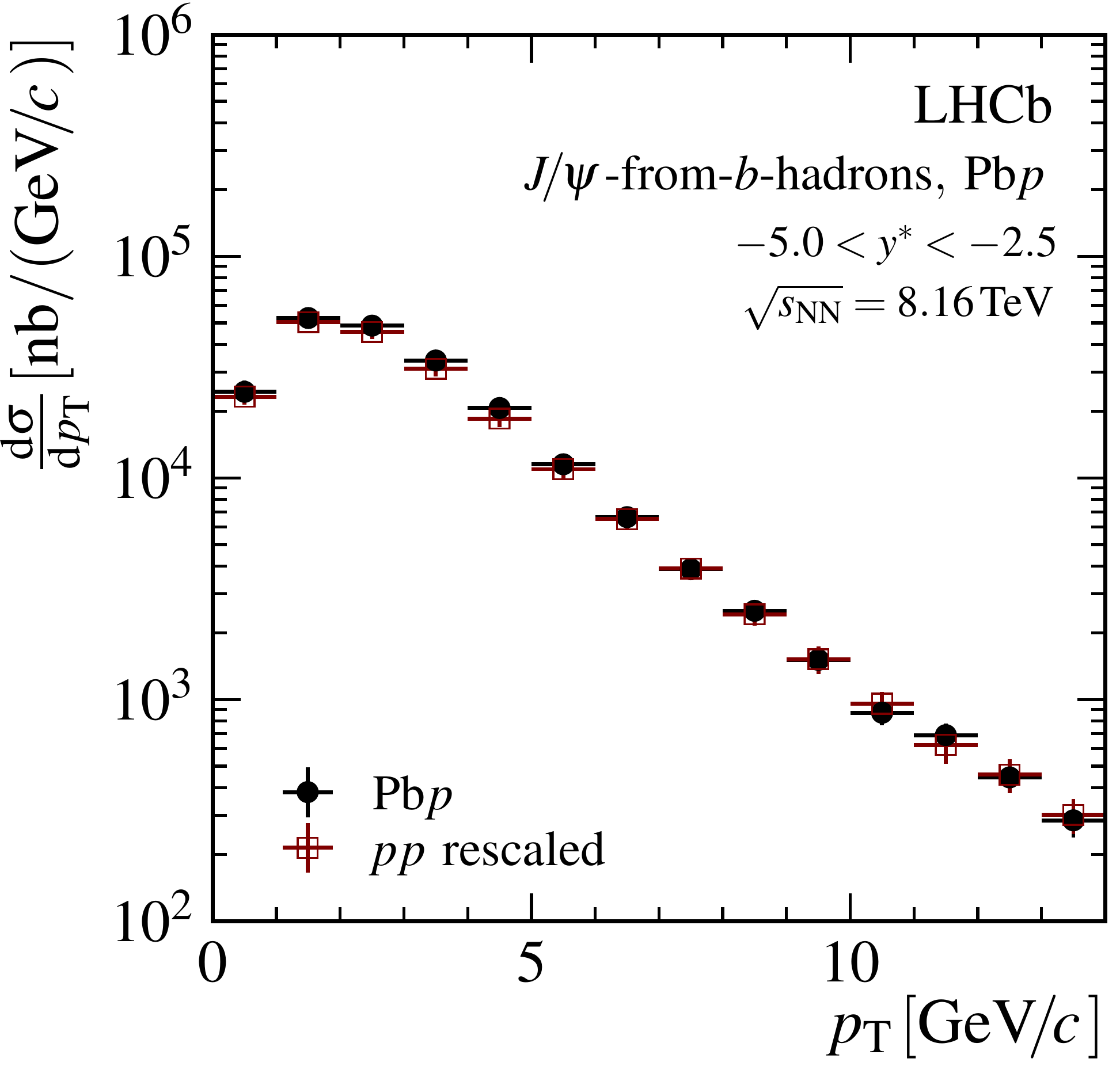}
\caption{\small Absolute production cross-sections of (top left) prompt \jpsi in $p$Pb, (top right) prompt \jpsi in Pb$p$, 
(bottom left) \jpsi-from-$b$-hadrons in Pb$p$ and (bottom left) \jpsi-from-$b$-hadrons in Pb$p$, as a function of $\pt$ and integrated
over the rapidity range of the analysis. The black circles are the $p$Pb and Pb$p$ values and the red open squares the values for $pp$ collisions
at the same energy, multiplied by the Pb  mass number $A=208$, integrated over the same rapidity ranges. 
The horizontal error bars are the bin widths and vertical error bars the total uncertainties.}
\label{fig:integrated_pt_withpp}
\end{center}
\end{figure}

The cross-section as a function of $y^*$, integrated over \pt in the range $0<\pt<14\gevc$ in $p$Pb and Pb$p$ collisions, 
is shown in Fig.~\ref{fig:integrated_y_withpp}. The cross-section is compared with the reference cross-section for prompt \jpsi and \jpsi-from-$b$-hadrons production in $pp$ collisions at $\sqrt{s}=8.16\,{\rm TeV}$, multiplied by the Pb  mass number $A=208$. 
The total relative uncertainties on the $pp$ cross-section range between $3$\% and $11$\% and are  largest in the 
bins based on extrapolations. 
The cross-sections as a function of \pt, integrated over the range $1.5<y^*<4.0$ for $p$Pb and 
$-5.0<y^*<-2.5$ for Pb$p$, and the corresponding scaled $pp$ cross-sections are represented in Fig.~\ref{fig:integrated_pt_withpp}.  In this case, the total relative uncertainties on the $pp$ reference cross-section vary between $3$\% and $18$\%.

\subsection{Nuclear modification factors}

The nuclear modification factor $R_{p{\rm Pb}}$ defined in Eq.~\eqref{eq:rpa} is computed from the prompt \jpsi and \jpsi-from-$b$-hadrons production cross-sections in $pp$ and $p$Pb or Pb$p$ collisions. The systematic uncertainties are 
 assumed to be uncorrelated between the measurements in proton-lead and in $pp$ collisions.
The  nuclear modification factors 
for prompt \jpsi and \jpsi-from-$b$-hadrons production as functions of \pt or $y^*$, integrating over the other variable, are shown in 
Figs.~\ref{fig:rpAintegrated_pt} and \ref{fig:rpAintegrated_y}, respectively. The numerical values are available in Appendix~\ref{app:rpa}.
The results at
$\sqrt{s_{NN}}=5\,{\rm TeV}$~\cite{LHCb-PAPER-2013-052} 
are also depicted on Fig.~\ref{fig:rpAintegrated_y} and are in good agreement with the new and more precise results at $\sqrt{s_{NN}}=8.16\,{\rm TeV}$. 

\begin{figure}[!t]
\begin{center}
\includegraphics[width=7.8cm]{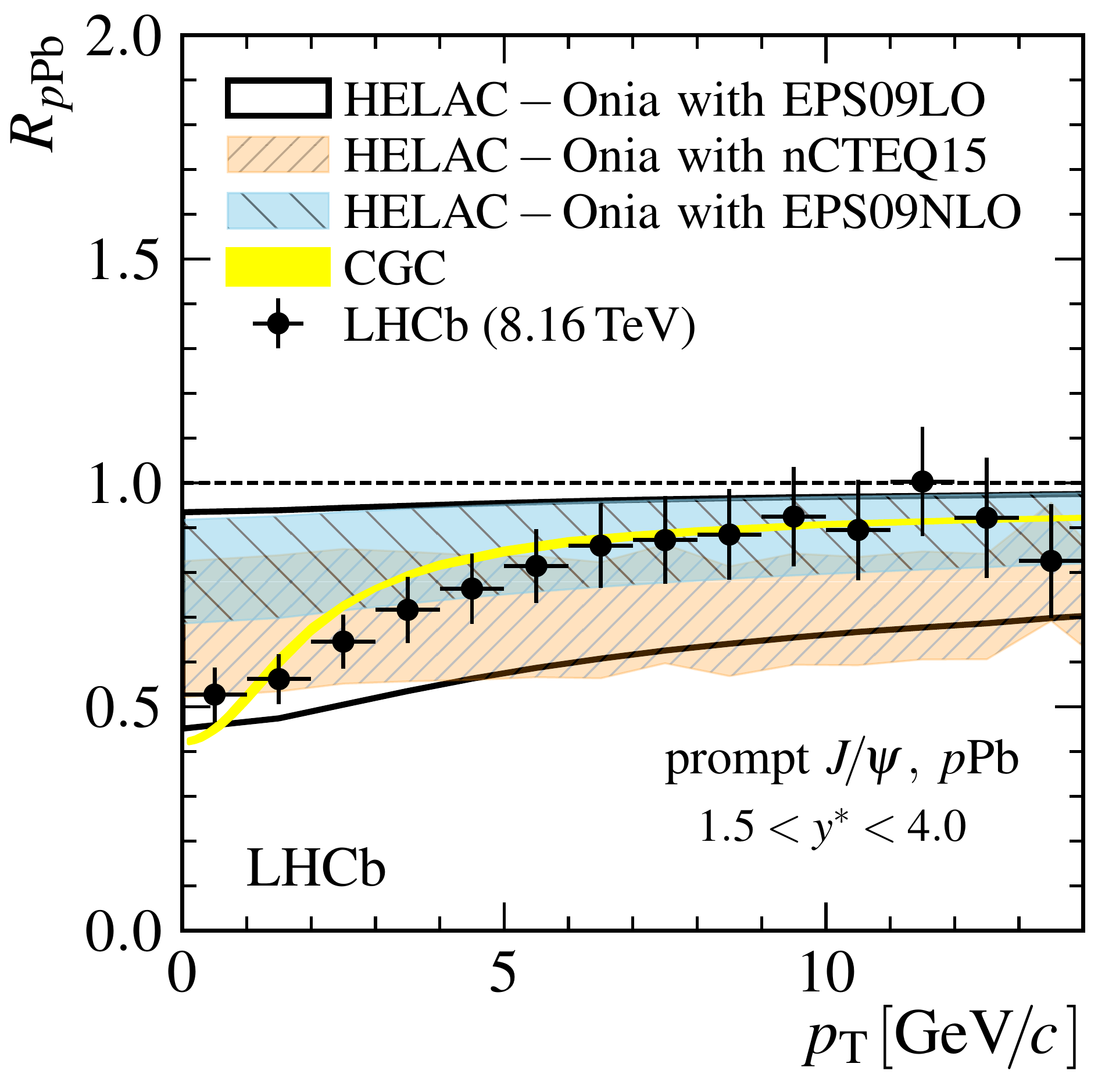}
\includegraphics[width=7.8cm]{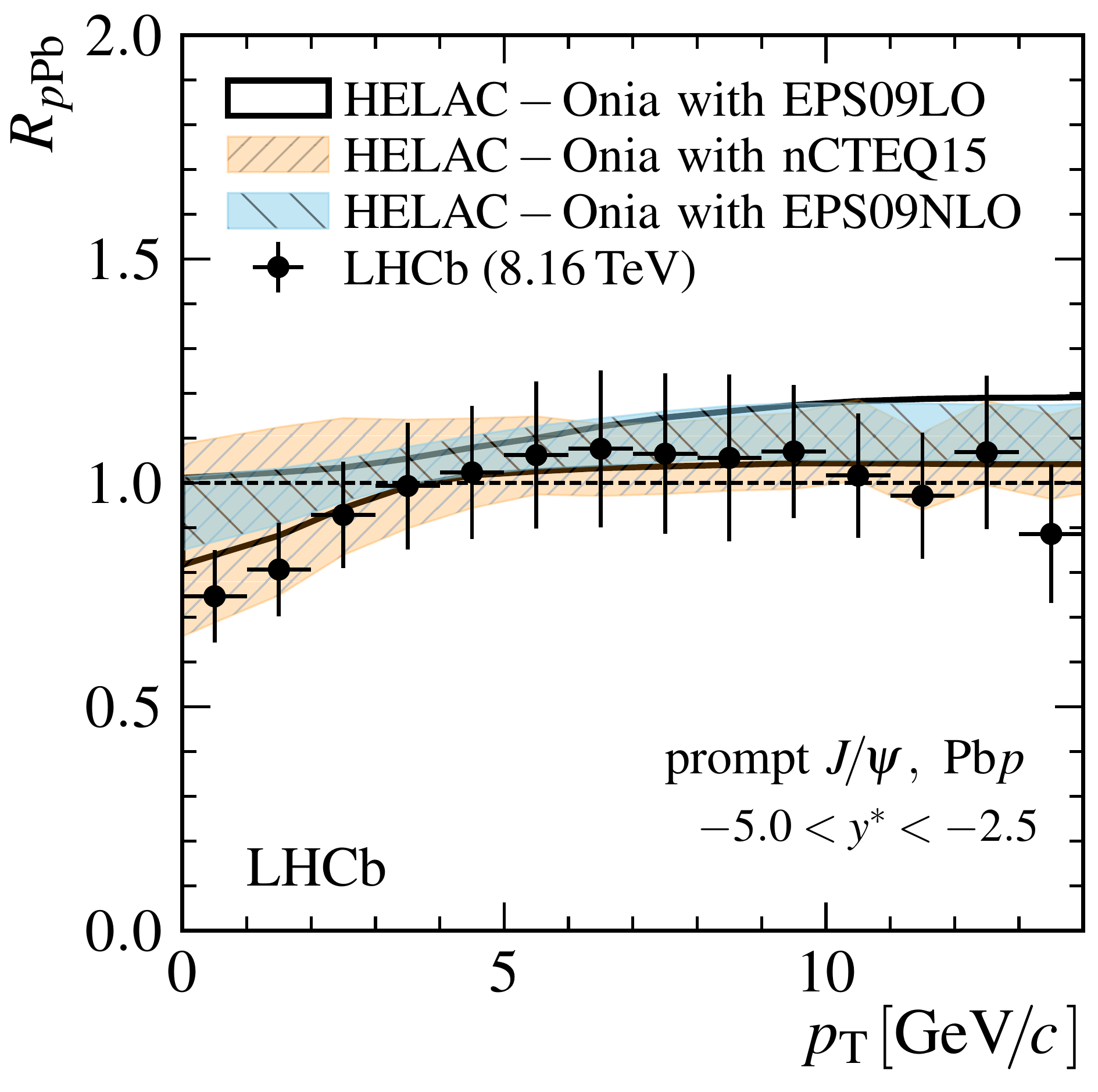}
\includegraphics[width=7.8cm]{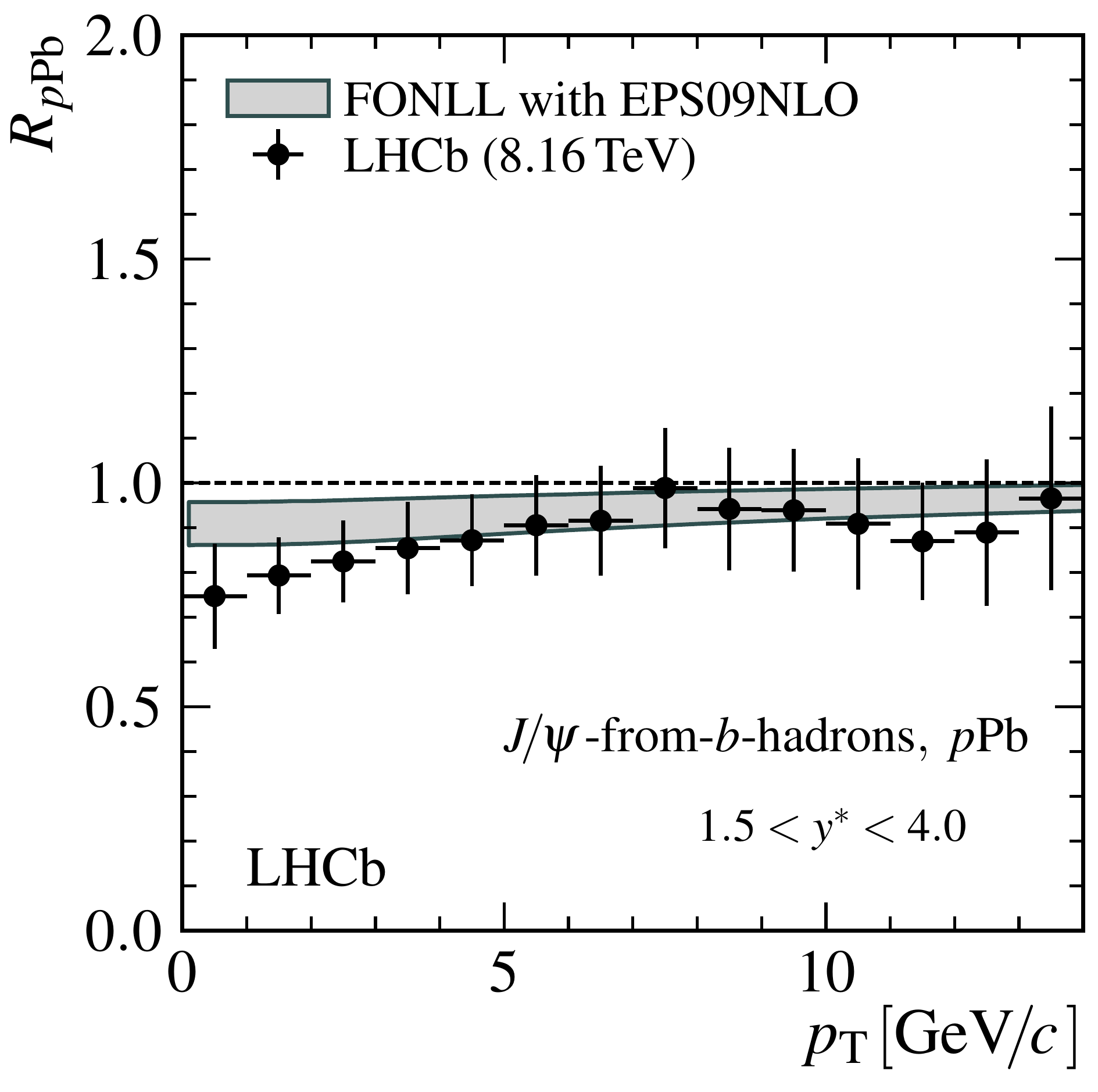}
\includegraphics[width=7.8cm]{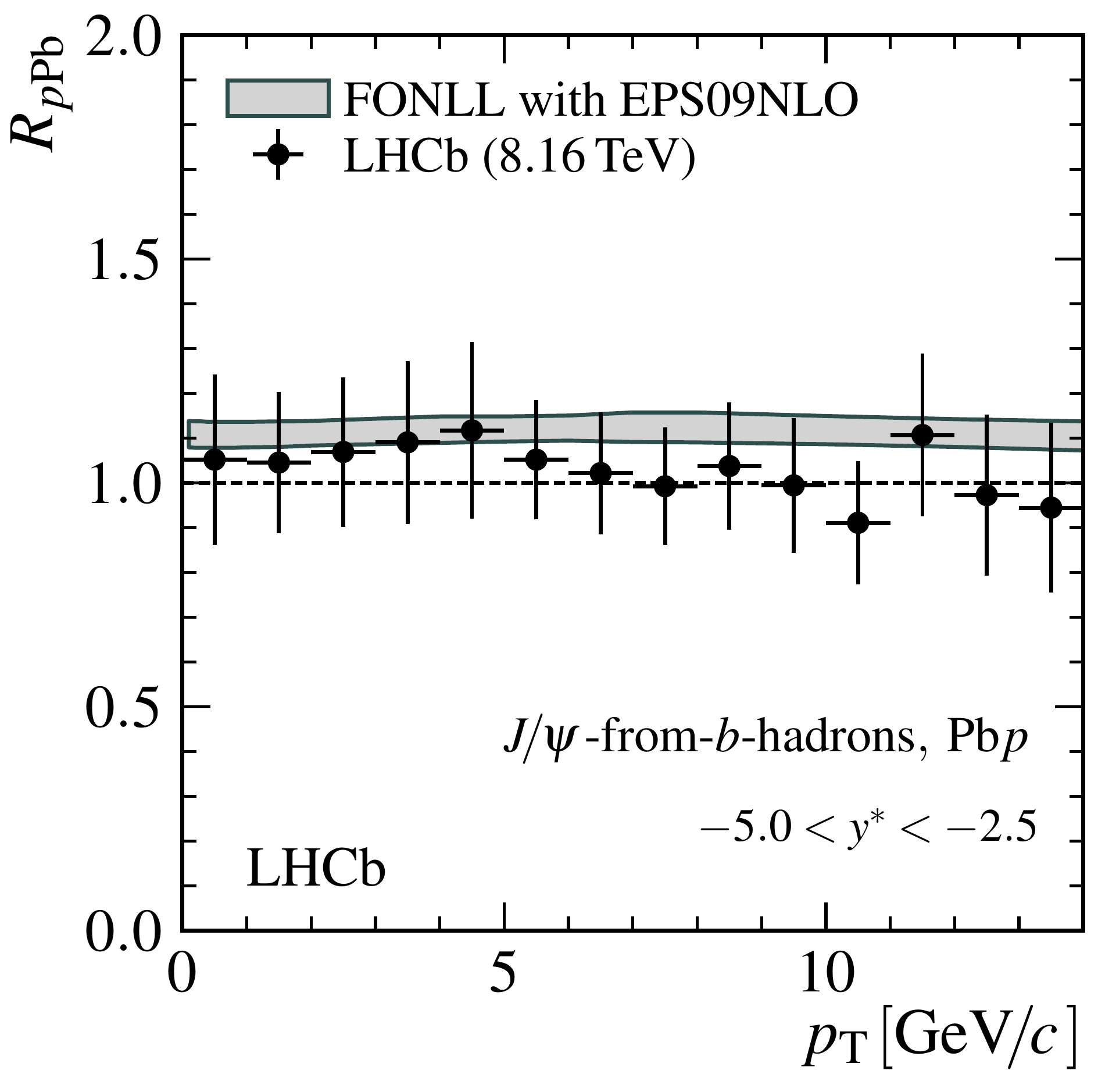}
\caption{\small  \jpsi nuclear modification factor, $R_{p{\rm Pb}}$, integrated over $y^*$ in the analysis range,                                                                                                                                                             
as a function of \pt for (top left) prompt \jpsi in $p$Pb, (bottom left) \jpsi-from-$b$-hadrons in $p$Pb, (top right) prompt \jpsi in Pb$p$
and (bottom right) \jpsi-from-$b$-hadrons in Pb$p$. Horizontal error bars are the bin widths, vertical error bars the total uncertainties.
The black circles are the values measured in this letter and the coloured areas the theoretical predictions from the models
detailed in the text with their uncertainties.}
\label{fig:rpAintegrated_pt}
\end{center}
\end{figure}

At forward rapidity, $1.5<y^{*}<4.0$, a strong suppression of up to 50\%  is observed in the case of prompt 
\jpsi production at low \pt (Fig.~\ref{fig:rpAintegrated_pt}). This behaviour results in a strong suppression in the nuclear modification factor as a function of rapidity shown in Fig.~\ref{fig:rpAintegrated_y}. 
With increasing \pt, $R_{p{\rm Pb}}$ approaches unity and the suppression is stronger at more forward rapidities.  
The production of \jpsi-from-$b$-hadrons is also suppressed compared to that in $pp$ collisions at forward rapidities, although to a lesser degree, 
as shown in Fig.~\ref{fig:rpAintegrated_y}. No dependence as a function of rapidity can be observed within the experimental uncertainties. 
The dependence as a function of the transverse momentum is weaker for \jpsi-from-$b$-hadrons compared to prompt \jpsi, but the nuclear modification factor is also approaching unity 
at high transverse momentum.

At backward rapidity, $-5.0<y^{*}<-2.5$, a weaker suppression of prompt \jpsi production at low \pt is observed, of up to 25\%. 
Similarly to the forward-rapidity region, the suppression is weakening and the nuclear modification factor is approaching values consistent 
with unity at high transverse momentum. The  nuclear modification factor as a function of rapidity shows a weak suppression with no visible rapidity 
dependence within experimental uncertainties. The nuclear modification factor of \jpsi-from-$b$-hadrons at backward rapidity is consistent with unity over the full 
kinematic region.

\begin{figure}[!t]
\begin{center}
\includegraphics[width=7.6cm]{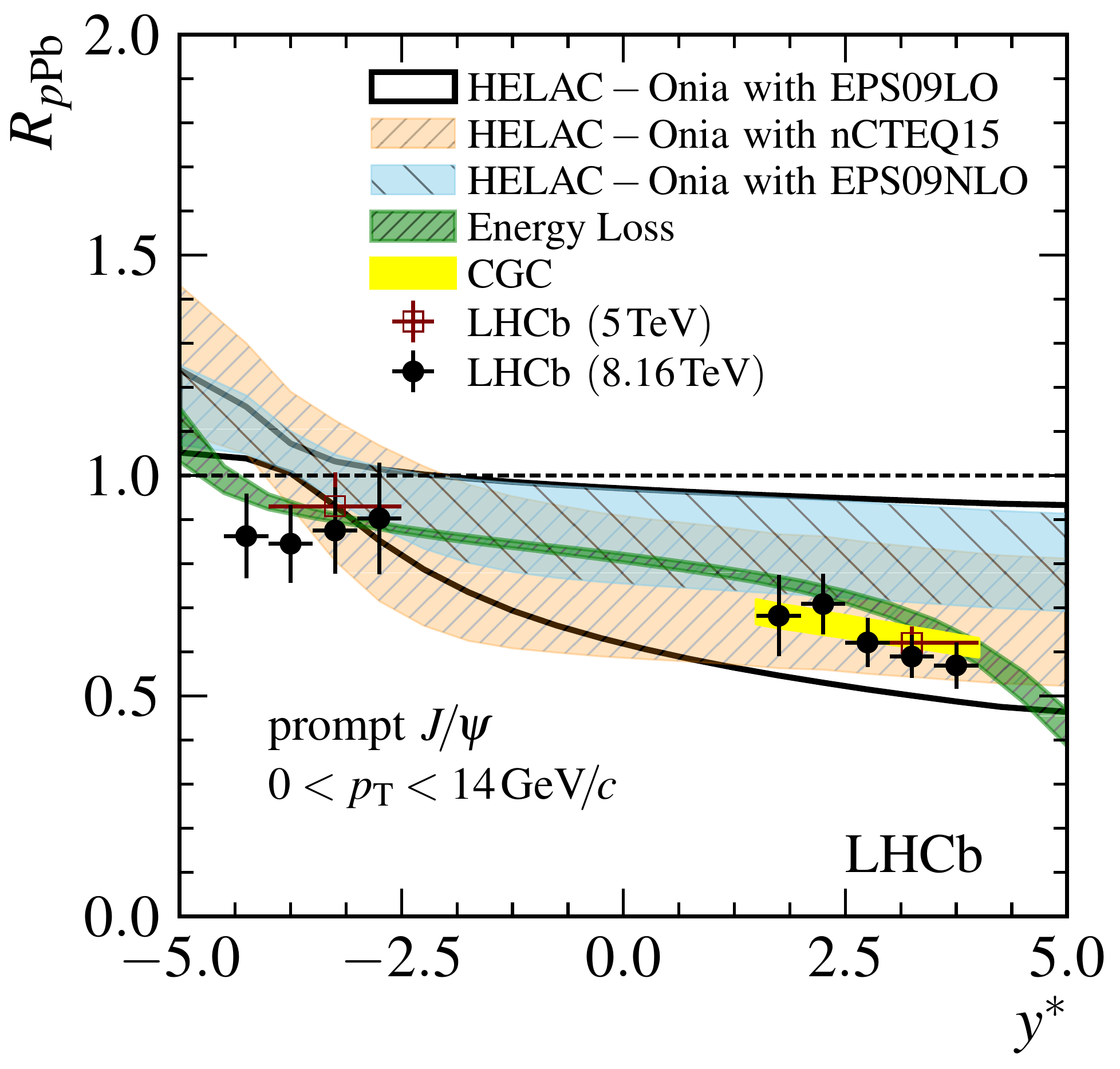}
\includegraphics[width=7.6cm]{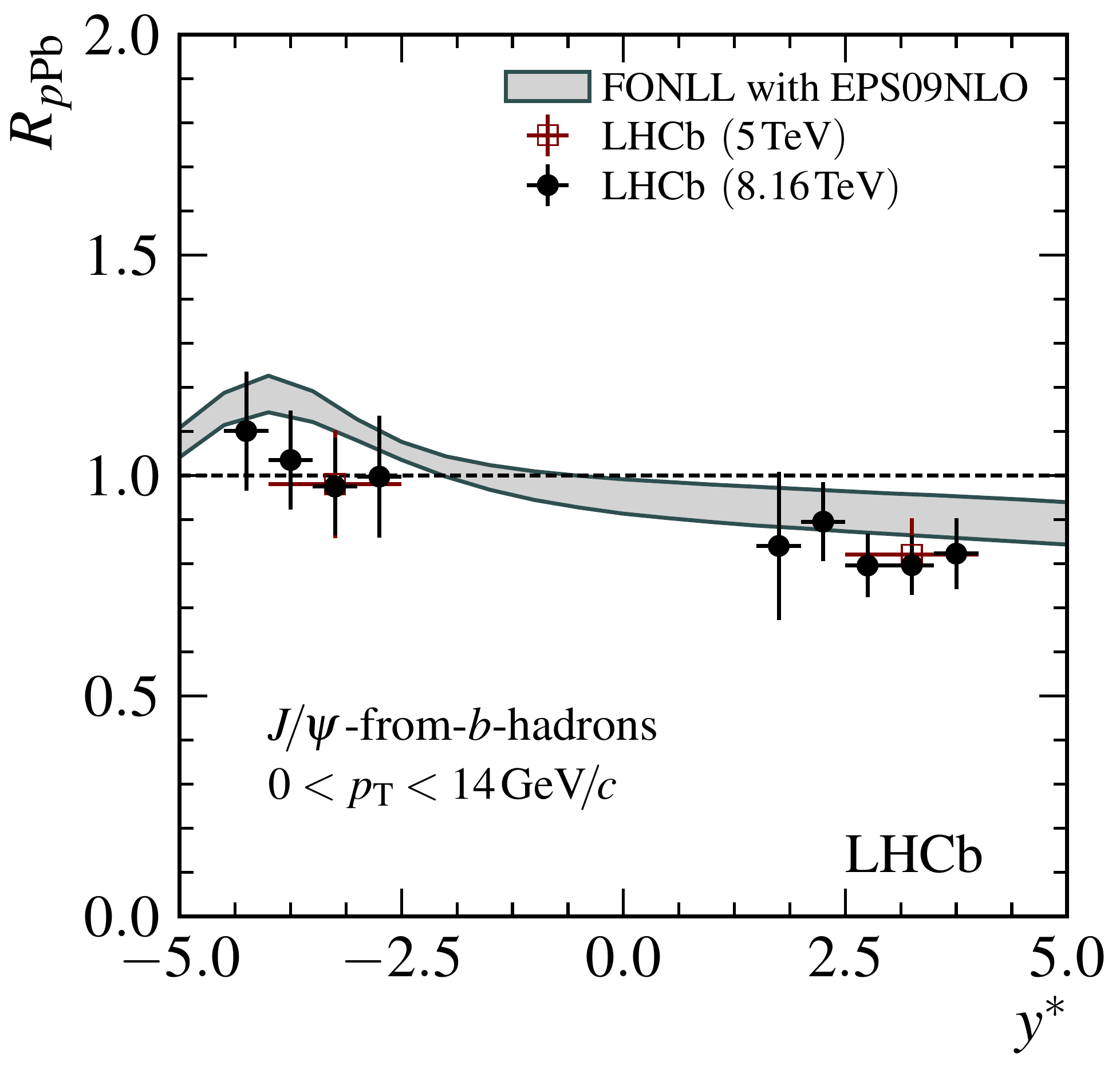}
\caption{\small \jpsi nuclear modification factor, $R_{p{\rm Pb}}$, integrated over \pt in the range $0<\pt<14\gevc$,                                                                                                                                                 
as a function of $y^*$ for (left) prompt \jpsi and (right) \jpsi-from-$b$-hadrons. The horizontal error bars are the bin widths and vertical error bars the 
total uncertainties. The black circles are the values measured in this letter, the red squares the values measured at
$\sqrt{s_{NN}}=5\,{\rm TeV}$ from Ref.~\cite{LHCb-PAPER-2013-052} and the coloured areas the theoretical computations from the models
detailed in the text, with their uncertainties.}
\label{fig:rpAintegrated_y}
\end{center}
\end{figure}

The measurements of prompt \jpsi nuclear modification factors are compared in 
Figs.~\ref{fig:rpAintegrated_pt} and \ref{fig:rpAintegrated_y} with three groups of calculations:
\begin{enumerate}
\item collinear factorisation using different nPDFs~\cite{Shao:2012iz,Shao:2015vga} 
(labelled ``HELAC-Onia with EPS09LO", ``HELAC-Onia with nCTEQ15" 
and ``HELAC-Onia with EPS09NLO" on the figures), 
\item CGC effective field theory  in the dilute-dense approximation taking into account the dense nature
of the Pb nucleus, but approximating the proton as a dilute parton source~\cite{Ducloue:2015gfa,Ducloue:2016pqr} (labelled ``CGC"),
\item coherent energy loss calculating the impact of low angle coherent gluon radiation during the crossing of the nucleus~\cite{Arleo:2012rs} (labelled ``Energy Loss").
\end{enumerate}

The CGC calculations~\cite{Ducloue:2015gfa,Ducloue:2016pqr} describe well the behaviour of the prompt \jpsi data at forward rapidity. At 
backward rapidity, this approach is not available due to the breakdown of the dilute approximation for the partons in the proton. 
The uncertainties take into account the variation of the charm-quark mass and the factorisation scale. These uncertainties  largely cancel in this ratio of cross-sections.
The collinear calculations are based on the HELAC-Onia event generator~\cite{Shao:2012iz,Shao:2015vga}, tuned to 
reproduce prompt \jpsi cross-section measurements in $pp$ collision~\cite{Lansberg:2016deg} and combined with different sets of 
nPDFs: nCTEQ15~\cite{Kovarik:2015cma} and EPS09 at leading (LO) and at next-to-leading order (NLO)~\cite{Eskola:2009uj}.
However, the large uncertainties reveal the missing experimental constraints on the gluon density in the nucleus at low $x$ probed 
by the measurements in the LHCb detector acceptance.  
At backward rapidities, the experimental points are found at the lower bound or slightly below the theoretical uncertainty bands  
and exhibit a different rapidity shape from the calculations.
The coherent energy loss model~\cite{Arleo:2012rs} is able to provide the overall shape of the suppression, but overestimates 
the experimental data at forward rapidities. The uncertainty of this calculation reflects the allowed variation of the parameterisation of $pp$ data used in the model and the allowed variation of the only free model parameter from fits to other measurements. 

The measurements of $\jpsi$-from-$b$-hadrons nuclear modification factors are compared in Figs.~\ref{fig:rpAintegrated_pt} and \ref{fig:rpAintegrated_y} with a perturbative QCD calculation at fixed-order next-to-leading-logarithms (FONLL)~\cite{Cacciari:1998it,Cacciari:2001td} coupled with the EPS09 nPDF set at next-to-leading order~\cite{Eskola:2009uj} (labelled ``FONLL with EPS09NLO" on the figures). The displayed uncertainties correspond to the uncertainties from the nPDF, which are of similar size to or smaller than the total experimental uncertainties.  The \pt dependence of the experimental data is described within uncertainties by the model. However, the calculation tends to show larger nuclear modification factors than the data. This tendency is confirmed by the nuclear modification factor as a function of rapidity, where the most precise experimental data points are  below the model uncertainty band. Furthermore, at backward rapidity, the slope of the theoretical curve is not seen in the experimental data.

Finally, recent measurements have shown that long-range collective effects, which have previously been observed in relatively large nucleus-nucleus collision systems, may also be present in smaller collision systems at large charged-particle multiplicites~\cite{CMS:2012qk,Abelev:2012ola,Aad:2012gla,Adare:2013piz}. If these effects have a hydrodynamic origin, momentum anisotropies at the quark level can arise and may modify the distribution of observed heavy-quark hadrons~\cite{Beraudo:2015wsd}. However, the expected magnitude of these effects on prompt \jpsi~or $\jpsi$-from-$b$-hadrons production has not yet been calculated. Since the measurements in this letter are integrated over charged-particle multiplicity, potential modifications in high-multiplicity events are diluted.

\subsection{Forward-to-backward ratios}

\begin{figure}[!t]
\begin{center}
\includegraphics[width=7.4cm]{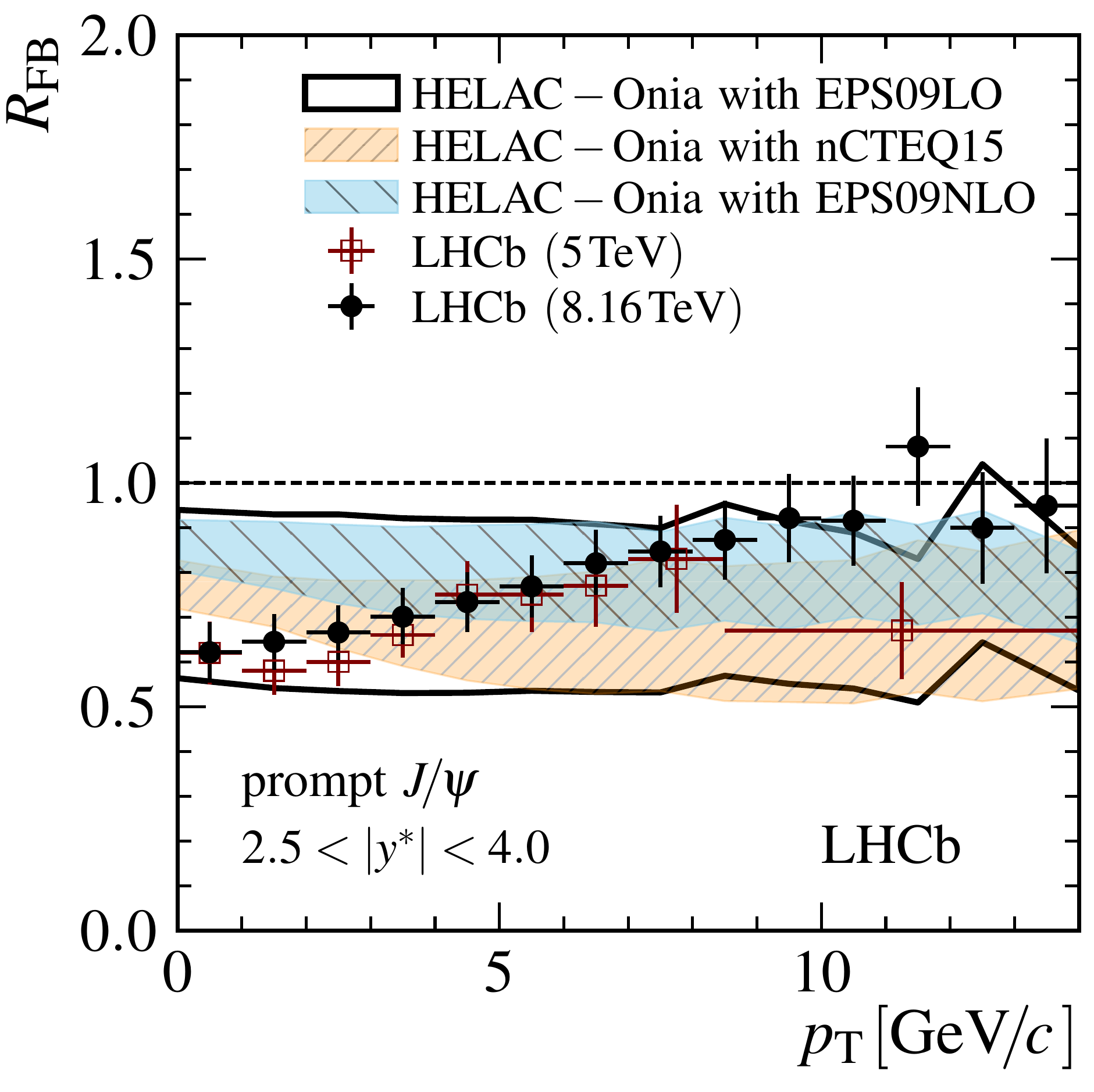}
\includegraphics[width=7.4cm]{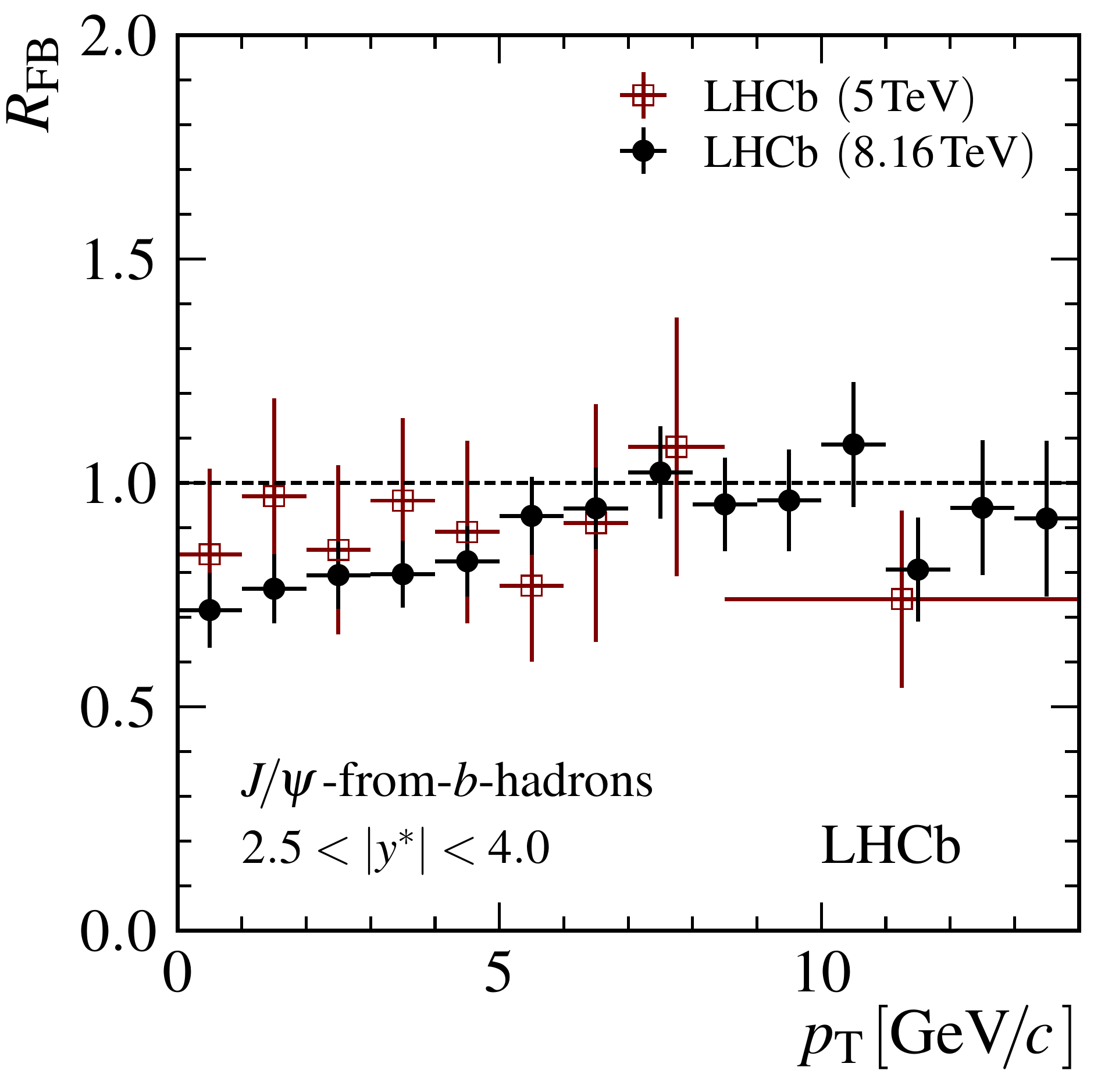}
\caption{\small Forward-to-backward ratios, $R_{{\rm FB}}$, integrated over the common rapidity range $2.5<|y^*|<4.0$          
as a function of \pt for (left) prompt \jpsi and (right) \jpsi-from-$b$-hadrons.                                               
The horizontal error bars are the bin widths and the vertical error                                                            
bars the total uncertainties. The black circles are the values measured in this letter, the red squares the values measured at
$\sqrt{s_{NN}}=5\,{\rm TeV}$ from Ref.~\cite{LHCb-PAPER-2013-052} and the coloured areas the theoretical computations from the
models detailed in the text, with their uncertainties.}
\label{fig:rfb_pt}
\end{center}
\end{figure}

\begin{figure}[!t]
\begin{center}
\includegraphics[width=7.8cm]{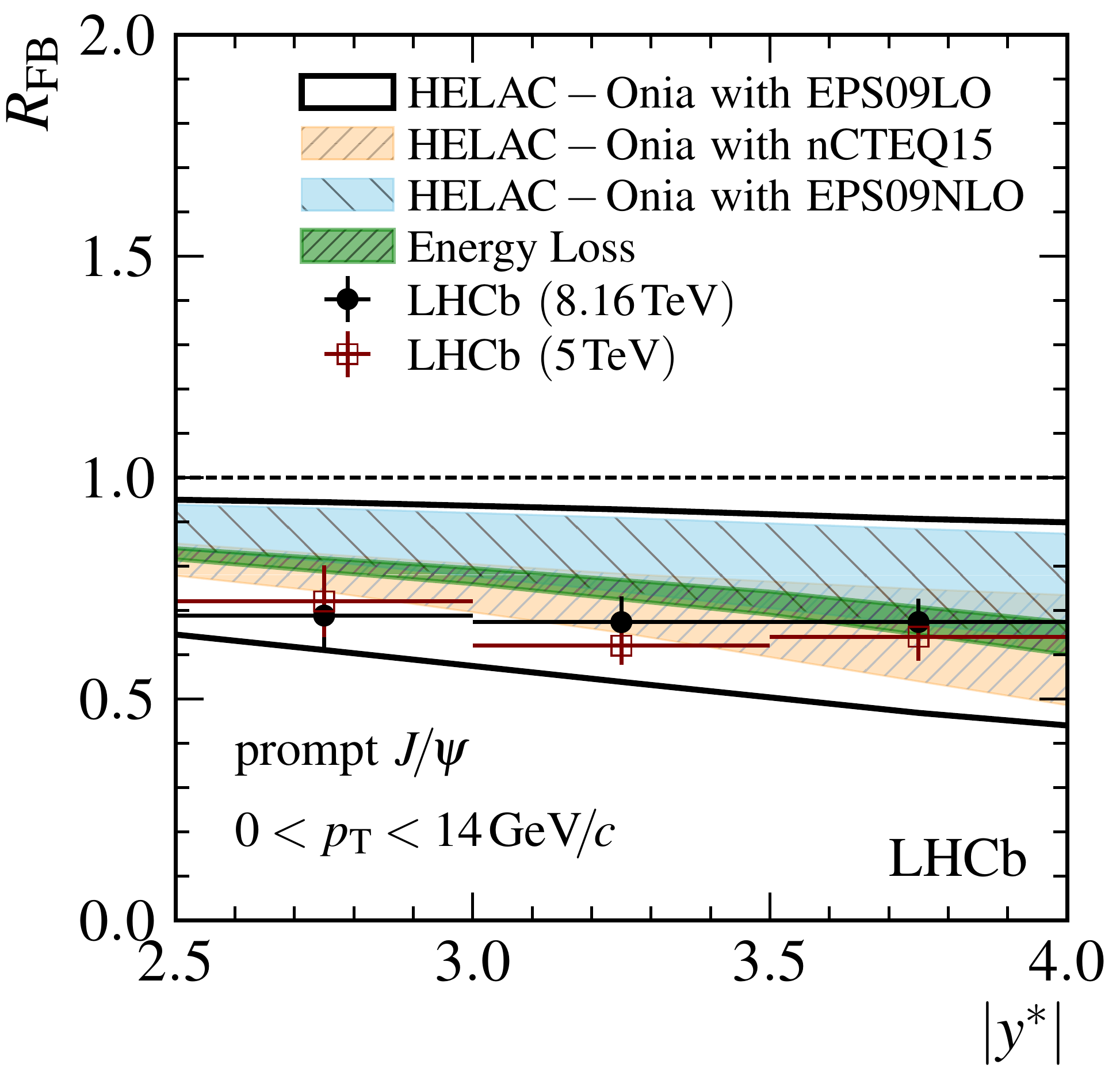}
\includegraphics[width=7.8cm]{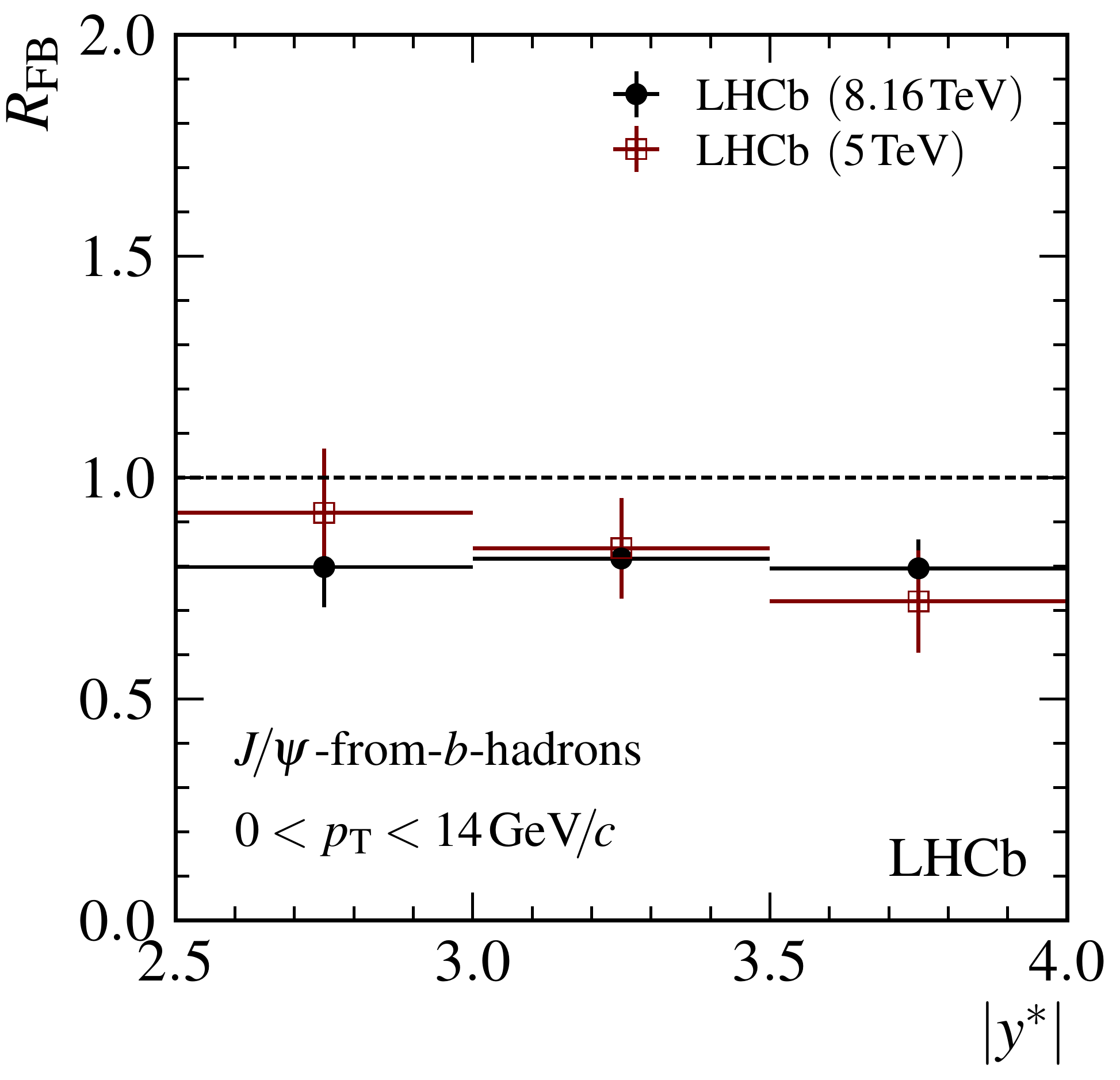}
\caption{\small Forward-to-backward ratios, $R_{{\rm FB}}$, integrated over \pt in the range $0<\pt<14\gevc$
as a function of $|y^{*}|$ for (left) prompt \jpsi and (right) \jpsi-from-$b$-hadrons. The horizontal error bars are the bin widths and the vertical error 
bars the total uncertainties. The black circles are the values measured in this letter, the red squares the values measured at 
$\sqrt{s_{NN}}=5\,{\rm TeV}$ from Ref.~\cite{LHCb-PAPER-2013-052} and the coloured areas the theoretical computations from the
models detailed in the text, with their uncertainties.}
\label{fig:rfb_rap}
\end{center}
\end{figure}

Figures~\ref{fig:rfb_pt} and \ref{fig:rfb_rap} show the forward-to-backward ratio, $R_{\rm FB}$, of the production of prompt \jpsi 
and \jpsi-from-$b$-hadrons, in the overlapping acceptance between the two beam configurations, as functions of transverse momentum and 
rapidity, respectively. 
The numerical results are listed in Appendix~\ref{app:rfb}.
In the $R_{\rm FB}$ ratio, most of the systematic uncertainties cancel. The measurements of $R_{\rm FB}$ at 
$\sqrt{s_{NN}}=5\,{\rm TeV}$~\cite{LHCb-PAPER-2013-052} are compared with the measurements at $8.16\,{\rm TeV}$ and are
found to be in agreement. They are compared with the theoretical computations based on collinear factorisation with different
nPDFs described in the previous section.

The calculations with different nPDFs do not fully cover the experimental points within uncertainties in particular at low \pt with the exception of the EPS09LO combination, which has considerably larger uncertainties. However, a detailed analysis of theoretical correlations in the \pt-dependent $R_{\rm FB}$ may be interesting for future studies in order to quantify more precisely the discrepancies. The coherent energy loss calculation is compared with the rapidity dependence of the experimental data points in Fig.~\ref{fig:rfb_rap}. It shows within its small uncertainties a slightly different slope from the experimental data points and predicts larger values in the bin at smallest $|y^{*}|$.

The $R_{\rm FB}$ ratio of $\jpsi$-from-b-hadrons in Fig.~\ref{fig:rfb_pt} shows a rising trend as a function of transverse momentum starting from a value 0.7 at low \pt towards values consistent with unity at high \pt. The rapidity dependence of $R_{\rm FB}$ in Fig.~\ref{fig:rfb_rap} is consistent with a flat behaviour with a central value of 0.8.

\section{Conclusions}

The differential production cross-sections of prompt \jpsi and \jpsi-from-$b$-hadrons in $p$Pb and Pb$p$ collisions at 
$\sqrt{s_{NN}}=8.16\,{\rm TeV}$ are measured in the range $0<\pt <14\gevc$. 
The nuclear modification factors are similar to the findings at a collision energy of $\sqrt{s_{NN}}= 5\,{\rm TeV}$, 
but with increased precision thanks to 10 and 40 times larger data sets in $p$Pb and Pb$p$ collisions, respectively. 
A suppression of prompt \jpsi production compared to $pp$ collisions of up to 50\% (25\%) in $p$Pb (Pb$p$) at the lowest transverse momentum is observed. 
 In both configurations, 
the nuclear modification factor approaches unity asymptotically at the highest \pt. 
Theoretical calculations for the nuclear modification factor based on collinear factorisation 
with different nuclear parton distribution functions, coherent energy loss as well as 
the colour glass condensate model can account for the majority of the observed dependences. 
For the first time, beauty-hadron production is measured precisely down to $\pt=0$ at the LHC in $p$Pb and Pb$p$ collisions. 
In $p$Pb, a weak suppression at the lowest transverse 
momenta is observed, whereas in Pb$p$ no significant deviation from unity in the nuclear modification factor is found. 
This weak modification of beauty production in proton-ion collisions is an important ingredient for the investigation 
of the modifications of beauty production in heavy-ion collisions.
Although the presented measurements have improved precision, it is not possible to single out the main nuclear modification mechanism 
between different phenomenological approaches for charmonium production in proton-lead collisions at the TeV scale. 
This measurement of \jpsi production is the first step towards measurements of other charmonium states as well as complementary 
observables like Drell-Yan production, to improve the understanding of quantum chromodynamics at low $x$ and in dense nuclear environments.


\section*{Acknowledgements}

We thank G.~Bruno, B.~Duclou\'e, H.~Shao, J.-Ph.~Lansberg and F.~Arl\'eo for providing theoretical predictions
of \jpsi production in $p$Pb and Pb$p$ collisions in the LHCb acceptance range.
 We express our gratitude to our colleagues in the CERN
accelerator departments for the excellent performance of the LHC. We
thank the technical and administrative staff at the LHCb
institutes. We acknowledge support from CERN and from the national
agencies: CAPES, CNPq, FAPERJ and FINEP (Brazil); MOST and NSFC (China);
CNRS/IN2P3 (France); BMBF, DFG and MPG (Germany); INFN (Italy); 
NWO (The Netherlands); MNiSW and NCN (Poland); MEN/IFA (Romania); 
MinES and FASO (Russia); MinECo (Spain); SNSF and SER (Switzerland); 
NASU (Ukraine); STFC (United Kingdom); NSF (USA).
We acknowledge the computing resources that are provided by CERN, IN2P3 (France), KIT and DESY (Germany), INFN (Italy), SURF (The 
Netherlands), PIC (Spain), GridPP (United Kingdom), RRCKI and Yandex LLC (Russia), CSCS (Switzerland), IFIN-HH (Romania), CBPF 
(Brazil), PL-GRID (Poland) and OSC (USA). We are indebted to the communities behind the multiple open 
source software packages on which we depend.
Individual groups or members have received support from AvH Foundation (Germany),
EPLANET, Marie Sk\l{}odowska-Curie Actions and ERC (European Union), 
Conseil G\'{e}n\'{e}ral de Haute-Savoie, Labex ENIGMASS and OCEVU, 
R\'{e}gion Auvergne (France), RFBR and Yandex LLC (Russia), GVA, XuntaGal and GENCAT (Spain), Herchel Smith Fund, The Royal Society, 
Royal Commission for the Exhibition of 1851 and the Leverhulme Trust (United Kingdom).

\clearpage

{\noindent\normalfont\bfseries\Large Appendices}

\appendix

\section{Cross-section numerical results}\label{app:crosssection}

\subsection{\texorpdfstring{$\bm{\frac{{\rm d}^2\sigma}{{\rm d}p_{\rm T}{\rm d}y^*}}$}{$\frac{{\rm d}^2\sigma}{{\rm d}p_{\rm T}{\rm d}y^*}$} for prompt \texorpdfstring{$\bm{J} \mskip -3mu/\mskip -2mu\bm{\psi}\mskip 2mu$}{\jpsi} in \texorpdfstring{$\bm{p}$}{$p$}Pb}\label{app:crosssectionpPbprompt}

\begin{table}[h!]
\caption{\small Prompt \jpsi absolute production cross-section in $p$Pb, as a function of \pt and $y^*$.
The quoted uncertainties are the total uncertainties, and the breakdown into statistical uncertainties, and correlated and 
uncorrelated systematic uncertainties.}
\label{tab:promptcrosssectionresultspPb}
\small
\centering
\renewcommand{\arraystretch}{0.96}
\begin{tabular}{@{}l@{$\,<\,\pt<\,$}ll@{$\,<\,y^*<\,$}llr@{$\,\pm\,$}@{}rlrrr@{}}
\toprule
\multicolumn{2}{@{}l}{\pt bin (\gevc)}  & \multicolumn{2}{c}{$y^*$ bin} & \multicolumn{4}{c}{$\frac{{\rm d}^2\sigma}{{\rm d}p_{\rm T}{\rm d}y^*}$  [nb/(\gevc)]} & stat. & corr. & uncorr. \\
\midrule
 \phantom{1}0& \phantom{1}1& $1.5$ & $2.0$ & & 108\,700& 16\,000& & 2\,700& 15\,700& 1\,700 \\ 
\phantom{1}0& \phantom{1}1& $2.0$ & $2.5$ & & 94\,300& 8\,900& & 1\,400& 8\,800& 700 \\ 
\phantom{1}0& \phantom{1}1& $2.5$ & $3.0$ & & 79\,700& 5\,400& & 1\,100& 5\,200& 500 \\ 
\phantom{1}0& \phantom{1}1& $3.0$ & $3.5$ & & 69\,800& 4\,300& & 1\,000& 4\,200& 400 \\ 
\phantom{1}0& \phantom{1}1& $3.5$ & $4.0$ & & 64\,000& 4\,100& & 1\,100& 3\,900& 500 \\ 
\phantom{1}1& \phantom{1}2& $1.5$ & $2.0$ & & 212\,200& 18\,100& & 3\,300& 17\,700& 2\,000 \\ 
\phantom{1}1& \phantom{1}2& $2.0$ & $2.5$ & & 194\,000& 12\,000& & 2\,000& 12\,000& 1\,000 \\ 
\phantom{1}1& \phantom{1}2& $2.5$ & $3.0$ & & 166\,400& 14\,300& & 1\,500& 14\,200& 700 \\ 
\phantom{1}1& \phantom{1}2& $3.0$ & $3.5$ & & 144\,800& 7\,900& & 1\,400& 7\,700& 600 \\ 
\phantom{1}1& \phantom{1}2& $3.5$ & $4.0$ & & 126\,100& 8\,800& & 1\,500& 8\,700& 700 \\ 
\phantom{1}2& \phantom{1}3& $1.5$ & $2.0$ & & 192\,600& 14\,800& & 2\,800& 14\,400& 1\,900 \\ 
\phantom{1}2& \phantom{1}3& $2.0$ & $2.5$ & & 180\,000& 11\,000& & 2\,000& 11\,000& 1\,000 \\ 
\phantom{1}2& \phantom{1}3& $2.5$ & $3.0$ & & 157\,800& 8\,900& & 1\,400& 8\,800& 800 \\ 
\phantom{1}2& \phantom{1}3& $3.0$ & $3.5$ & & 131\,400& 7\,300& & 1\,300& 7\,100& 700 \\ 
\phantom{1}2& \phantom{1}3& $3.5$ & $4.0$ & & 107\,500& 7\,600& & 1\,400& 7\,400& 800 \\ 
\phantom{1}3& \phantom{1}4& $1.5$ & $2.0$ & & 133\,100& 12\,200& & 2\,100& 11\,900& 1\,500 \\ 
\phantom{1}3& \phantom{1}4& $2.0$ & $2.5$ & & 123\,800& 8\,700& & 1\,200& 8\,600& 800 \\ 
\phantom{1}3& \phantom{1}4& $2.5$ & $3.0$ & & 108\,500& 7\,300& & 1\,000& 7\,200& 600 \\ 
\phantom{1}3& \phantom{1}4& $3.0$ & $3.5$ & & 88\,200& 5\,700& & 900& 5\,600& 500 \\ 
\phantom{1}3& \phantom{1}4& $3.5$ & $4.0$ & & 67\,900& 5\,300& & 900& 5\,200& 600 \\ 
\phantom{1}4& \phantom{1}5& $1.5$ & $2.0$ & & 78\,800& 6\,700& & 1\,400& 6\,500& 1\,000 \\ 
\phantom{1}4& \phantom{1}5& $2.0$ & $2.5$ & & 74\,600& 4\,800& & 800& 4\,700& 500 \\ 
\phantom{1}4& \phantom{1}5& $2.5$ & $3.0$ & & 64\,400& 3\,900& & 700& 3\,800& 400 \\ 
\phantom{1}4& \phantom{1}5& $3.0$ & $3.5$ & & 52\,500& 3\,200& & 600& 3\,200& 400 \\ 
\phantom{1}4& \phantom{1}5& $3.5$ & $4.0$ & & 37\,700& 2\,800& & 700& 2\,700& 400 \\ 
\phantom{1}5& \phantom{1}6& $1.5$ & $2.0$ & & 45\,600& 3\,700& & 900& 3\,500& 700 \\ 
\phantom{1}5& \phantom{1}6& $2.0$ & $2.5$ & & 42\,500& 2\,700& & 500& 2\,600& 400 \\ 
\phantom{1}5& \phantom{1}6& $2.5$ & $3.0$ & & 34\,750& 2\,080& & 460& 2\,010& 280 \\ 
\phantom{1}5& \phantom{1}6& $3.0$ & $3.5$ & & 29\,790& 1\,940& & 440& 1\,870& 260 \\ 
\phantom{1}5& \phantom{1}6& $3.5$ & $4.0$ & & 21\,100& 1\,680& & 460& 1\,600& 250 \\ 
\phantom{1}6& \phantom{1}7& $1.5$ & $2.0$ & & 25\,200& 2\,100& & 600& 2\,000& 400 \\ 
\phantom{1}6& \phantom{1}7& $2.0$ & $2.5$ & & 23\,940& 1\,680& & 380& 1\,620& 250 \\ 
\phantom{1}6& \phantom{1}7& $2.5$ & $3.0$ & & 19\,050& 1\,350& & 320& 1\,300& 190 \\ 
\phantom{1}6& \phantom{1}7& $3.0$ & $3.5$ & & 15\,500& 1\,110& & 300& 1\,050& 170 \\ 
\phantom{1}6& \phantom{1}7& $3.5$ & $4.0$ & & 12\,230& 1\,090& & 340& 1\,020& 190 \\ 
 
\bottomrule
\end{tabular}
\end{table}

\begin{table}[h!]
\caption{\small Prompt \jpsi absolute production cross-section in $p$Pb, as a function of \pt and $y^*$.
The quoted uncertainties are the total uncertainties, and the breakdown into statistical uncertainties, and correlated and 
uncorrelated systematic uncertainties.}
\label{tab:promptcrosssectionresultspPb1}
\small
\centering
\renewcommand{\arraystretch}{0.96}
\begin{tabular}{@{}l@{$\,<\,\pt<\,$}ll@{$\,<\,y^*<\,$}llr@{$\,\pm\,$}@{}rlrrr@{}}
\toprule
\multicolumn{2}{@{}l}{\pt bin (\gevc)}  & \multicolumn{2}{c}{$y^*$ bin} & \multicolumn{4}{c}{$\frac{{\rm d}^2\sigma}{{\rm d}p_{\rm T}{\rm d}y^*}$  [nb/(\gevc)]} & stat. & corr. & uncorr. \\
\midrule
 \phantom{1}7& \phantom{1}8& $1.5$ & $2.0$ & & 14\,410& 1\,170& & 440& 1\,030& 330 \\ 
\phantom{1}7& \phantom{1}8& $2.0$ & $2.5$ & & 12\,660& 800& & 260& 740& 160 \\ 
\phantom{1}7& \phantom{1}8& $2.5$ & $3.0$ & & 10\,260& 680& & 230& 630& 130 \\ 
\phantom{1}7& \phantom{1}8& $3.0$ & $3.5$ & & 8\,870& 660& & 230& 600& 130 \\ 
\phantom{1}7& \phantom{1}8& $3.5$ & $4.0$ & & 6\,310& 660& & 240& 600& 120 \\ 
\phantom{1}8& \phantom{1}9& $1.5$ & $2.0$ & & 7\,700& 620& & 290& 500& 210 \\ 
\phantom{1}8& \phantom{1}9& $2.0$ & $2.5$ & & 7\,440& 490& & 190& 430& 120 \\ 
\phantom{1}8& \phantom{1}9& $2.5$ & $3.0$ & & 6\,060& 410& & 170& 360& 100 \\ 
\phantom{1}8& \phantom{1}9& $3.0$ & $3.5$ & & 4\,640& 360& & 160& 310& 90 \\ 
\phantom{1}8& \phantom{1}9& $3.5$ & $4.0$ & & 3\,700& 400& & 200& 400& 100 \\ 
\phantom{1}9& 10& $1.5$ & $2.0$ & & 4\,810& 420& & 220& 320& 160 \\ 
\phantom{1}9& 10& $2.0$ & $2.5$ & & 4\,270& 300& & 140& 240& 90 \\ 
\phantom{1}9& 10& $2.5$ & $3.0$ & & 3\,360& 260& & 130& 210& 70 \\ 
\phantom{1}9& 10& $3.0$ & $3.5$ & & 2\,680& 240& & 120& 190& 70 \\ 
\phantom{1}9& 10& $3.5$ & $4.0$ & & 2\,200& 280& & 130& 230& 70 \\ 
10& 11& $1.5$ & $2.0$ & & 2\,630& 240& & 160& 150& 100 \\ 
10& 11& $2.0$ & $2.5$ & & 2\,620& 200& & 110& 150& 60 \\ 
10& 11& $2.5$ & $3.0$ & & 2\,230& 180& & 100& 130& 60 \\ 
10& 11& $3.0$ & $3.5$ & & 1\,490& 150& & 80& 110& 40 \\ 
10& 11& $3.5$ & $4.0$ & & 1\,130& 170& & 90& 140& 40 \\ 
11& 12& $1.5$ & $2.0$ & & 1\,840& 190& & 120& 110& 90 \\ 
11& 12& $2.0$ & $2.5$ & & 1\,600& 130& & 90& 90& 50 \\ 
11& 12& $2.5$ & $3.0$ & & 1\,300& 120& & 80& 80& 50 \\ 
11& 12& $3.0$ & $3.5$ & & 1\,000& 110& & 70& 80& 40 \\ 
11& 12& $3.5$ & $4.0$ & & 750& 110& & 80& 70& 40 \\ 
12& 13& $1.5$ & $2.0$ & & 1\,190& 140& & 100& 80& 70 \\ 
12& 13& $2.0$ & $2.5$ & & 958& 94& & 64& 59& 33 \\ 
12& 13& $2.5$ & $3.0$ & & 779& 82& & 58& 49& 31 \\ 
12& 13& $3.0$ & $3.5$ & & 531& 71& & 51& 41& 26 \\ 
12& 13& $3.5$ & $4.0$ & & 436& 80& & 47& 61& 21 \\ 
13& 14& $1.5$ & $2.0$ & & 740& 100& & 70& 40& 50 \\ 
13& 14& $2.0$ & $2.5$ & & 596& 65& & 49& 34& 25 \\ 
13& 14& $2.5$ & $3.0$ & & 476& 59& & 45& 27& 24 \\ 
13& 14& $3.0$ & $3.5$ & & 349& 47& & 35& 27& 15 \\ 
13& 14& $3.5$ & $4.0$ & & 241& 47& & 38& 21& 16 \\ 
 
\bottomrule
\end{tabular}
\end{table}

\clearpage

\subsection{\texorpdfstring{$\bm{\frac{{\rm d}^2\sigma}{{\rm d}p_{\rm T}{\rm d}y^*}}$}{$\frac{{\rm d}^2\sigma}{{\rm d}p_{\rm T}{\rm d}y^*}$} for \texorpdfstring{$\bm{J} \mskip -3mu/\mskip -2mu\bm{\psi}\mskip 2mu$}{\jpsi}-from-\texorpdfstring{$\bm{b}$}{$b$}-hadrons in \texorpdfstring{$\bm{p}$}{$p$}Pb}\label{app:crosssectionpPbb}

\begin{table}[h!]
\caption{\small  \jpsi-from-$b$-hadrons absolute production cross-section in $p$Pb, as a function of \pt and $y^*$.
The quoted uncertainties are the total uncertainties, and the breakdown into statistical uncertainties, and correlated and 
uncorrelated systematic uncertainties.}
\label{tab:bcrosssectionresultspPb}
\small
\centering
\renewcommand{\arraystretch}{0.96}
\begin{tabular}{@{}l@{$\,<\,\pt<\,$}ll@{$\,<\,y^*<\,$}llr@{$\,\pm\,$}@{}rlrrr@{}}
\toprule
\multicolumn{2}{@{}l}{\pt bin (\gevc)}  & \multicolumn{2}{c}{$y^*$ bin} & \multicolumn{4}{c}{$\frac{{\rm d}^2\sigma}{{\rm d}p_{\rm T}{\rm d}y^*}$  [nb/(\gevc)]} & stat. & corr. & uncorr. \\
\midrule
 \phantom{1}0& \phantom{1}1& $1.5$ & $2.0$ & & 15\,580& 2\,480& & 1\,020& 2\,250& 240 \\ 
\phantom{1}0& \phantom{1}1& $2.0$ & $2.5$ & & 13\,400& 1\,300& & 500& 1\,300& 100 \\ 
\phantom{1}0& \phantom{1}1& $2.5$ & $3.0$ & & 10\,320& 780& & 370& 680& 60 \\ 
\phantom{1}0& \phantom{1}1& $3.0$ & $3.5$ & & 8\,940& 640& & 350& 540& 50 \\ 
\phantom{1}0& \phantom{1}1& $3.5$ & $4.0$ & & 7\,330& 600& & 400& 440& 60 \\ 
\phantom{1}1& \phantom{1}2& $1.5$ & $2.0$ & & 32\,950& 3\,050& & 1\,290& 2\,740& 320 \\ 
\phantom{1}1& \phantom{1}2& $2.0$ & $2.5$ & & 29\,550& 1\,980& & 670& 1\,850& 150 \\ 
\phantom{1}1& \phantom{1}2& $2.5$ & $3.0$ & & 24\,530& 2\,170& & 540& 2\,100& 110 \\ 
\phantom{1}1& \phantom{1}2& $3.0$ & $3.5$ & & 20\,390& 1\,200& & 510& 1\,090& 90 \\ 
\phantom{1}1& \phantom{1}2& $3.5$ & $4.0$ & & 17\,180& 1\,330& & 600& 1\,180& 100 \\ 
\phantom{1}2& \phantom{1}3& $1.5$ & $2.0$ & & 32\,980& 2\,740& & 1\,160& 2\,460& 320 \\ 
\phantom{1}2& \phantom{1}3& $2.0$ & $2.5$ & & 30\,480& 2\,010& & 650& 1\,900& 170 \\ 
\phantom{1}2& \phantom{1}3& $2.5$ & $3.0$ & & 25\,420& 1\,520& & 530& 1\,420& 120 \\ 
\phantom{1}2& \phantom{1}3& $3.0$ & $3.5$ & & 21\,100& 1\,260& & 500& 1\,150& 110 \\ 
\phantom{1}2& \phantom{1}3& $3.5$ & $4.0$ & & 14\,440& 1\,140& & 550& 1\,000& 100 \\ 
\phantom{1}3& \phantom{1}4& $1.5$ & $2.0$ & & 24\,320& 2\,370& & 900& 2\,180& 280 \\ 
\phantom{1}3& \phantom{1}4& $2.0$ & $2.5$ & & 23\,150& 1\,690& & 510& 1\,600& 140 \\ 
\phantom{1}3& \phantom{1}4& $2.5$ & $3.0$ & & 18\,590& 1\,300& & 410& 1\,230& 100 \\ 
\phantom{1}3& \phantom{1}4& $3.0$ & $3.5$ & & 14\,810& 1\,030& & 390& 940& 90 \\ 
\phantom{1}3& \phantom{1}4& $3.5$ & $4.0$ & & 10\,030& 860& & 390& 760& 80 \\ 
\phantom{1}4& \phantom{1}5& $1.5$ & $2.0$ & & 14\,650& 1\,370& & 620& 1\,210& 190 \\ 
\phantom{1}4& \phantom{1}5& $2.0$ & $2.5$ & & 15\,480& 1\,050& & 380& 970& 110 \\ 
\phantom{1}4& \phantom{1}5& $2.5$ & $3.0$ & & 12\,160& 790& & 300& 720& 80 \\ 
\phantom{1}4& \phantom{1}5& $3.0$ & $3.5$ & & 9\,410& 640& & 290& 570& 70 \\ 
\phantom{1}4& \phantom{1}5& $3.5$ & $4.0$ & & 6\,250& 540& & 310& 440& 60 \\ 
\phantom{1}5& \phantom{1}6& $1.5$ & $2.0$ & & 10\,090& 910& & 460& 770& 150 \\ 
\phantom{1}5& \phantom{1}6& $2.0$ & $2.5$ & & 9\,270& 630& & 270& 560& 80 \\ 
\phantom{1}5& \phantom{1}6& $2.5$ & $3.0$ & & 7\,560& 500& & 220& 440& 60 \\ 
\phantom{1}5& \phantom{1}6& $3.0$ & $3.5$ & & 6\,080& 440& & 210& 380& 50 \\ 
\phantom{1}5& \phantom{1}6& $3.5$ & $4.0$ & & 3\,710& 350& & 210& 280& 40 \\ 
\phantom{1}6& \phantom{1}7& $1.5$ & $2.0$ & & 6\,560& 630& & 330& 520& 120 \\ 
\phantom{1}6& \phantom{1}7& $2.0$ & $2.5$ & & 5\,600& 430& & 190& 380& 60 \\ 
\phantom{1}6& \phantom{1}7& $2.5$ & $3.0$ & & 4\,630& 360& & 160& 320& 50 \\ 
\phantom{1}6& \phantom{1}7& $3.0$ & $3.5$ & & 3\,620& 290& & 160& 250& 40 \\ 
\phantom{1}6& \phantom{1}7& $3.5$ & $4.0$ & & 2\,114& 240& & 161& 176& 32 \\ 
 
\bottomrule
\end{tabular}
\end{table}

\begin{table}[h!]
\caption{\small \jpsi-from-$b$-hadrons absolute production cross-section in $p$Pb, as a function of \pt and $y^*$.
The quoted uncertainties are the total uncertainties, and the breakdown into statistical uncertainties, and correlated and 
uncorrelated systematic uncertainties.}
\label{tab:bcrosssectionresultspPb1}
\small
\centering
\renewcommand{\arraystretch}{0.96}
\begin{tabular}{@{}l@{$\,<\,\pt<\,$}ll@{$\,<\,y^*<\,$}llr@{$\,\pm\,$}@{}rlrrr@{}}
\toprule
\multicolumn{2}{@{}l}{\pt bin (\gevc)}  & \multicolumn{2}{c}{$y^*$ bin} & \multicolumn{4}{c}{$\frac{{\rm d}^2\sigma}{{\rm d}p_{\rm T}{\rm d}y^*}$  [nb/(\gevc)]} & stat. & corr. & uncorr. \\
\midrule
 \phantom{1}7& \phantom{1}8& $1.5$ & $2.0$ & & 4\,610& 430& & 250& 330& 100 \\ 
\phantom{1}7& \phantom{1}8& $2.0$ & $2.5$ & & 3\,650& 260& & 140& 210& 50 \\ 
\phantom{1}7& \phantom{1}8& $2.5$ & $3.0$ & & 3\,010& 230& & 130& 180& 40 \\ 
\phantom{1}7& \phantom{1}8& $3.0$ & $3.5$ & & 2\,327& 203& & 123& 157& 34 \\ 
\phantom{1}7& \phantom{1}8& $3.5$ & $4.0$ & & 1\,432& 183& & 120& 135& 28 \\ 
\phantom{1}8& \phantom{1}9& $1.5$ & $2.0$ & & 2\,590& 260& & 180& 170& 70 \\ 
\phantom{1}8& \phantom{1}9& $2.0$ & $2.5$ & & 2\,300& 180& & 110& 130& 40 \\ 
\phantom{1}8& \phantom{1}9& $2.5$ & $3.0$ & & 1\,859& 152& & 99& 110& 31 \\ 
\phantom{1}8& \phantom{1}9& $3.0$ & $3.5$ & & 1\,273& 126& & 89& 85& 24 \\ 
\phantom{1}8& \phantom{1}9& $3.5$ & $4.0$ & & 1\,002& 139& & 92& 100& 26 \\ 
\phantom{1}9& 10& $1.5$ & $2.0$ & & 1\,770& 190& & 140& 120& 60 \\ 
\phantom{1}9& 10& $2.0$ & $2.5$ & & 1\,529& 127& & 87& 87& 30 \\ 
\phantom{1}9& 10& $2.5$ & $3.0$ & & 1\,142& 107& & 75& 72& 23 \\ 
\phantom{1}9& 10& $3.0$ & $3.5$ & & 864& 95& & 69& 61& 21 \\ 
\phantom{1}9& 10& $3.5$ & $4.0$ & & 544& 90& & 67& 57& 17 \\ 
10& 11& $1.5$ & $2.0$ & & 1\,070& 130& & 100& 60& 40 \\ 
10& 11& $2.0$ & $2.5$ & & 917& 88& & 66& 52& 22 \\ 
10& 11& $2.5$ & $3.0$ & & 804& 83& & 64& 47& 21 \\ 
10& 11& $3.0$ & $3.5$ & & 477& 63& & 51& 35& 14 \\ 
10& 11& $3.5$ & $4.0$ & & 397& 73& & 51& 50& 15 \\ 
11& 12& $1.5$ & $2.0$ & & 800& 100& & 80& 50& 40 \\ 
11& 12& $2.0$ & $2.5$ & & 678& 72& & 57& 38& 21 \\ 
11& 12& $2.5$ & $3.0$ & & 446& 60& & 50& 27& 15 \\ 
11& 12& $3.0$ & $3.5$ & & 386& 57& & 47& 29& 14 \\ 
11& 12& $3.5$ & $4.0$ & & 162& 41& & 37& 15& 8 \\ 
12& 13& $1.5$ & $2.0$ & & 526& 80& & 65& 33& 32 \\ 
12& 13& $2.0$ & $2.5$ & & 474& 57& & 45& 29& 16 \\ 
12& 13& $2.5$ & $3.0$ & & 370& 48& & 39& 23& 15 \\ 
12& 13& $3.0$ & $3.5$ & & 231& 40& & 34& 18& 11 \\ 
12& 13& $3.5$ & $4.0$ & & 153& 37& & 29& 21& 7 \\ 
13& 14& $1.5$ & $2.0$ & & 419& 67& & 56& 24& 28 \\ 
13& 14& $2.0$ & $2.5$ & & 387& 48& & 39& 22& 16 \\ 
13& 14& $2.5$ & $3.0$ & & 224& 35& & 31& 13& 11 \\ 
13& 14& $3.0$ & $3.5$ & & 151& 27& & 23& 11& 6 \\ 
13& 14& $3.5$ & $4.0$ & & 100& 27& & 24& 9& 7 \\ 
 
\bottomrule
\end{tabular}
\end{table}

\clearpage

\subsection{\texorpdfstring{$\bm{\frac{{\rm d}^2\sigma}{{\rm d}p_{\rm T}{\rm d}y^*}}$}{$\frac{{\rm d}^2\sigma}{{\rm d}p_{\rm T}{\rm d}y^*}$} for prompt \texorpdfstring{$\bm{J} \mskip -3mu/\mskip -2mu\bm{\psi}\mskip 2mu$}{\jpsi} in Pb\texorpdfstring{$\bm{p}$}{$p$}}\label{app:crosssectionPbpprompt}

\begin{table}[h!]
\caption{\small Prompt \jpsi absolute production cross-section in Pb$p$, as a function of \pt and $y^*$.
The quoted uncertainties are the total uncertainties, and the breakdown into statistical uncertainties, and correlated and 
uncorrelated systematic uncertainties.}
\label{tab:promptcrosssectionresultsPbp}
\small
\centering
\renewcommand{\arraystretch}{0.96}
\begin{tabular}{@{}l@{$\,<\,\pt<\,$}ll@{$\,<\,y^*<\,$}llr@{$\,\pm\,$}@{}rlrrr@{}}
\toprule
\multicolumn{2}{@{}l}{\pt bin (\gevc)}  & \multicolumn{2}{c}{$y^*$ bin} & \multicolumn{4}{c}{$\frac{{\rm d}^2\sigma}{{\rm d}p_{\rm T}{\rm d}y^*}$  [nb/(\gevc)]} & stat. & corr. & uncorr. \\
\midrule
 \phantom{1}0& \phantom{1}1& $-3.0$ & $-2.5$ & & $132\,900$& $23\,100$& & 2\,300& 22\,800& 2\,500 \\ 
\phantom{1}0& \phantom{1}1& $-3.5$ & $-3.0$ & & $114\,000$& $13\,100$& & 1\,300& 13\,000& 1\,200 \\ 
\phantom{1}0& \phantom{1}1& $-4.0$ & $-3.5$ & & $96\,600$& $9\,300$& & 1\,200& 9\,200& 900 \\ 
\phantom{1}0& \phantom{1}1& $-4.5$ & $-4.0$ & & $83\,600$& $6\,900$& & 1\,200& 6\,800& 800 \\ 
\phantom{1}0& \phantom{1}1& $-5.0$ & $-4.5$ & & $70\,500$& $6\,600$& & 1\,400& 6\,300& 1\,000 \\ 
\phantom{1}1& \phantom{1}2& $-3.0$ & $-2.5$ & & $263\,000$& $34\,800$& & 2\,900& 34\,600& 2\,800 \\ 
\phantom{1}1& \phantom{1}2& $-3.5$ & $-3.0$ & & $226\,900$& $21\,600$& & 1\,800& 21\,500& 1\,400 \\ 
\phantom{1}1& \phantom{1}2& $-4.0$ & $-3.5$ & & $188\,300$& $15\,900$& & 1\,500& 15\,800& 1\,100 \\ 
\phantom{1}1& \phantom{1}2& $-4.5$ & $-4.0$ & & $161\,400$& $12\,500$& & 1\,500& 12\,300& 1\,000 \\ 
\phantom{1}1& \phantom{1}2& $-5.0$ & $-4.5$ & & $135\,700$& $16\,300$& & 1\,800& 16\,100& 1\,200 \\ 
\phantom{1}2& \phantom{1}3& $-3.0$ & $-2.5$ & & $230\,900$& $27\,400$& & 2\,400& 27\,200& 2\,400 \\ 
\phantom{1}2& \phantom{1}3& $-3.5$ & $-3.0$ & & $198\,600$& $18\,400$& & 1\,600& 18\,300& 1\,300 \\ 
\phantom{1}2& \phantom{1}3& $-4.0$ & $-3.5$ & & $167\,000$& $13\,000$& & 1\,000& 13\,000& 1\,000 \\ 
\phantom{1}2& \phantom{1}3& $-4.5$ & $-4.0$ & & $128\,700$& $10\,600$& & 1\,300& 10\,400& 900 \\ 
\phantom{1}2& \phantom{1}3& $-5.0$ & $-4.5$ & & $98\,400$& $14\,700$& & 1\,500& 14\,600& 1\,000 \\ 
\phantom{1}3& \phantom{1}4& $-3.0$ & $-2.5$ & & $144\,600$& $18\,400$& & 1\,700& 18\,300& 1\,700 \\ 
\phantom{1}3& \phantom{1}4& $-3.5$ & $-3.0$ & & $128\,400$& $13\,000$& & 1\,100& 12\,900& 900 \\ 
\phantom{1}3& \phantom{1}4& $-4.0$ & $-3.5$ & & $104\,600$& $9\,300$& & 900& 9\,200& 700 \\ 
\phantom{1}3& \phantom{1}4& $-4.5$ & $-4.0$ & & $77\,600$& $8\,100$& & 900& 8\,000& 600 \\ 
\phantom{1}3& \phantom{1}4& $-5.0$ & $-4.5$ & & $55\,300$& $9\,400$& & 1\,000& 9\,300& 700 \\ 
\phantom{1}4& \phantom{1}5& $-3.0$ & $-2.5$ & & $83\,600$& $9\,600$& & 1\,100& 9\,500& 1\,200 \\ 
\phantom{1}4& \phantom{1}5& $-3.5$ & $-3.0$ & & $71\,400$& $6\,600$& & 700& 6\,500& 600 \\ 
\phantom{1}4& \phantom{1}5& $-4.0$ & $-3.5$ & & $55\,900$& $4\,600$& & 600& 4\,600& 500 \\ 
\phantom{1}4& \phantom{1}5& $-4.5$ & $-4.0$ & & $40\,500$& $4\,900$& & 500& 4\,800& 400 \\ 
\phantom{1}4& \phantom{1}5& $-5.0$ & $-4.5$ & & $25\,600$& $4\,600$& & 600& 4\,600& 400 \\ 
\phantom{1}5& \phantom{1}6& $-3.0$ & $-2.5$ & & $46\,600$& $5\,000$& & 700& 4\,900& 800 \\ 
\phantom{1}5& \phantom{1}6& $-3.5$ & $-3.0$ & & $37\,100$& $3\,300$& & 400& 3\,300& 400 \\ 
\phantom{1}5& \phantom{1}6& $-4.0$ & $-3.5$ & & $27\,810$& $2\,400$& & 350& 2\,350& 330 \\ 
\phantom{1}5& \phantom{1}6& $-4.5$ & $-4.0$ & & $19\,990$& $2\,770$& & 320& 2\,730& 290 \\ 
\phantom{1}5& \phantom{1}6& $-5.0$ & $-4.5$ & & $13\,540$& $2\,660$& & 380& 2\,620& 320 \\ 
\phantom{1}6& \phantom{1}7& $-3.0$ & $-2.5$ & & $22\,500$& $2\,500$& & 400& 2\,400& 500 \\ 
\phantom{1}6& \phantom{1}7& $-3.5$ & $-3.0$ & & $19\,950$& $1\,830$& & 290& 1\,780& 320 \\ 
\phantom{1}6& \phantom{1}7& $-4.0$ & $-3.5$ & & $14\,620$& $1\,440$& & 240& 1\,400& 260 \\ 
\phantom{1}6& \phantom{1}7& $-4.5$ & $-4.0$ & & $10\,330$& $1\,630$& & 220& 1\,600& 230 \\ 
\phantom{1}6& \phantom{1}7& $-5.0$ & $-4.5$ & & $5\,670$& $1\,260$& & 240& 1\,220& 200 \\ 
 
\bottomrule
\end{tabular}
\end{table}

\begin{table}[h!]
\caption{\small Prompt \jpsi absolute production cross-section in Pb$p$, as a function of \pt and $y^*$.
The quoted uncertainties are the total uncertainties, and the breakdown into statistical uncertainties, and correlated and 
uncorrelated systematic uncertainties.}
\label{tab:promptcrosssectionresultsPbp1}
\small
\centering
\renewcommand{\arraystretch}{0.96}
\begin{tabular}{@{}l@{$\,<\,\pt<\,$}ll@{$\,<\,y^*<\,$}llr@{$\,\pm\,$}@{}rlrrr@{}}
\toprule
\multicolumn{2}{@{}l}{\pt bin (\gevc)}  & \multicolumn{2}{c}{$y^*$ bin} & \multicolumn{4}{c}{$\frac{{\rm d}^2\sigma}{{\rm d}p_{\rm T}{\rm d}y^*}$  [nb/(\gevc)]} & stat. & corr. & uncorr. \\
\midrule
 \phantom{1}7& \phantom{1}8& $-3.0$ & $-2.5$ & & $12\,260$& $1\,220$& & 290& 1\,130& 350 \\ 
\phantom{1}7& \phantom{1}8& $-3.5$ & $-3.0$ & & $10\,320$& $970$& & 190& 920& 220 \\ 
\phantom{1}7& \phantom{1}8& $-4.0$ & $-3.5$ & & $7\,480$& $790$& & 170& 760& 180 \\ 
\phantom{1}7& \phantom{1}8& $-4.5$ & $-4.0$ & & $4\,760$& $820$& & 140& 800& 150 \\ 
\phantom{1}7& \phantom{1}8& $-5.0$ & $-4.5$ & & $2\,930$& $730$& & 160& 700& 140 \\ 
\phantom{1}8& \phantom{1}9& $-3.0$ & $-2.5$ & & $6\,690$& $700$& & 210& 610& 270 \\ 
\phantom{1}8& \phantom{1}9& $-3.5$ & $-3.0$ & & $5\,850$& $560$& & 150& 500& 180 \\ 
\phantom{1}8& \phantom{1}9& $-4.0$ & $-3.5$ & & $3\,970$& $510$& & 120& 480& 130 \\ 
\phantom{1}8& \phantom{1}9& $-4.5$ & $-4.0$ & & $2\,320$& $440$& & 90& 420& 100 \\ 
\phantom{1}8& \phantom{1}9& $-5.0$ & $-4.5$ & & $1\,110$& $310$& & 100& 290& 60 \\ 
\phantom{1}9& 10& $-3.0$ & $-2.5$ & & $4\,050$& $450$& & 150& 370& 210 \\ 
\phantom{1}9& 10& $-3.5$ & $-3.0$ & & $3\,000$& $300$& & 100& 300& 100 \\ 
\phantom{1}9& 10& $-4.0$ & $-3.5$ & & $1\,940$& $290$& & 80& 260& 80 \\ 
\phantom{1}9& 10& $-4.5$ & $-4.0$ & & $1\,290$& $270$& & 70& 250& 80 \\ 
\phantom{1}9& 10& $-5.0$ & $-4.5$ & & $670$& $200$& & 70& 170& 70 \\ 
10& 11& $-3.0$ & $-2.5$ & & $2\,230$& $240$& & 100& 180& 130 \\ 
10& 11& $-3.5$ & $-3.0$ & & $1\,890$& $210$& & 80& 160& 100 \\ 
10& 11& $-4.0$ & $-3.5$ & & $1\,180$& $190$& & 60& 160& 70 \\ 
10& 11& $-4.5$ & $-4.0$ & & $590$& $130$& & 40& 120& 40 \\ 
10& 11& $-5.0$ & $-4.5$ & & $297$& $99$& & 42& 82& 35 \\ 
11& 12& $-3.0$ & $-2.5$ & & $1\,300$& $160$& & 80& 100& 100 \\ 
11& 12& $-3.5$ & $-3.0$ & & $930$& $110$& & 50& 90& 50 \\ 
11& 12& $-4.0$ & $-3.5$ & & $600$& $110$& & 50& 80& 50 \\ 
11& 12& $-4.5$ & $-4.0$ & & $420$& $110$& & 40& 90& 40 \\ 
11& 12& $-5.0$ & $-4.5$ & & $210$& $80$& & 40& 50& 40 \\ 
12& 13& $-3.0$ & $-2.5$ & & $980$& $140$& & 70& 90& 90 \\ 
12& 13& $-3.5$ & $-3.0$ & & $660$& $100$& & 50& 60& 60 \\ 
12& 13& $-4.0$ & $-3.5$ & & $313$& $64$& & 32& 46& 29 \\ 
12& 13& $-4.5$ & $-4.0$ & & $229$& $67$& & 27& 50& 34 \\ 
12& 13& $-5.0$ & $-4.5$ & & $140$& $70$& & 40& 40& 50 \\ 
13& 14& $-3.0$ & $-2.5$ & & $550$& $100$& & 60& 40& 60 \\ 
13& 14& $-3.5$ & $-3.0$ & & $328$& $60$& & 40& 32& 30 \\ 
13& 14& $-4.0$ & $-3.5$ & & $248$& $64$& & 27& 37& 43 \\ 
13& 14& $-4.5$ & $-4.0$ & & $135$& $54$& & 23& 32& 36 \\ 
13& 14& $-5.0$ & $-4.5$ & & $29$& $14$& & 10& 7& 6 \\ 
 
\bottomrule
\end{tabular}
\end{table}

\clearpage

\subsection{\texorpdfstring{$\bm{\frac{{\rm d}^2\sigma}{{\rm d}p_{\rm T}{\rm d}y^*}}$}{$\frac{{\rm d}^2\sigma}{{\rm d}p_{\rm T}{\rm d}y^*}$} for \texorpdfstring{$\bm{J} \mskip -3mu/\mskip -2mu\bm{\psi}\mskip 2mu$}{\jpsi}-from-\texorpdfstring{$\bm{b}$}{$b$}-hadrons in Pb\texorpdfstring{$\bm{p}$}{$p$}}\label{app:crosssectionPbpb}

\begin{table}[h!]
\caption{\small  \jpsi-from-$b$-hadrons absolute production cross-section in Pb$p$, as a function of \pt and $y^*$.
The quoted uncertainties are the total uncertainties, and the breakdown into statistical uncertainties, and correlated and 
uncorrelated systematic uncertainties.}
\label{tab:bcrosssectionresultsPbp}
\small
\centering
\renewcommand{\arraystretch}{0.96}
\begin{tabular}{@{}l@{$\,<\,\pt<\,$}ll@{$\,<\,y^*<\,$}llr@{$\,\pm\,$}@{}rlrrr@{}}
\toprule
\multicolumn{2}{@{}l}{\pt bin (\gevc)}  & \multicolumn{2}{c}{$y^*$ bin} & \multicolumn{4}{c}{$\frac{{\rm d}^2\sigma}{{\rm d}p_{\rm T}{\rm d}y^*}$  [nb/(\gevc)]} & stat. & corr. & uncorr. \\
\midrule
 \phantom{1}0& \phantom{1}1& $-3.0$ & $-2.5$ & & $16\,120$ & $2\,890$& & 770& 2\,770& 300 \\ 
\phantom{1}0& \phantom{1}1& $-3.5$ & $-3.0$ & & $11\,760$ & $1\,400$& & 400& 1\,340& 120 \\ 
\phantom{1}0& \phantom{1}1& $-4.0$ & $-3.5$ & & $9\,270$ & $950$& & 330& 880& 80 \\ 
\phantom{1}0& \phantom{1}1& $-4.5$ & $-4.0$ & & $7\,000$ & $650$& & 320& 570& 70 \\ 
\phantom{1}0& \phantom{1}1& $-5.0$ & $-4.5$ & & $4\,750$ & $580$& & 390& 430& 70 \\ 
\phantom{1}1& \phantom{1}2& $-3.0$ & $-2.5$ & & $35\,000$ & $4\,700$& & 1\,000& 4\,600& 400 \\ 
\phantom{1}1& \phantom{1}2& $-3.5$ & $-3.0$ & & $26\,050$ & $2\,540$& & 560& 2\,470& 170 \\ 
\phantom{1}1& \phantom{1}2& $-4.0$ & $-3.5$ & & $20\,280$ & $1\,770$& & 450& 1\,710& 120 \\ 
\phantom{1}1& \phantom{1}2& $-4.5$ & $-4.0$ & & $14\,270$ & $1\,170$& & 420& 1\,090& 90 \\ 
\phantom{1}1& \phantom{1}2& $-5.0$ & $-4.5$ & & $9\,630$ & $1\,250$& & 490& 1\,140& 90 \\ 
\phantom{1}2& \phantom{1}3& $-3.0$ & $-2.5$ & & $31\,420$ & $3\,810$& & 850& 3\,700& 320 \\ 
\phantom{1}2& \phantom{1}3& $-3.5$ & $-3.0$ & & $25\,560$ & $2\,410$& & 510& 2\,350& 160 \\ 
\phantom{1}2& \phantom{1}3& $-4.0$ & $-3.5$ & & $19\,830$ & $1\,640$& & 420& 1\,580& 110 \\ 
\phantom{1}2& \phantom{1}3& $-4.5$ & $-4.0$ & & $12\,690$ & $1\,100$& & 370& 1\,030& 80 \\ 
\phantom{1}2& \phantom{1}3& $-5.0$ & $-4.5$ & & $7\,760$ & $1\,230$& & 400& 1\,150& 80 \\ 
\phantom{1}3& \phantom{1}4& $-3.0$ & $-2.5$ & & $21\,880$ & $2\,840$& & 630& 2\,760& 250 \\ 
\phantom{1}3& \phantom{1}4& $-3.5$ & $-3.0$ & & $19\,200$ & $1\,980$& & 390& 1\,940& 140 \\ 
\phantom{1}3& \phantom{1}4& $-4.0$ & $-3.5$ & & $13\,490$ & $1\,230$& & 310& 1\,190& 90 \\ 
\phantom{1}3& \phantom{1}4& $-4.5$ & $-4.0$ & & $8\,720$ & $940$& & 270& 900& 70 \\ 
\phantom{1}3& \phantom{1}4& $-5.0$ & $-4.5$ & & $4\,420$ & $800$& & 280& 750& 60 \\ 
\phantom{1}4& \phantom{1}5& $-3.0$ & $-2.5$ & & $14\,340$ & $1\,700$& & 440& 1\,630& 200 \\ 
\phantom{1}4& \phantom{1}5& $-3.5$ & $-3.0$ & & $11\,200$ & $1\,060$& & 260& 1\,020& 100 \\ 
\phantom{1}4& \phantom{1}5& $-4.0$ & $-3.5$ & & $8\,210$ & $710$& & 220& 670& 70 \\ 
\phantom{1}4& \phantom{1}5& $-4.5$ & $-4.0$ & & $4\,920$ & $620$& & 180& 590& 50 \\ 
\phantom{1}4& \phantom{1}5& $-5.0$ & $-4.5$ & & $2\,660$ & $520$& & 190& 480& 50 \\ 
\phantom{1}5& \phantom{1}6& $-3.0$ & $-2.5$ & & $7\,640$ & $870$& & 290& 800& 140 \\ 
\phantom{1}5& \phantom{1}6& $-3.5$ & $-3.0$ & & $6\,730$ & $630$& & 180& 590& 80 \\ 
\phantom{1}5& \phantom{1}6& $-4.0$ & $-3.5$ & & $4\,370$ & $400$& & 140& 370& 50 \\ 
\phantom{1}5& \phantom{1}6& $-4.5$ & $-4.0$ & & $2\,740$ & $400$& & 120& 380& 40 \\ 
\phantom{1}5& \phantom{1}6& $-5.0$ & $-4.5$ & & $1\,550$ & $330$& & 130& 300& 40 \\ 
\phantom{1}6& \phantom{1}7& $-3.0$ & $-2.5$ & & $4\,400$ & $500$& & 200& 500& 100 \\ 
\phantom{1}6& \phantom{1}7& $-3.5$ & $-3.0$ & & $3\,920$ & $380$& & 130& 350& 60 \\ 
\phantom{1}6& \phantom{1}7& $-4.0$ & $-3.5$ & & $2\,650$ & $280$& & 110& 250& 50 \\ 
\phantom{1}6& \phantom{1}7& $-4.5$ & $-4.0$ & & $1\,640$ & $270$& & 90& 250& 40 \\ 
\phantom{1}6& \phantom{1}7& $-5.0$ & $-4.5$ & & $668$ & $170$& & 89& 143& 23 \\ 
 
\bottomrule
\end{tabular}
\end{table}

\begin{table}[h!]
\caption{\small \jpsi-from-$b$-hadrons absolute production cross-section in Pb$p$, as a function of \pt and $y^*$.
The quoted uncertainties are the total uncertainties, and the breakdown into statistical uncertainties, and correlated and 
uncorrelated systematic uncertainties.}
\label{tab:bcrosssectionresultsPbp1}
\small
\centering
\renewcommand{\arraystretch}{0.96}
\begin{tabular}{@{}l@{$\,<\,\pt<\,$}ll@{$\,<\,y^*<\,$}llr@{$\,\pm\,$}@{}rlrrr@{}}
\toprule
\multicolumn{2}{@{}l}{\pt bin (\gevc)}  & \multicolumn{2}{c}{$y^*$ bin} & \multicolumn{4}{c}{$\frac{{\rm d}^2\sigma}{{\rm d}p_{\rm T}{\rm d}y^*}$  [nb/(\gevc)]} & stat. & corr. & uncorr. \\
\midrule
 \phantom{1}7& \phantom{1}8& $-3.0$ & $-2.5$ & & $2\,860$ & $310$& & 150& 260& 80 \\ 
\phantom{1}7& \phantom{1}8& $-3.5$ & $-3.0$ & & $2\,340$ & $230$& & 90& 210& 50 \\ 
\phantom{1}7& \phantom{1}8& $-4.0$ & $-3.5$ & & $1\,414$ & $165$& & 77& 142& 34 \\ 
\phantom{1}7& \phantom{1}8& $-4.5$ & $-4.0$ & & $748$ & $140$& & 58& 125& 22 \\ 
\phantom{1}7& \phantom{1}8& $-5.0$ & $-4.5$ & & $374$ & $105$& & 52& 89& 17 \\ 
\phantom{1}8& \phantom{1}9& $-3.0$ & $-2.5$ & & $2\,090$ & $240$& & 120& 190& 80 \\ 
\phantom{1}8& \phantom{1}9& $-3.5$ & $-3.0$ & & $1\,450$ & $150$& & 70& 120& 40 \\ 
\phantom{1}8& \phantom{1}9& $-4.0$ & $-3.5$ & & $812$ & $114$& & 54& 97& 25 \\ 
\phantom{1}8& \phantom{1}9& $-4.5$ & $-4.0$ & & $474$ & $99$& & 44& 86& 19 \\ 
\phantom{1}8& \phantom{1}9& $-5.0$ & $-4.5$ & & $207$ & $66$& & 36& 53& 12 \\ 
\phantom{1}9& 10& $-3.0$ & $-2.5$ & & $1\,370$ & $170$& & 90& 120& 70 \\ 
\phantom{1}9& 10& $-3.5$ & $-3.0$ & & $794$ & $97$& & 53& 76& 29 \\ 
\phantom{1}9& 10& $-4.0$ & $-3.5$ & & $488$ & $80$& & 40& 66& 21 \\ 
\phantom{1}9& 10& $-4.5$ & $-4.0$ & & $266$ & $63$& & 33& 51& 16 \\ 
\phantom{1}9& 10& $-5.0$ & $-4.5$ & & $94$ & $37$& & 26& 23& 9 \\ 
10& 11& $-3.0$ & $-2.5$ & & $700$ & $90$& & 60& 60& 40 \\ 
10& 11& $-3.5$ & $-3.0$ & & $479$ & $65$& & 44& 41& 24 \\ 
10& 11& $-4.0$ & $-3.5$ & & $372$ & $65$& & 35& 50& 22 \\ 
10& 11& $-4.5$ & $-4.0$ & & $152$ & $39$& & 22& 30& 9 \\ 
10& 11& $-5.0$ & $-4.5$ & & $42$ & $19$& & 14& 11& 5 \\ 
11& 12& $-3.0$ & $-2.5$ & & $670$ & $90$& & 60& 50& 50 \\ 
11& 12& $-3.5$ & $-3.0$ & & $373$ & $53$& & 35& 34& 21 \\ 
11& 12& $-4.0$ & $-3.5$ & & $191$ & $40$& & 26& 25& 15 \\ 
11& 12& $-4.5$ & $-4.0$ & & $57$ & $20$& & 14& 12& 5 \\ 
11& 12& $-5.0$ & $-4.5$ & & $88$ & $36$& & 22& 23& 17 \\ 
12& 13& $-3.0$ & $-2.5$ & & $440$ & $70$& & 40& 40& 40 \\ 
12& 13& $-3.5$ & $-3.0$ & & $247$ & $43$& & 29& 22& 23 \\ 
12& 13& $-4.0$ & $-3.5$ & & $107$ & $27$& & 20& 15& 10 \\ 
12& 13& $-4.5$ & $-4.0$ & & $70$ & $24$& & 16& 15& 10 \\ 
12& 13& $-5.0$ & $-4.5$ & & $19$ & $15$& & 12& 5& 7 \\ 
13& 14& $-3.0$ & $-2.5$ & & $257$ & $53$& & 39& 19& 30 \\ 
13& 14& $-3.5$ & $-3.0$ & & $146$ & $33$& & 27& 14& 13 \\ 
13& 14& $-4.0$ & $-3.5$ & & $114$ & $32$& & 19& 17& 20 \\ 
13& 14& $-4.5$ & $-4.0$ & & $40$ & $19$& & 12& 9& 10 \\ 
13& 14& $-5.0$ & $-4.5$ & & $10$ & $7$& & 6& 2& 2 \\ 
 
\bottomrule
\end{tabular}
\end{table}

\clearpage

\subsection{Fraction of \texorpdfstring{$\bm{J} \mskip -3mu/\mskip -2mu\bm{\psi}\mskip 2mu$}{\jpsi}-from-\texorpdfstring{$\bm{b}$}{$b$}-hadrons in \texorpdfstring{$\bm{p}$}{$p$}Pb}\label{app:fb}

\begin{table}[h!]
\caption{\small Fraction of \jpsi-from-$b$-hadrons, $f_b$, in $p$Pb in bins of \pt and $y^*$. The uncertainty is the quadratic sum of statistical and
systematic uncertainties.}
\label{tab:bfractionresultspPb}
\small
\centering
\renewcommand{\arraystretch}{0.96}
\setlength{\tabcolsep}{18pt}
\begin{tabular}{@{}l@{$\,<\,\pt<\,$}ll@{$\,<\,y^*<\,$}lr@{$\,\pm\,$}r@{}}
\toprule
\multicolumn{2}{@{}l}{\pt bin (\gevc)}  & \multicolumn{2}{c}{$y^*$ bin} & \multicolumn{2}{c}{$f_b$} \\
\midrule
 \phantom{1}0& \phantom{1}1& $1.5$ & $2.0$ & 0.13& 0.01 \\ 
\phantom{1}0& \phantom{1}1& $2.0$ & $2.5$ & 0.12& 0.01 \\ 
\phantom{1}0& \phantom{1}1& $2.5$ & $3.0$ & 0.11& 0.01 \\ 
\phantom{1}0& \phantom{1}1& $3.0$ & $3.5$ & 0.11& 0.01 \\ 
\phantom{1}0& \phantom{1}1& $3.5$ & $4.0$ & 0.10& 0.01 \\ 
\phantom{1}1& \phantom{1}2& $1.5$ & $2.0$ & 0.13& 0.01 \\ 
\phantom{1}1& \phantom{1}2& $2.0$ & $2.5$ & 0.13& 0.01 \\ 
\phantom{1}1& \phantom{1}2& $2.5$ & $3.0$ & 0.13& 0.01 \\ 
\phantom{1}1& \phantom{1}2& $3.0$ & $3.5$ & 0.12& 0.01 \\ 
\phantom{1}1& \phantom{1}2& $3.5$ & $4.0$ & 0.12& 0.01 \\ 
\phantom{1}2& \phantom{1}3& $1.5$ & $2.0$ & 0.15& 0.01 \\ 
\phantom{1}2& \phantom{1}3& $2.0$ & $2.5$ & 0.15& 0.01 \\ 
\phantom{1}2& \phantom{1}3& $2.5$ & $3.0$ & 0.14& 0.01 \\ 
\phantom{1}2& \phantom{1}3& $3.0$ & $3.5$ & 0.14& 0.01 \\ 
\phantom{1}2& \phantom{1}3& $3.5$ & $4.0$ & 0.12& 0.01 \\ 
\phantom{1}3& \phantom{1}4& $1.5$ & $2.0$ & 0.15& 0.01 \\ 
\phantom{1}3& \phantom{1}4& $2.0$ & $2.5$ & 0.16& 0.01 \\ 
\phantom{1}3& \phantom{1}4& $2.5$ & $3.0$ & 0.15& 0.01 \\ 
\phantom{1}3& \phantom{1}4& $3.0$ & $3.5$ & 0.14& 0.01 \\ 
\phantom{1}3& \phantom{1}4& $3.5$ & $4.0$ & 0.13& 0.01 \\ 
\phantom{1}4& \phantom{1}5& $1.5$ & $2.0$ & 0.16& 0.01 \\ 
\phantom{1}4& \phantom{1}5& $2.0$ & $2.5$ & 0.17& 0.01 \\ 
\phantom{1}4& \phantom{1}5& $2.5$ & $3.0$ & 0.16& 0.01 \\ 
\phantom{1}4& \phantom{1}5& $3.0$ & $3.5$ & 0.15& 0.01 \\ 
\phantom{1}4& \phantom{1}5& $3.5$ & $4.0$ & 0.14& 0.01 \\ 
\phantom{1}5& \phantom{1}6& $1.5$ & $2.0$ & 0.18& 0.01 \\ 
\phantom{1}5& \phantom{1}6& $2.0$ & $2.5$ & 0.18& 0.01 \\ 
\phantom{1}5& \phantom{1}6& $2.5$ & $3.0$ & 0.18& 0.01 \\ 
\phantom{1}5& \phantom{1}6& $3.0$ & $3.5$ & 0.17& 0.01 \\ 
\phantom{1}5& \phantom{1}6& $3.5$ & $4.0$ & 0.15& 0.01 \\ 
\phantom{1}6& \phantom{1}7& $1.5$ & $2.0$ & 0.21& 0.01 \\ 
\phantom{1}6& \phantom{1}7& $2.0$ & $2.5$ & 0.19& 0.01 \\ 
\phantom{1}6& \phantom{1}7& $2.5$ & $3.0$ & 0.20& 0.01 \\ 
\phantom{1}6& \phantom{1}7& $3.0$ & $3.5$ & 0.19& 0.01 \\ 
\phantom{1}6& \phantom{1}7& $3.5$ & $4.0$ & 0.15& 0.01 \\ 
 
\bottomrule
\end{tabular}
\end{table}

\begin{table}[h!]
\caption{\small Fraction of \jpsi-from-$b$-hadrons, $f_b$, in $p$Pb in bins of \pt and $y^*$.  The uncertainty is the quadratic sum of statistical and systematic uncertainties.}
\label{tab:bfractionresultspPb1}
\small
\centering
\renewcommand{\arraystretch}{0.96}
\setlength{\tabcolsep}{18pt}
\begin{tabular}{@{}l@{$\,<\,\pt<\,$}ll@{$\,<\,y^*<\,$}lr@{$\,\pm\,$}r@{}}
\toprule
\multicolumn{2}{@{}l}{\pt bin (\gevc)}  & \multicolumn{2}{c}{$y^*$ bin} & \multicolumn{2}{c}{$f_b$} \\\midrule
 \phantom{1}7& \phantom{1}8& $1.5$ & $2.0$ & 0.24& 0.01 \\ 
\phantom{1}7& \phantom{1}8& $2.0$ & $2.5$ & 0.22& 0.01 \\ 
\phantom{1}7& \phantom{1}8& $2.5$ & $3.0$ & 0.23& 0.01 \\ 
\phantom{1}7& \phantom{1}8& $3.0$ & $3.5$ & 0.21& 0.01 \\ 
\phantom{1}7& \phantom{1}8& $3.5$ & $4.0$ & 0.19& 0.01 \\ 
\phantom{1}8& \phantom{1}9& $1.5$ & $2.0$ & 0.25& 0.02 \\ 
\phantom{1}8& \phantom{1}9& $2.0$ & $2.5$ & 0.24& 0.01 \\ 
\phantom{1}8& \phantom{1}9& $2.5$ & $3.0$ & 0.23& 0.01 \\ 
\phantom{1}8& \phantom{1}9& $3.0$ & $3.5$ & 0.22& 0.01 \\ 
\phantom{1}8& \phantom{1}9& $3.5$ & $4.0$ & 0.21& 0.02 \\ 
\phantom{1}9& 10& $1.5$ & $2.0$ & 0.27& 0.02 \\ 
\phantom{1}9& 10& $2.0$ & $2.5$ & 0.26& 0.01 \\ 
\phantom{1}9& 10& $2.5$ & $3.0$ & 0.25& 0.01 \\ 
\phantom{1}9& 10& $3.0$ & $3.5$ & 0.24& 0.02 \\ 
\phantom{1}9& 10& $3.5$ & $4.0$ & 0.20& 0.02 \\ 
10& 11& $1.5$ & $2.0$ & 0.29& 0.02 \\ 
10& 11& $2.0$ & $2.5$ & 0.26& 0.02 \\ 
10& 11& $2.5$ & $3.0$ & 0.26& 0.02 \\ 
10& 11& $3.0$ & $3.5$ & 0.24& 0.02 \\ 
10& 11& $3.5$ & $4.0$ & 0.26& 0.03 \\ 
11& 12& $1.5$ & $2.0$ & 0.30& 0.03 \\ 
11& 12& $2.0$ & $2.5$ & 0.30& 0.02 \\ 
11& 12& $2.5$ & $3.0$ & 0.25& 0.03 \\ 
11& 12& $3.0$ & $3.5$ & 0.28& 0.03 \\ 
11& 12& $3.5$ & $4.0$ & 0.18& 0.04 \\ 
12& 13& $1.5$ & $2.0$ & 0.31& 0.03 \\ 
12& 13& $2.0$ & $2.5$ & 0.33& 0.03 \\ 
12& 13& $2.5$ & $3.0$ & 0.32& 0.03 \\ 
12& 13& $3.0$ & $3.5$ & 0.30& 0.04 \\ 
12& 13& $3.5$ & $4.0$ & 0.26& 0.04 \\ 
13& 14& $1.5$ & $2.0$ & 0.36& 0.04 \\ 
13& 14& $2.0$ & $2.5$ & 0.39& 0.03 \\ 
13& 14& $2.5$ & $3.0$ & 0.32& 0.04 \\ 
13& 14& $3.0$ & $3.5$ & 0.30& 0.04 \\ 
13& 14& $3.5$ & $4.0$ & 0.29& 0.06 \\ 
 
\bottomrule
\end{tabular}
\end{table}

\clearpage
\subsection{Fraction of \texorpdfstring{$\bm{J} \mskip -3mu/\mskip -2mu\bm{\psi}\mskip 2mu$}{\jpsi}-from-\texorpdfstring{$\bm{b}$}{$b$}-hadrons in Pb\texorpdfstring{$\bm{p}$}{$p$}}\label{app:fb1}

\begin{table}[h!]
\caption{\small Fraction of \jpsi-from-$b$-hadrons, $f_b$, in Pb$p$ in bins of \pt and $y^*$.  The uncertainty is the quadratic sum of statistical and systematic uncertainties.}
\label{tab:bfractionresultsPbp}
\small
\centering
\renewcommand{\arraystretch}{0.96}
\setlength{\tabcolsep}{18pt}
\begin{tabular}{@{}l@{$\,<\,\pt<\,$}ll@{$\,<\,y^*<\,$}lr@{$\,\pm\,$}r@{}}
\toprule
\multicolumn{2}{@{}l}{\pt bin (\gevc)}  & \multicolumn{2}{c}{$y^*$ bin} & \multicolumn{2}{c}{$f_b$} \\
\midrule
 \phantom{1}0& \phantom{1}1& $-3.0$ & $-2.5$ & 0.11& 0.01 \\ 
\phantom{1}0& \phantom{1}1& $-3.5$ & $-3.0$ & 0.09& 0.01 \\ 
\phantom{1}0& \phantom{1}1& $-4.0$ & $-3.5$ & 0.09& 0.01 \\ 
\phantom{1}0& \phantom{1}1& $-4.5$ & $-4.0$ & 0.08& 0.01 \\ 
\phantom{1}0& \phantom{1}1& $-5.0$ & $-4.5$ & 0.06& 0.01 \\ 
\phantom{1}1& \phantom{1}2& $-3.0$ & $-2.5$ & 0.12& 0.01 \\ 
\phantom{1}1& \phantom{1}2& $-3.5$ & $-3.0$ & 0.10& 0.01 \\ 
\phantom{1}1& \phantom{1}2& $-4.0$ & $-3.5$ & 0.10& 0.01 \\ 
\phantom{1}1& \phantom{1}2& $-4.5$ & $-4.0$ & 0.08& 0.01 \\ 
\phantom{1}1& \phantom{1}2& $-5.0$ & $-4.5$ & 0.07& 0.01 \\ 
\phantom{1}2& \phantom{1}3& $-3.0$ & $-2.5$ & 0.12& 0.01 \\ 
\phantom{1}2& \phantom{1}3& $-3.5$ & $-3.0$ & 0.11& 0.01 \\ 
\phantom{1}2& \phantom{1}3& $-4.0$ & $-3.5$ & 0.11& 0.01 \\ 
\phantom{1}2& \phantom{1}3& $-4.5$ & $-4.0$ & 0.09& 0.01 \\ 
\phantom{1}2& \phantom{1}3& $-5.0$ & $-4.5$ & 0.07& 0.01 \\ 
\phantom{1}3& \phantom{1}4& $-3.0$ & $-2.5$ & 0.13& 0.01 \\ 
\phantom{1}3& \phantom{1}4& $-3.5$ & $-3.0$ & 0.13& 0.01 \\ 
\phantom{1}3& \phantom{1}4& $-4.0$ & $-3.5$ & 0.11& 0.01 \\ 
\phantom{1}3& \phantom{1}4& $-4.5$ & $-4.0$ & 0.10& 0.01 \\ 
\phantom{1}3& \phantom{1}4& $-5.0$ & $-4.5$ & 0.07& 0.01 \\ 
\phantom{1}4& \phantom{1}5& $-3.0$ & $-2.5$ & 0.15& 0.01 \\ 
\phantom{1}4& \phantom{1}5& $-3.5$ & $-3.0$ & 0.14& 0.01 \\ 
\phantom{1}4& \phantom{1}5& $-4.0$ & $-3.5$ & 0.13& 0.01 \\ 
\phantom{1}4& \phantom{1}5& $-4.5$ & $-4.0$ & 0.11& 0.01 \\ 
\phantom{1}4& \phantom{1}5& $-5.0$ & $-4.5$ & 0.09& 0.01 \\ 
\phantom{1}5& \phantom{1}6& $-3.0$ & $-2.5$ & 0.14& 0.01 \\ 
\phantom{1}5& \phantom{1}6& $-3.5$ & $-3.0$ & 0.15& 0.01 \\ 
\phantom{1}5& \phantom{1}6& $-4.0$ & $-3.5$ & 0.14& 0.01 \\ 
\phantom{1}5& \phantom{1}6& $-4.5$ & $-4.0$ & 0.12& 0.01 \\ 
\phantom{1}5& \phantom{1}6& $-5.0$ & $-4.5$ & 0.10& 0.01 \\ 
\phantom{1}6& \phantom{1}7& $-3.0$ & $-2.5$ & 0.16& 0.01 \\ 
\phantom{1}6& \phantom{1}7& $-3.5$ & $-3.0$ & 0.16& 0.01 \\ 
\phantom{1}6& \phantom{1}7& $-4.0$ & $-3.5$ & 0.15& 0.01 \\ 
\phantom{1}6& \phantom{1}7& $-4.5$ & $-4.0$ & 0.14& 0.01 \\ 
\phantom{1}6& \phantom{1}7& $-5.0$ & $-4.5$ & 0.11& 0.01 \\ 
 
\bottomrule
\end{tabular}
\end{table}

\begin{table}[h!]
\caption{\small Fraction of \jpsi-from-$b$-hadrons, $f_b$, in Pb$p$ in bins of \pt and $y^*$.  The uncertainty is the quadratic sum of statistical and systematic uncertainties.}
\label{tab:bfractionresultsPbp1}
\small
\centering
\renewcommand{\arraystretch}{0.96}
\setlength{\tabcolsep}{18pt}
\begin{tabular}{@{}l@{$\,<\,\pt<\,$}ll@{$\,<\,y^*<\,$}lr@{$\,\pm\,$}r@{}}
\toprule
\multicolumn{2}{@{}l}{\pt bin (\gevc)}  & \multicolumn{2}{c}{$y^*$ bin} & \multicolumn{2}{c}{$f_b$} \\
\midrule
 \phantom{1}7& \phantom{1}8& $-3.0$ & $-2.5$ & 0.19& 0.01 \\ 
\phantom{1}7& \phantom{1}8& $-3.5$ & $-3.0$ & 0.18& 0.01 \\ 
\phantom{1}7& \phantom{1}8& $-4.0$ & $-3.5$ & 0.16& 0.01 \\ 
\phantom{1}7& \phantom{1}8& $-4.5$ & $-4.0$ & 0.14& 0.01 \\ 
\phantom{1}7& \phantom{1}8& $-5.0$ & $-4.5$ & 0.11& 0.02 \\ 
\phantom{1}8& \phantom{1}9& $-3.0$ & $-2.5$ & 0.24& 0.01 \\ 
\phantom{1}8& \phantom{1}9& $-3.5$ & $-3.0$ & 0.20& 0.01 \\ 
\phantom{1}8& \phantom{1}9& $-4.0$ & $-3.5$ & 0.17& 0.01 \\ 
\phantom{1}8& \phantom{1}9& $-4.5$ & $-4.0$ & 0.17& 0.01 \\ 
\phantom{1}8& \phantom{1}9& $-5.0$ & $-4.5$ & 0.16& 0.03 \\ 
\phantom{1}9& 10& $-3.0$ & $-2.5$ & 0.25& 0.01 \\ 
\phantom{1}9& 10& $-3.5$ & $-3.0$ & 0.21& 0.01 \\ 
\phantom{1}9& 10& $-4.0$ & $-3.5$ & 0.20& 0.02 \\ 
\phantom{1}9& 10& $-4.5$ & $-4.0$ & 0.17& 0.02 \\ 
\phantom{1}9& 10& $-5.0$ & $-4.5$ & 0.12& 0.03 \\ 
10& 11& $-3.0$ & $-2.5$ & 0.24& 0.02 \\ 
10& 11& $-3.5$ & $-3.0$ & 0.20& 0.02 \\ 
10& 11& $-4.0$ & $-3.5$ & 0.24& 0.02 \\ 
10& 11& $-4.5$ & $-4.0$ & 0.21& 0.03 \\ 
10& 11& $-5.0$ & $-4.5$ & 0.13& 0.04 \\ 
11& 12& $-3.0$ & $-2.5$ & 0.34& 0.02 \\ 
11& 12& $-3.5$ & $-3.0$ & 0.29& 0.02 \\ 
11& 12& $-4.0$ & $-3.5$ & 0.24& 0.03 \\ 
11& 12& $-4.5$ & $-4.0$ & 0.12& 0.03 \\ 
11& 12& $-5.0$ & $-4.5$ & 0.30& 0.07 \\ 
12& 13& $-3.0$ & $-2.5$ & 0.31& 0.03 \\ 
12& 13& $-3.5$ & $-3.0$ & 0.27& 0.03 \\ 
12& 13& $-4.0$ & $-3.5$ & 0.25& 0.04 \\ 
12& 13& $-4.5$ & $-4.0$ & 0.24& 0.05 \\ 
12& 13& $-5.0$ & $-4.5$ & 0.12& 0.07 \\ 
13& 14& $-3.0$ & $-2.5$ & 0.32& 0.05 \\ 
13& 14& $-3.5$ & $-3.0$ & 0.31& 0.06 \\ 
13& 14& $-4.0$ & $-3.5$ & 0.32& 0.04 \\ 
13& 14& $-4.5$ & $-4.0$ & 0.23& 0.06 \\ 
13& 14& $-5.0$ & $-4.5$ & 0.27& 0.14 \\ 
 
\bottomrule
\end{tabular}
\end{table}

\clearpage

\section{Nuclear modification factor numerical results}\label{app:rpa}

\subsection{\texorpdfstring{$\bm{R_{p{\rm Pb}}}$}{$R_{p{\rm Pb}}$} for prompt \texorpdfstring{$\bm{J} \mskip -3mu/\mskip -2mu\bm{\psi}\mskip 2mu$}{\jpsi}}

\begin{table}[h!]
\caption{\small Prompt \jpsi nuclear modification factor, $R_{p{\rm Pb}}$, in $p$Pb and Pb$p$ as a function of \pt integrated over $y^*$
in the range $1.5<y^*<4.0$ for $p$Pb and $-5.0<y^*<-2.5$ for Pb$p$.
The quoted uncertainties are the quadratic sums of statistical and systematic uncertainties.}
\label{tab:rpAppPb}
\small
\centering
\renewcommand{\arraystretch}{1.2}
\begin{tabular}{@{}l@{$\,<\,\pt<\,$}lr@{$\,\pm\,$}lr@{$\,\pm\,$}l@{}}
\toprule
\multicolumn{2}{@{}c}{\pt bin (\gevc)}  &  \multicolumn{2}{c}{$R_{p{\rm Pb}}$ in $p$Pb} & \multicolumn{2}{c}{$R_{p{\rm Pb}}$ in Pb$p$} \\
\midrule
 \phantom{1}0&\phantom{1}1&0.53&0.06&0.75&0.10 \\
\phantom{1}1&\phantom{1}2&0.56&0.06&0.81&0.10 \\
\phantom{1}2&\phantom{1}3&0.65&0.06&0.93&0.12 \\
\phantom{1}3&\phantom{1}4&0.72&0.07&0.99&0.14 \\
\phantom{1}4&\phantom{1}5&0.76&0.08&1.02&0.15 \\
\phantom{1}5&\phantom{1}6&0.81&0.08&1.06&0.16 \\
\phantom{1}6&\phantom{1}7&0.86&0.09&1.08&0.18 \\
\phantom{1}7&\phantom{1}8&0.87&0.10&1.06&0.18 \\
\phantom{1}8&\phantom{1}9&0.88&0.10&1.06&0.19 \\
\phantom{1}9&10&0.92&0.11&1.07&0.15 \\
10&11&0.89&0.11&1.02&0.14 \\
11&12&1.00&0.12&0.97&0.14 \\
12&13&0.92&0.13&1.07&0.17 \\
13&14&0.83&0.13&0.89&0.15 \\
 
\bottomrule
\end{tabular}
\end{table}

\begin{table}[h!]
\caption{\small Prompt \jpsi nuclear modification factor, $R_{p{\rm Pb}}$, in $p$Pb and Pb$p$ as a function of $y^*$ integrated over \pt
in the range $0<\pt<14\gevc$.
The quoted uncertainties are the quadratic sums of statistical and systematic uncertainties.}
\label{tab:rpAppPb1}
\small
\centering
\renewcommand{\arraystretch}{1.2}
\begin{tabular}{@{}l@{$\,<\,y^*<\,$}lr@{$\,\pm\,$}r@{}}
\toprule
\multicolumn{2}{c}{$y^*$ bin} & \multicolumn{2}{c}{$R_{p{\rm Pb}}$} \\
\midrule
 $-4.5$ & $-4.0$ & 0.86& 0.10 \\ 
$-4.0$ & $-3.5$ & 0.84& 0.09 \\ 
$-3.5$ & $-3.0$ & 0.87& 0.10 \\ 
$-3.0$ & $-2.5$ & 0.90& 0.13 \\ 
$\phantom{-}1.5$ & $\phantom{-}2.0$ & 0.68& 0.09 \\ 
$\phantom{-}2.0$ & $\phantom{-}2.5$ & 0.71& 0.07 \\ 
$\phantom{-}2.5$ & $\phantom{-}3.0$ & 0.62& 0.06 \\ 
$\phantom{-}3.0$ & $\phantom{-}3.5$ & 0.59& 0.05 \\ 
$\phantom{-}3.5$ & $\phantom{-}4.0$ & 0.57& 0.05 \\ 
 
\bottomrule
\end{tabular}
\end{table}

\clearpage

\subsection{\texorpdfstring{$\bm{R_{p{\rm Pb}}}$}{$R_{p{\rm Pb}}$} for \texorpdfstring{$\bm{J} \mskip -3mu/\mskip -2mu\bm{\psi}\mskip 2mu$}{\jpsi}-from-\texorpdfstring{$\bm{b}$}{$b$}-hadrons}

\begin{table}[h!]
\caption{\small \jpsi-from-$b$-hadrons nuclear modification factor, $R_{p{\rm Pb}}$, in $p$Pb and Pb$p$ as a function of \pt integrated over $y^*$
in the range $1.5<y^*<4.0$ for $p$Pb and $-5.0<y^*<-2.5$ for Pb$p$. The quoted uncertainties are the quadratic sums of statistical and systematic uncertainties.}
\label{tab:rpAppPb_b}
\small
\centering
\renewcommand{\arraystretch}{1.2}
\begin{tabular}{@{}l@{$\,<\,\pt<\,$}lr@{$\,\pm\,$}lr@{$\,\pm\,$}l@{}}
\toprule
\multicolumn{2}{@{}c}{\pt bin (\gevc)}  &  \multicolumn{2}{c}{$R_{p{\rm Pb}}$ in $p$Pb} & \multicolumn{2}{c}{$R_{p{\rm Pb}}$ in Pb$p$} \\
\midrule
 \phantom{1}0&\phantom{1}1&0.75&0.12&1.05&0.19 \\
\phantom{1}1&\phantom{1}2&0.79&0.09&1.05&0.16 \\
\phantom{1}2&\phantom{1}3&0.82&0.09&1.07&0.17 \\
\phantom{1}3&\phantom{1}4&0.85&0.10&1.09&0.18 \\
\phantom{1}4&\phantom{1}5&0.87&0.10&1.12&0.20 \\
\phantom{1}5&\phantom{1}6&0.91&0.11&1.05&0.13 \\
\phantom{1}6&\phantom{1}7&0.91&0.12&1.02&0.14 \\
\phantom{1}7&\phantom{1}8&0.99&0.13&0.99&0.13 \\
\phantom{1}8&\phantom{1}9&0.94&0.14&1.04&0.14 \\
\phantom{1}9&10&0.94&0.14&0.99&0.15 \\
10&11&0.91&0.15&0.91&0.14 \\
11&12&0.87&0.13&1.11&0.18 \\
12&13&0.89&0.16&0.97&0.18 \\
13&14&0.96&0.21&0.94&0.19 \\
 
\bottomrule
\end{tabular}
\end{table}

\begin{table}[h!]
\caption{\small \jpsi-from-$b$-hadrons nuclear modification factor, $R_{p{\rm Pb}}$, in $p$Pb as a function of $y^*$ integrated over \pt
in the range $0<\pt<14\gevc$.
The quoted uncertainties are the quadratic sums of statistical and systematic uncertainties.}
\label{tab:rpAppPb1_b}
\small
\centering
\renewcommand{\arraystretch}{1.2}
\begin{tabular}{@{}l@{$\,<\,y^*<\,$}lr@{$\,\pm\,$}r@{}}
\toprule
\multicolumn{2}{c}{$y^*$ bin} & \multicolumn{2}{c}{$R_{p{\rm Pb}}$} \\
\midrule
 $-4.5$ & $-4.0$ & 1.10& 0.13 \\ 
$-4.0$ & $-3.5$ & 1.03& 0.11 \\ 
$-3.5$ & $-3.0$ & 0.97& 0.11 \\ 
$-3.0$ & $-2.5$ & 1.00& 0.14 \\ 
$\phantom{-}1.5$ & $\phantom{-}2.0$ & 0.84& 0.17 \\ 
$\phantom{-}2.0$ & $\phantom{-}2.5$ & 0.89& 0.09 \\ 
$\phantom{-}2.5$ & $\phantom{-}3.0$ & 0.80& 0.07 \\ 
$\phantom{-}3.0$ & $\phantom{-}3.5$ & 0.80& 0.07 \\ 
$\phantom{-}3.5$ & $\phantom{-}4.0$ & 0.82& 0.08 \\ 
 
\bottomrule
\end{tabular}
\end{table}

\clearpage

\section{Forward-to-backward ratios numerical results}\label{app:rfb}

\subsection{\texorpdfstring{$\bm{R_{{\rm FB}}}$}{$R_{{\rm FB}}$} for prompt \texorpdfstring{$\bm{J} \mskip -3mu/\mskip -2mu\bm{\psi}\mskip 2mu$}{\jpsi}}

\begin{table}[h!]
\caption{\small Prompt \jpsi forward-to-backward ratio, $R_{{\rm FB}}$, as a function of \pt integrated over $|y^*|$
in the range $2.5<|y^*|<4.0$.  
The quoted uncertainties are the quadratic sums of statistical and systematic uncertainties.}
\label{tab:rfb}
\small
\centering
\renewcommand{\arraystretch}{1.3}
\begin{tabular}{@{}l@{$\,<\,\pt<\,$}lr@{$\,\pm\,$}r@{}}
\toprule
\multicolumn{2}{@{}c}{\pt bin (\gevc)}  & \multicolumn{2}{c}{$R_{{\rm FB}}$} \\
\midrule
 \phantom{1}0& \phantom{1}1& 0.62& 0.07 \\ 
\phantom{1}1& \phantom{1}2& 0.64& 0.06 \\ 
\phantom{1}2& \phantom{1}3& 0.67& 0.06 \\ 
\phantom{1}3& \phantom{1}4& 0.70& 0.06 \\ 
\phantom{1}4& \phantom{1}5& 0.73& 0.07 \\ 
\phantom{1}5& \phantom{1}6& 0.77& 0.07 \\ 
\phantom{1}6& \phantom{1}7& 0.82& 0.07 \\ 
\phantom{1}7& \phantom{1}8& 0.85& 0.08 \\ 
\phantom{1}8& \phantom{1}9& 0.87& 0.09 \\ 
\phantom{1}9& 10& 0.92& 0.10 \\ 
10& 11& 0.92& 0.10 \\ 
11& 12& 1.08& 0.13 \\ 
12& 13& 0.90& 0.12 \\ 
13& 14& 0.95& 0.15 \\ 
 
\bottomrule
\end{tabular}
\end{table}

\begin{table}[h!]
\caption{\small Prompt \jpsi forward-to-backward ratio, $R_{{\rm FB}}$, as a function of $y^*$ integrated over \pt
in the range $0<\pt<14\gevc$.
The quoted uncertainties are the quadratic sums of statistical and systematic uncertainties.}
\label{tab:rfb1}
\small
\centering
\renewcommand{\arraystretch}{1.3}
\begin{tabular}{@{}l@{$\,<\,y^*<\,$}lr@{$\,\pm\,$}r@{}}
\toprule
\multicolumn{2}{c}{$y^*$ bin}  & \multicolumn{2}{c}{$R_{{\rm FB}}$} \\
\midrule
 2.5 & 3.0 & 0.69& 0.08 \\ 
3.0 & 3.5 & 0.67& 0.06 \\ 
3.5 & 4.0 & 0.67& 0.05 \\ 
 
\bottomrule
\end{tabular}
\end{table}

\clearpage
\subsection{\texorpdfstring{$\bm{R_{{\rm FB}}}$}{$R_{{\rm FB}}$} for \texorpdfstring{$\bm{J} \mskip -3mu/\mskip -2mu\bm{\psi}\mskip 2mu$}{\jpsi}-from-\texorpdfstring{$\bm{b}$}{$b$}-hadrons}

\begin{table}[h!]
\caption{\small \jpsi-from-$b$-hadrons forward-to-backward ratio, $R_{{\rm FB}}$, as a function of \pt integrated over $|y^*|$
in the range $2.5<|y^*|<4.0$.
The quoted uncertainties are the quadratic sums of statistical and systematic uncertainties.}
\label{tab:rfbb}
\small
\centering
\renewcommand{\arraystretch}{1.3}
\begin{tabular}{@{}l@{$\,<\,\pt<\,$}lr@{$\,\pm\,$}r@{}}
\toprule
\multicolumn{2}{@{}c}{\pt bin (\gevc)}  & \multicolumn{2}{c}{$R_{{\rm FB}}$} \\
\midrule
 \phantom{1}0& \phantom{1}1& 0.72& 0.08 \\ 
\phantom{1}1& \phantom{1}2& 0.76& 0.08 \\ 
\phantom{1}2& \phantom{1}3& 0.79& 0.07 \\ 
\phantom{1}3& \phantom{1}4& 0.80& 0.08 \\ 
\phantom{1}4& \phantom{1}5& 0.82& 0.08 \\ 
\phantom{1}5& \phantom{1}6& 0.93& 0.09 \\ 
\phantom{1}6& \phantom{1}7& 0.94& 0.09 \\ 
\phantom{1}7& \phantom{1}8& 1.02& 0.10 \\ 
\phantom{1}8& \phantom{1}9& 0.95& 0.10 \\ 
\phantom{1}9& 10& 0.96& 0.11 \\ 
10& 11& 1.09& 0.14 \\ 
11& 12& 0.81& 0.12 \\ 
12& 13& 0.94& 0.15 \\ 
13& 14& 0.92& 0.17 \\ 
 
\bottomrule
\end{tabular}
\end{table}

\begin{table}[h!]
\caption{\small \jpsi-from-$b$-hadrons forward-to-backward ratio, $R_{{\rm FB}}$, as a function of $y^*$ integrated over \pt
in the range $0<\pt<14\gevc$.
The quoted uncertainties are the quadratic sums of statistical and systematic uncertainties.}
\label{tab:rfbb1}
\small
\centering
\renewcommand{\arraystretch}{1.3}
\begin{tabular}{@{}l@{$\,<\,y^*<\,$}lr@{$\,\pm\,$}r@{}}
\toprule
\multicolumn{2}{c}{$y^*$ bin}  & \multicolumn{2}{c}{$R_{{\rm FB}}$} \\
\midrule
 2.5 & 3.0 & 0.80& 0.09 \\ 
3.0 & 3.5 & 0.82& 0.07 \\ 
3.5 & 4.0 & 0.79& 0.07 \\ 
 
\bottomrule
\end{tabular}
\end{table}


\clearpage
\addcontentsline{toc}{section}{References}
\bibliographystyle{LHCb}
\bibliography{main,LHCb-PAPER,LHCb-CONF,LHCb-DP,LHCb-TDR,LHCb-Phys}


 
\newpage
\centerline{\large\bf LHCb collaboration}
\begin{flushleft}
\small
R.~Aaij$^{40}$,
B.~Adeva$^{39}$,
M.~Adinolfi$^{48}$,
Z.~Ajaltouni$^{5}$,
S.~Akar$^{59}$,
J.~Albrecht$^{10}$,
F.~Alessio$^{40}$,
M.~Alexander$^{53}$,
A.~Alfonso~Albero$^{38}$,
S.~Ali$^{43}$
% ...

{\footnotesize \it
% ...
$ ^{5}$Clermont Universit{\'e}, Universit{\'e} Blaise Pascal, CNRS/IN2P3, LPC, Clermont-Ferrand, France\\
% ...
$ ^{10}$Fakult{\"a}t Physik, Technische Universit{\"a}t Dortmund, Dortmund, Germany\\
% ...
$ ^{38}$ICCUB, Universitat de Barcelona, Barcelona, Spain\\
$ ^{39}$Universidad de Santiago de Compostela, Santiago de Compostela, Spain\\
$ ^{40}$European Organization for Nuclear Research (CERN), Geneva, Switzerland\\
% ...
$ ^{43}$Nikhef National Institute for Subatomic Physics, Amsterdam, The Netherlands\\
% ...
$ ^{48}$H.H. Wills Physics Laboratory, University of Bristol, Bristol, United Kingdom\\
% ...
$ ^{53}$School of Physics and Astronomy, University of Glasgow, Glasgow, United Kingdom\\
% ...
$ ^{59}$University of Cincinnati, Cincinnati, OH, United States\\
% ...
}
\end{flushleft}

\end{document}